\documentclass[twocolumn]{aastex631}

\usepackage{graphicx}
\usepackage[caption=false]{subfig}	
\usepackage{longtable}
\usepackage{latexsym}
\usepackage{amsmath}
\usepackage{comment}
\usepackage{scalerel}

\newcommand{\wise}{\textit{WISE}}
\newcommand{\WISE}{\textit{WISE}}
\newcommand{\MIR}{mid-IR}

\newcommand{\Mstar}{M$_{\textrm{stellar}}$}

\renewcommand{\micron}{\,$\mu$m}

\newcommand{\arcs}{\arcsec }

\newcommand{\Lw}{L$_{\rm W1}$}

\newcommand{\WIWII}{$\rm W1{-}\rm W2$}

\newcommand{\WIWIII}{$\rm W1{-}\rm W3$}

\newcommand{\spitzerg}{S$^4$G}

\shorttitle{S$^4$G-WISE}
\shortauthors{Cluver, Jarrett et al.}

\graphicspath{{ }}

\begin{document}

\title{The S$^4$G-WISE View of Global Star Formation in the Nearby Universe
}

\author[0000-0002-9871-6490]{M.E. Cluver}
\affiliation{Centre for Astrophysics and Supercomputing, Swinburne University of Technology,
John Street, Hawthorn, 3122, Australia}
\affiliation{Department of Physics and Astronomy, University of the Western Cape, Robert Sobukwe Road, Bellville, 7535, South Africa}

\author[0000-0002-4939-734X]{T.H. Jarrett}
\affiliation{Institute for Astronomy, University of Hawaii at Hilo, 640 N Aohoku Pl 209, Hilo, HI 96720, USA}
\affiliation{Department of Astronomy, University of Cape Town, Rondebosch,
 South Africa}
\affiliation{Centre for Astrophysics and Supercomputing, Swinburne University of Technology,
John Street, Hawthorn, 3122, Australia}

\author[0000-0002-5782-9093]{D.A. Dale}
\affiliation{Department of Physics and Astronomy, University of Wyoming, Laramie, WY 82071, USA}

\author[0000-0003-1545-5078]{J.-D.T. Smith}
\affiliation{Ritter Astrophysical Research Center, University of Toledo, Toledo, OH 43606, USA}

\author[0000-0002-1207-9137]{M.J.I. Brown}
\affiliation{School of Physics and Astronomy, Monash University, Clayton 3800, VIC, Australia}

\author[0009-0009-8499-1326]{W. van Kempen}
\affiliation{Centre for Astrophysics and Supercomputing, Swinburne University of Technology,
John Street, Hawthorn, 3122, Australia}

\author[0009-0006-0106-9124]{E. Lengerer}
\affiliation{Centre for Astrophysics and Supercomputing, Swinburne University of Technology,
John Street, Hawthorn, 3122, Australia}

\author[0009-0003-3702-0881]{R. Incoll}
\affiliation{Centre for Astrophysics and Supercomputing, Swinburne University of Technology,
John Street, Hawthorn, 3122, Australia}

\author[0009-0000-8764-616X]{C. Davey}
\affiliation{Centre for Astrophysics and Supercomputing, Swinburne University of Technology,
John Street, Hawthorn, 3122, Australia}

\author[0009-0008-3021-1710]{R. Holloway}
\affiliation{Centre for Astrophysics and Supercomputing, Swinburne University of Technology,
John Street, Hawthorn, 3122, Australia}

\author[0009-0007-3146-654X]{J. Cameron}
\affiliation{Centre for Astrophysics and Supercomputing, Swinburne University of Technology,
John Street, Hawthorn, 3122, Australia}

\author[0000-0002-5496-4118]{K. Sheth}
\affiliation{NASA Headquarters, 300 Hidden Figures Way SW, Washington, D.C.}

\correspondingauthor{Michelle Cluver}
\email{michelle.cluver@gmail.com}

\accepted{17 October 2024}

\submitjournal{ApJ}

\begin{abstract}

In this work we present source-tailored \textit{WISE} mid-infrared photometry (at 3.4\micron, 4.6\micron, 12\micron, and 23\micron) of 2812 galaxies in the extended \textit{Spitzer} Survey of Stellar Structure in Galaxies (\spitzerg) sample, and characterise the mid-infrared colors and dust properties of this legacy nearby galaxy data set. Informed by the relative emission between W3 (12\micron) and W4 (23\micron), we re-derive star formation rate (SFR) scaling relations calibrated to L$_{\rm TIR}$, which results in improved agreement between the two tracers. By inverse-variance weighting the W3 and W4-derived SFRs, we generate a combined mid-infrared SFR that is a broadly robust measure of star formation activity in dusty, star-forming galaxies in the nearby Universe.  In addition, we investigate the use of a W3-derived dust density metric, $\Sigma_{\rm 12\mu m}$ (L$_\odot$/kpc$^2$), to estimate the SFR deficit of low mass, low dust galaxies. This is achieved by combining \wise\ with existing GALEX ultraviolet (UV) photometry, which we further use to explore the relationship between dust and UV emission as a function of morphology. Finally, we use our derived SFR prescriptions to examine the location of galaxies in the log\,SFR -- log\,\Mstar\ plane, as a function of morphological type, which underscores the complexity of dust-derived properties seen in galaxies of progressively earlier type.

\end{abstract}


\section{Introduction} \label{sec:intro}

The local Universe furnishes the benchmark that underpins our understanding of how galaxies have evolved over cosmic time. Here we can assemble a near complete view of a diverse zoo of galaxy specimens and, with homogeneous observations at multiple wavelengths, characterise the properties of different sub-populations of the nearby Universe.

One such benchmark sample is the \textit{Spitzer} Survey of Stellar Structure in Galaxies \citep[\spitzerg;][]{2010PASP..122.1397S}, created as a uniform legacy survey using the \textit{Spitzer Space Telescope} and its `warm' imaging bands (IRAC-1, 3.6\micron\ and IRAC-2, 4.5\micron). \spitzerg\ concentrated on the local volume within 40 Mpc, initially targeting $\sim$2300 galaxies of a wide variety of morphological types and stellar masses, but later expanded to over 2800 galaxies by including additional early-type systems. The survey was notably designed to study the stellar mass distribution of nearby galaxies since the 3.6\micron\ and 4.5\micron\ channels sensitively track the Rayleigh-Jeans tail of the stellar light distribution from luminous, evolved stars that dominate the baryonic mass of normal, evolved galaxies \citep[see studies by][]{JuanCarlos+2013,meidt+2014,que+2015}.  

These nearby sources can often be challenging to measure, not least because they are so large on the sky. But, equivalently, they provide the highest signal to noise measurements for both global and local (i.e. sub-galactic) studies of star formation \citep[see][and references within]{Kenn2012}. Where studies of star formation on sub-kpc scales provide insight into the physics and chemistry of this key process \citep[e.g.][]{Sun2023, Cal2024, Pelt2024}, global measures of star formation rate (SFR) are a fundamental property of large, statistical studies of galaxy evolution, particularly in combination with stellar mass \citep[e.g.][]{Noeske2007, Whit2014, Pop2019, MS_2024A&A}.

Multi-band photometry in combination with spectral energy distribution (SED) fitting is a powerful tool to extract stellar masses, SFRs, and dust properties \citep[e.g.][]{2008Magphys, Beag2016, Leja2017, Boq2019, prospect2020}, although the treatment of dust and its associated attenuation is far from trivial \citep[e.g.][]{Salim2020, Pac2023} requiring detailed radiative transfer modelling \citep[e.g.][]{2014deLooze, Via2017, Will2019}. However, the combination of GALEX \citep{GALEX} ultraviolet (UV) and \wise\ mid-infrared imaging, accounting for both unobscured and obscured star formation, has been shown to be a powerful hybrid SFR indicator in the nearby Universe \citep[e.g.][]{Salim+2016, Davies2016, leroy+2019, Belf23}, building on the work of, for example, \citet{Kenn1998}, \citet{Hao11}, and \citet{Boq2016}. This is well-illustrated in, for example, studies of the gas consumption and fueling of star formation in the Local Universe \citep[e.g.][]{Sain2017}.

Dust reprocesses the radiation from stars, but, for example, dust content, dust geometry, and the contributions from young versus evolved stellar populations will influence the resulting infrared spectrum \citep[see][and references within]{Cal2013}. As shown in \citet{Cluver17}, the power of \wise\ as a tracer of star formation activity lies in the fact that the 12\micron\ and 23\micron\ dust emission bands are well-correlated to total infrared luminosity \citep[$L_{\rm TIR}$;][]{Dale_Helou} where the particularly close correspondence of $L_{12\mu m}$ and $L_{\rm TIR}$ was noted by \citet{Spin1995} using the \textit{Infrared Astronomy Satellite (IRAS)} and further examined by \citet{Tak2005}. Hence, where $L_{\rm TIR}$ is expected to be a reliable measure of star formation (i.e. in typical, dusty star-forming galaxies), \wise\ 12\micron\ and 23\micron\ channels offer a stable measurement of recent star formation in the local Universe. However, it should be noted that at $z\sim0$, the \wise\ W3 and W4 bands measure two quite different parts of a galaxy's spectral energy distribution, with potentially complex provenance and hence dependencies. In addition, the limitations inherent in the use of a monochromatic star formation indicator should always be carefully considered and weighed against the convenience of \wise's all-sky coverage which allows for a uniform measure of star formation particularly useful in large area studies \citep[e.g.][]{cluver+2020}.

In this study, we present source-tailored mid-infrared measurements of the expanded \spitzerg\ sample derived using imaging from the \wise\ survey (Wright et al. 2010). Although the \wise\ 3.4\micron\ (W1) and 4.6\micron\ (W2) bands (with $\sim$\,6\arcsec resolution) do not offer any improvement on the \textit{Spitzer} observations, the \wise\ 12\micron\ (W3) and 23\micron\ (W4) bands allow us to examine the dust properties uniformly across the sample due to its all-sky coverage. Based on the observed dust emission characteristics, we re-derive the star formation rate calibrations of \citet{Cluver17}, finding much closer agreement between the W3 and W4 tracers.

The \spitzerg-\WISE\ sample covers the entire spectrum of galaxies observed in the local Universe; this diversity in morphological type and stellar mass allows us to examine the strengths and weaknesses of \WISE\ in support of large surveys. To this end, we introduce a dust density metric, $\Sigma_{\rm 12\mu m}$ (L$_\odot$/kpc$^2$), that can be used to estimate the deficit in dust-derived star formation rates for low mass, low dust galaxies. 

The all-sky coverage of the \wise\ survey fortifies its relevance as an ongoing resource for $z<0.1$ science, particularly in the Southern Hemisphere where mega-projects such as LSST \citep{LSST}, 4MOST \citep{4MOST}, and the Square Kilometer Array will create a rich scientific landscape in the next decade, and beyond.  The work presented here will be followed by a companion paper that will present the \WISE\ photometry and derived quantities (using the scaling relations presented in this current work) for a sample of $\sim$17,000 galaxies selected from the 2MASS Redshift Survey \citep{Huchra12}; this Ks brighter than 11th mag sample, plus the \spitzerg\ data presented here, will comprise the brightest and highest signal to noise (S/N) mid-infrared sample possible.

Throughout this paper mid-IR refers to the ``mid-infrared" wavelengths and all four \wise\ bands are considered to be in the mid-infrared. Monochromatic luminosities are defined as $\nu L_{\nu}$, i.e. $L_{12\mu m} \equiv \nu L_{\nu} (12\mu m)$, M$_{\rm stellar}$ is the global stellar mass for individual galaxies and likewise,  SFR is the global star formation rate.  The adopted cosmology is H$_0$ = 70\,km s$^{-1}$ Mpc$^{-1}$, $\Omega_{M}$ = 0.3 and $\Omega_{\Lambda}$= 0.7, but as detailed in the following section, robust redshift-independent and bulk flow corrected distances are used where available.  All magnitudes are in the Vega system \cite[$\textit{WISE}$ photometric calibration described in][]{Jarrett2011}. All linear fits are performed using the Hyper-fit package \citep{hyperfit}.

\section{Data} \label{sec:data}

Targeting the sample of galaxies that form part of \spitzerg\ \citep{2010PASP..122.1397S}, the
primary photometric and source characterization is drawn from the \WISE\ Extended Source Catalogue \citep[WXSC][]{Jarrett13}.
The sample and data are described below.

\subsection{The S\,$^4$G Sample}

\begin{figure*}[!t]
\begin{center}
\gridline{\fig{ 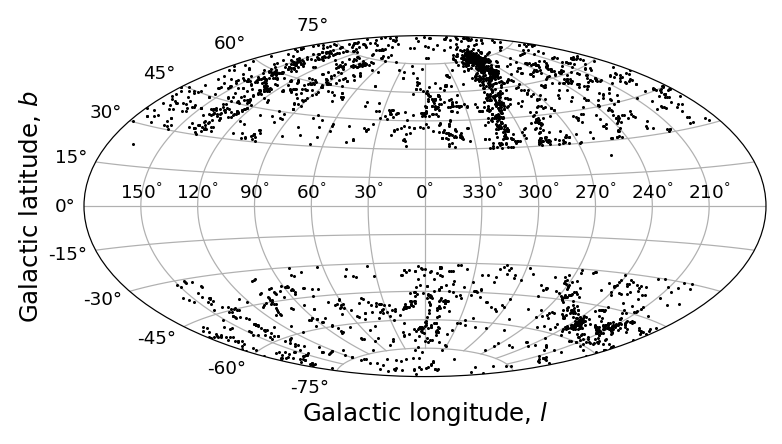}{0.6\textwidth}{(a)}
\fig{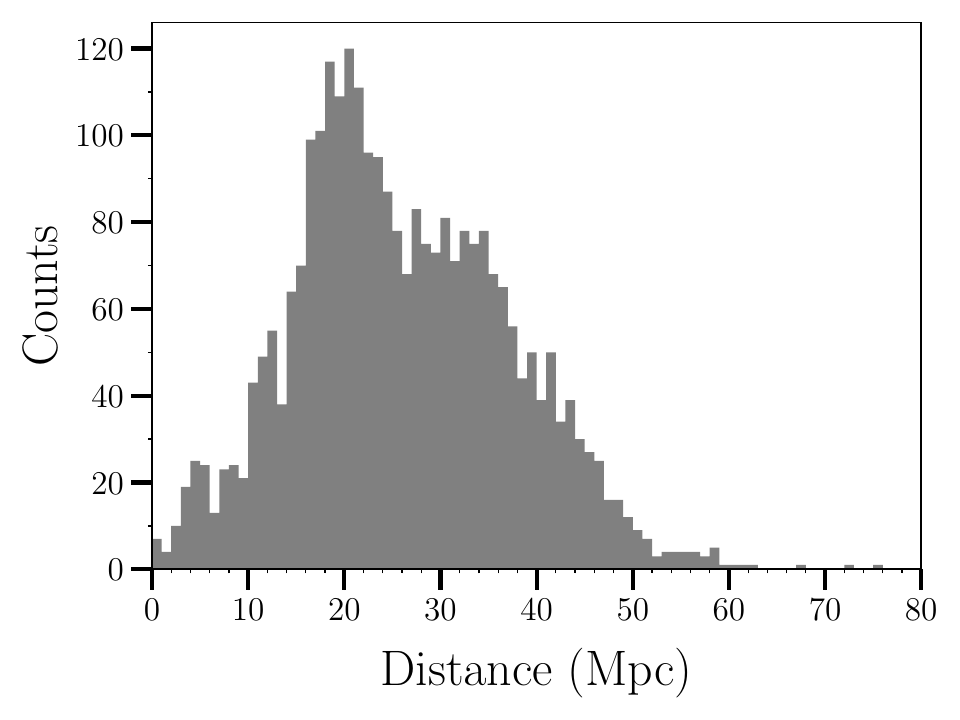}{0.4\textwidth}{(b)}}
\caption{(a) Galactic Projection on-sky distribution of the \spitzerg-\WISE\ sample (2812 galaxies). The Virgo Cluster gives rise to the overdensity in the North at $\sim l=280\deg$ and the Fornax-Eridanus structure in the South at $\sim l=240\deg$. (b) The distance distribution of sources shows that the majority of galaxies lie within 50 Mpc. }
\label{fig:S4G_summary}
\end{center}
\end{figure*}

\spitzerg\ was a legacy program conducted as part of the \textit{Spitzer} Space Telescope's post-cryogenic mission which performed imaging of 2331 galaxies using the Infrared Array Camera (IRAC) 3.6\micron\ and 4.5\micron\ bands providing a volume-, magnitude- and size-limited sample \citep{2010PASP..122.1397S} to a distance of 40 Mpc.

The original \spitzerg\ sample selection, however, suffered from a bias that notably excluded early-type galaxies. This was subsequently rectified by introducing a second sample, $\sim$500 galaxies, using morphology and distance information to sweep up any early type (bulge-dominated) galaxies within the volume. This latter sample is referred to as the \spitzerg-ETG;  details are provided in \cite{watkins+2022}. We have combined the `classic' sample with the ETG sample and measured their \wise\ counterparts to create the \spitzerg-\WISE\ sample consisting of 2812 galaxies. \spitzerg\ galaxies are typically bright and well-resolved in the \WISE\ imaging which allows us to relate morphological and structural characteristics to \WISE-derived physical properties of stellar mass and star formation rate. 

The \spitzerg\ sample comes with meta-information including distance information and morphological classifications. However, instead of using the published distances in the earlier \spitzerg\ catalogs, we make use of the latest distance information where available, notably taking advantage of redshift-independent compilations (NED-D V17.1.0) from the NASA Extragalactic Database (NED)\footnote{https://ned.ipac.caltech.edu/Library/Distances/}.  
We derive the distances in the following way:  (1) if the source has a high quality (at least 3 independent measurements) mean redshift-independent distance, it is adopted, but if not, (2) search for the source in the CosmicFlows-3 catalogue of galaxy distances \citep[]{Tully_2016}, otherwise (3) we use the redshift to compute the luminosity distance after correcting to the CMB frame of reference (with H$_0$ = 70\,km s$^{-1}$ Mpc$^{-1}$, $\Omega_{M}$ = 0.3). The majority of sources, 1525 ($\sim$54\%) have redshift-independent distances, a fraction of which have the highest quality distances from either Cepheid, Planetary Nebula Luminosity Function or Tip of the Red Giant Branch calibrations. Of these, 1174 come from CosmicFlows-3. The remainder of sources (1287) use distances derived from redshifts.  Consequently, we can expect the best quality luminosities from the majority of the sample.

The sources are distributed across the sky, as shown in Fig.~\ref{fig:S4G_summary}a, depicting the locations in Galactic projection.  The plane of the Milky Way is completely avoided, removing any uncertainties from extinction and crowd confusion.  The SuperGalactic Plane, dominated by the Virgo Cluster, is readily apparent in the sky distribution.  

The distribution of distances in the \spitzerg-\WISE\ sample is shown in the histogram of Fig.~\ref{fig:S4G_summary}b.  The vast majority of galaxies are within 50\,Mpc, consistent with the original 40\,Mpc selection of S4G, but with a handful of sources (13) beyond 60 Mpc.  The peak at $\sim$20\,Mpc is the Virgo Supercluster. We note that several Local Group Galaxies, D$<1$\,Mpc, are within the sample, including Andromeda (M31) and Triangulum (M33), and several low mass dwarfs that \WISE\ was able to detect and characterize due to their close proximity (e.g., ESO\,540-031).  At the other extreme, the sample includes starburst galaxies (e.g., NGC\,253 and M\,82) and powerful AGN (NGC\,1068) and other famous galaxies (e.g., M\,101). Most of the sample would be classified as ``normal" and hence, all in all, the \spitzerg\ sample well-represents the zoo of galaxy types that comprise our local Universe.

\subsection{\wise\ Imaging and Measurements}

For over a decade we have made a concerted effort to identify, extract and characterize resolved galaxies imaged by \WISE, building a rich database from which to draw catalogs.  Preliminary catalog releases of the 
 \WISE\ Extended Source Catalogue \citep[WXSC;][]{Jarrett13,jarrett+2019} describe the basic characterization methods, catalog features, and science results with the largest (angular size) galaxies in the sky.  A much larger and more comprehensive subset of the WXSC will be released following this current work (Jarrett et al., in preparation).  
 Here we briefly summarize the salient information relevant to this study.
 
Relative to the archival imaging and catalogs from the \WISE\ project 
 \citep[e.g., ALLWISE catalog][]{cutri+2012},
the WXSC endeavors to: (1) improve upon 4-band image co-addition (mosaics), using custom mosaic construction 
\citep{jarrett+2012} that includes the latest NEOWISE imaging and hence increases the sensitivity in the W1 and W2 bands,
and (2) full source characterization that the diversity of galaxies demand, ranging from large, bright and massive galaxies to faint low surface brightness dwarfs.
The W1 image mosaics reach  
1-sigma depths (typically) fainter than 23 mag\,arcsec$^{-2}$ (25.7 mag\,arcsec$^{-2}$ in AB mags), while  
native angular resolution (6\arcs\ in W1) is preserved in all four bands. We note that with axi-symmetry, the radial surface profiles reach W1 depths of 25 to 26 mag\,arcsec$^{-2}$ ($27.7-28.7$ mag\,arcsec$^{-2}$ in AB mags), similar depths as to those of \textit{Spitzer}-IRAC.

\begin{figure}[!th]
\begin{center}
\includegraphics[width=8.5cm]{ 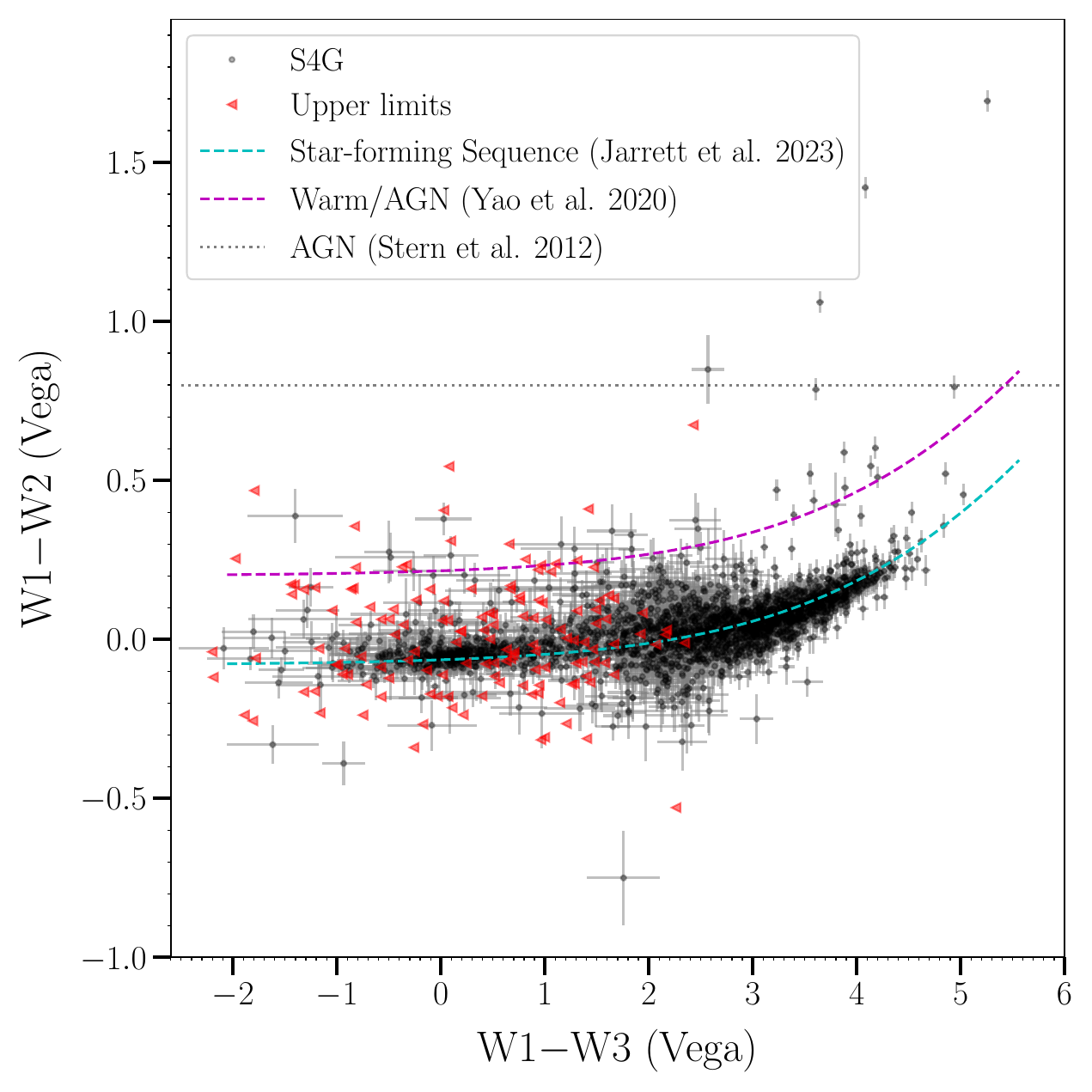}
\caption{The Color Sequence:   The \wise\ color-color plot for the \spitzerg-\wise\ sample shown as \WIWII\ versus \WIWIII, including the star-forming sequence from \citet{Jarrett2023}. The x-axis tracks the star formation history, from stellar-dominated to SF-dominated phases.  
Elevated \WIWII\ colors are a signature of dust heated by AGN; we include the relations of \citet{stern+2012} and \citet{yao+2020} above which the contribution of AGN activity should be taken into account.}
\label{fig:colors}
\end{center}
\end{figure}

It is non-trivial to measure the properties of resolved galaxies, and hence the WXSC characterization pipeline has evolved over time \citep[originating with 2MASS;][]{jarrett-2003} to accommodate the \WISE\ imaging strengths and limitations.  As such, for the target galaxy, basic source characterization consists of five primary steps: (1) identify and remove foreground stars and local/background galaxies, a critical step since the beam is large and source blending is a significant challenge
\citep[see e.g., the method for de-blending galaxy pairs in Appendix B of][]{bok+2020},
 (2) local ``sky" background estimation, (3) fixed 3-$\sigma$ elliptical shape and orientation metrics, (4) 1-$\sigma$ isophotal radial size and flux determination, (5) central surface brightness and radial profile fitting to separate bulge and disk components,  derive total fluxes and concentration indices, and other miscellaneous characterization.

For well-resolved galaxies,
axisymmetric radial profiles were constructed and fit with a double-S\'ersic function designed to separate the 
spheroidal and disk population distributions, as well as extrapolate to larger radii to determine total fluxes down to several disk scale lengths. Further details can be found in  \citet[][]{Jarrett13,jarrett+2019,Jarrett2023}. Photometry measurements for the \spitzerg\ sample are included in Table 1.

\subsection{Derived Quantities}

All fluxes and colors are rest-frame corrected using a large set of composite SED templates from \citet[]{Brown2014} and \textit{Spitzer}-SWIRE/GRASIL \cite[]{Polletta2006,Polletta2007,Silva1998}
which span early-type spheroidals to late-type spirals, and include active Seyferts and AGNs of differing orientation types.
Given the proximity of galaxies in \spitzerg-\wise, the conversion to rest-frame properties (or ``k-correction") introduces a relatively small uncertainty, less than 3 to 4\% \citep[see also][]{Jarrett2023}. \WISE\ is natively calibrated to Vega and all magnitudes are reported in this system \citep{wright+2010,Jarrett2011}.

The primary scaling parameters are the \MIR\ luminosities.  In the case of stellar mass, the in-band luminosity, \Lw, is the primary tracer of stellar mass \citep{Jarrett2023}. This luminosity is derived using 
 \begin{equation} \label{eq1}
 L_{\rm W1} [L_\odot] = 10^{-0.4(M-M_{\rm SUN})}
 \end{equation}
 where M is the absolute magnitude of the source in W1 -- determined from the W1 rest-frame flux, luminosity distance and the distance modulus -- and 3.24 mag is the W1 in-band Solar value 
 \citep{Jarrett13}. All stellar masses in this paper are derived using the prescription outlined in \citet{Jarrett2023} which assumes a \citet{chabrier+2003} IMF.

\begin{figure*}[!tb]
\begin{center}
\gridline{\fig{ 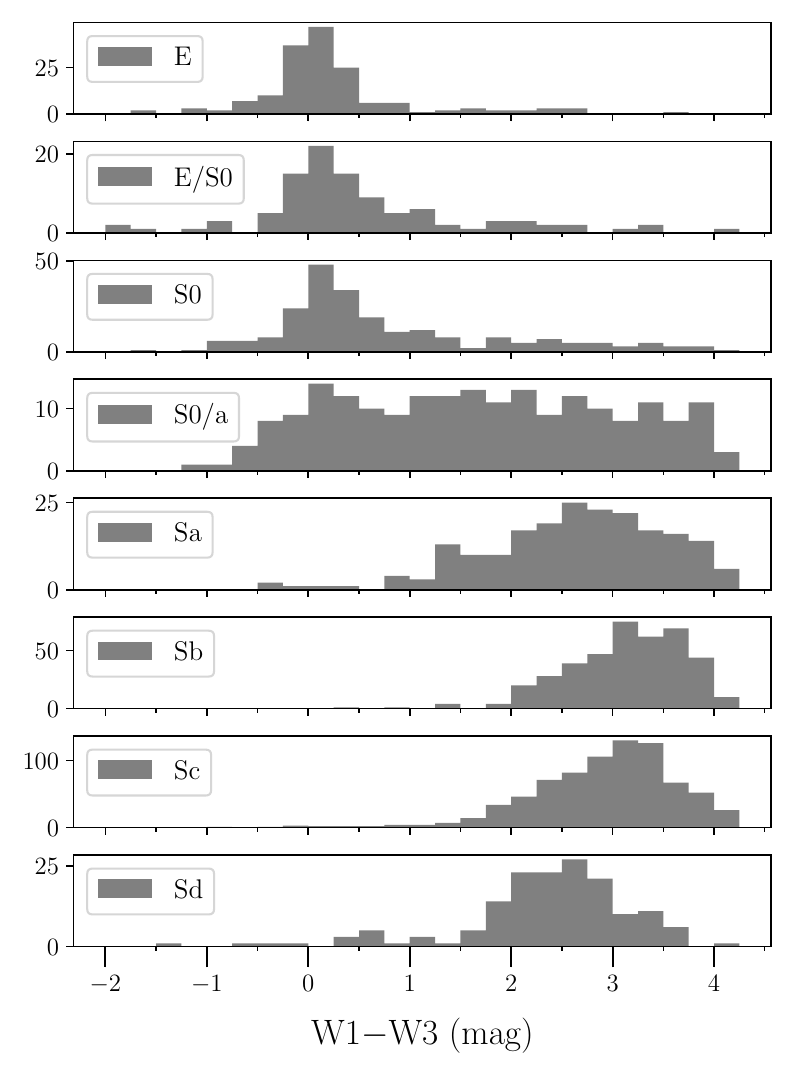}{0.5\textwidth}{(a)}
\fig{ 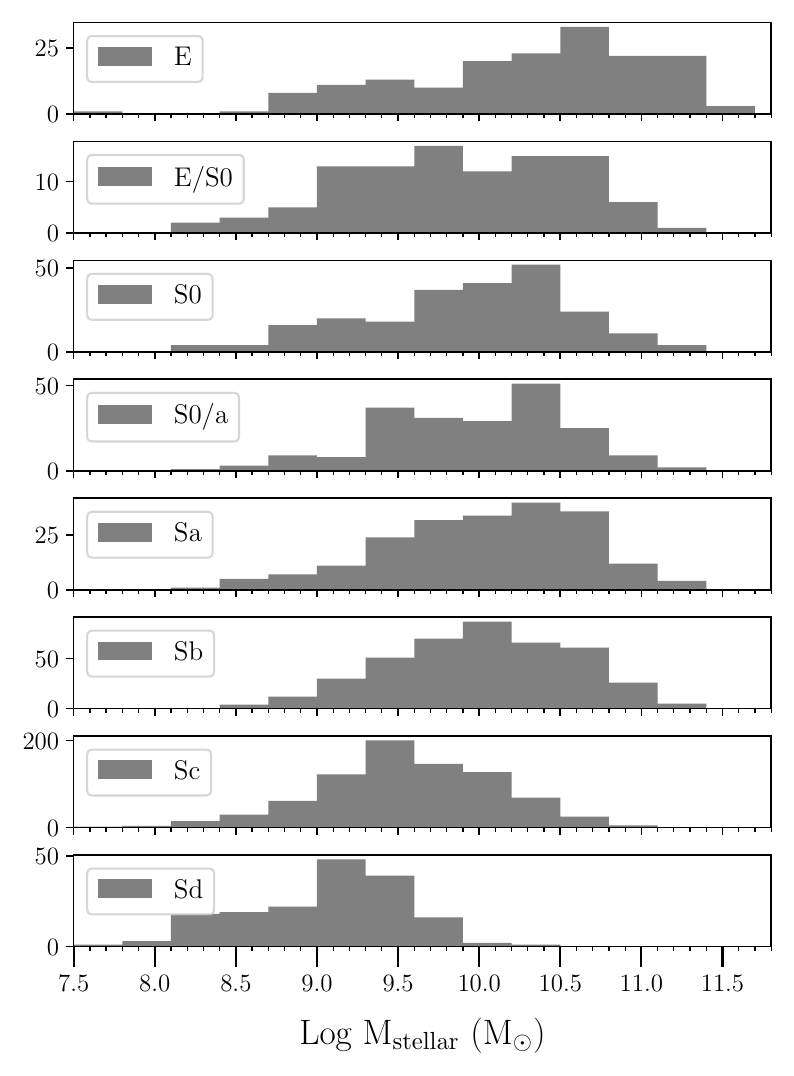}{0.5\textwidth}{(b)}
}
\caption{Galaxy Morphology Phase Space:  (a) The \WIWIII\ colors separated by morphology illustrating the complexity of mapping color to galaxy type. (b) The stellar mass distribution for different morphological types within the \spitzerg-\wise\ sample; we observe that a clear progression of stellar mass increasing as bulge prominence increases does not apply to all bulge-dominated systems (notably E and S0) which can be low mass.}
\label{fig:Type_Mass_color}
\end{center}
\end{figure*}
 
For measuring dust-obscured star formation we use the W3 and W4 spectral luminosities, $\nu$L$_\nu$(12\micron) and $\nu$L$_\nu$(23\micron), after correcting for the contribution from the stellar continuum \citep[using W1 as the proxy for the Rayleigh-Jeans model of the stellar emission, see][]{Cluver17,jarrett+2019}. In this paper, we update the SFR relations of \citet{Cluver17} using improved measurements, implementing more robust distance information, and correcting for low dust optical depths in dwarf galaxies; these details are in Section 3 and the Appendix.

Relevant to both \Mstar\ and SFR, the fundamental assumption is that W1 and W2 bands are tracing the Rayleigh-Jeans tail of the stellar light distribution. As discussed in \cite{Jarrett13,jarrett+2019}, 
we do not attempt to model or correct for any non-stellar emission in the W1 and W2 bands, such as the 3.3\,$\mu $m PAH feature, which is likely to be less than 10\% for most normal galaxies \citep[e.g., see][]{Lu2003}. In a future study using JWST spectroscopy of the M51a galaxy, it will be possible to assess the impact of the $3.3-3.4$\,$\mu $m PAH aliphatic feature -- now easily detected in nearby galaxies using JWST \citep[e.g.,][]{lai2023} -- on the integrated W1 emission for different SF environments.

Similarly, we do not attempt to account for any excess infrared emission in the mid-IR bands of W3 and W4, or the presence of silicate absorption. Hence for those cases where the \WISE\ bands have significant AGB dust emission (old stars in the asymptotic giant branch phase), 
or AGN-accretion emission (e.g. ultraluminous infrared galaxies), the \WISE\ luminosities may be contaminated and caution is recommended when interpreting the resulting \Mstar\ and SFR for extrema galaxies. Derived properties for the \spitzerg-\WISE\ sample are included in Table 2.

\subsection{Morphologies}

The proximity of galaxies in the \spitzerg-\wise\ sample furnishes us with largely robust morphological information. We adopt the morphological classification methodology used in \citet{bouquin+2018}, i.e. the RC2 classification scheme using numerical T-Types obtained from HyperLeda \citep{refId0}. This is summarized here for convenience: $-5\leq$ E $\leq -3.5$, $-3.5\le$ E/S0 $\leq-2.5$, $-2.5\le$ S0 $\leq-1.5$, $-1.5\le$ S0/a $\leq0.5$, $0.5\le$ Sa $\leq2.5$, $2.5\le$ Sb $\leq4.5$, $4.5\le$ Sc $\leq7.5$, $7.5\le$ Sd $\leq8.5$, $8.5\le$ Sm $\leq9.5$, and $9.5\le$ Irr $\leq999$.

\section{Results}

\subsection{\spitzerg-WISE: Colours and Mass by Morphological Type}

Mid-IR colors are a simple, yet powerful diagnostic tool which can help inform sample selection and population studies.
The \WISE\ color-color diagram for the entire sample is shown in Figure \ref{fig:colors}; the star-forming sequence from \citet{Jarrett2023} -- which roughly tracks star formation history through \WIWIII\ color, and excesses due to e.g. AGN through \WIWII\ color -- shows good agreement with the diverse \spitzerg\ sample. Also shown is the AGN delineation from \citet{stern+2012}; we find 8 galaxies with highly elevated \WIWII\ colors, namely NGC\,1068, NGC\,3094, NGC\,5506, UGC\,04704, NGC\,5253, NGC\,4151, ESO\,605-015, and NGC\,4355. The latter has the most extreme \wise\ colours (\WIWIII$=6.151$ and \WIWII$=0.814$\,mag) and is beyond the limits of the axes shown in Figure \ref{fig:colors}.

Also shown is the ``warm" line derived in \citet{yao+2020} above which sources likely have a component of AGN heating (based on optical emission line ratios). Elevated W1$-$W2 colors signal the presence of a hot dust component that will translate into an overestimation of stellar mass using the simple prescriptions we employ. This component will additionally boost the emission detected in the longer wavelength bands used to track star formation activity. We therefore exclude sources that lie above the \citet{yao+2020} line in the analysis that follows.  

\begin{figure}[!b]
\begin{center}
\includegraphics[width=8cm]{ 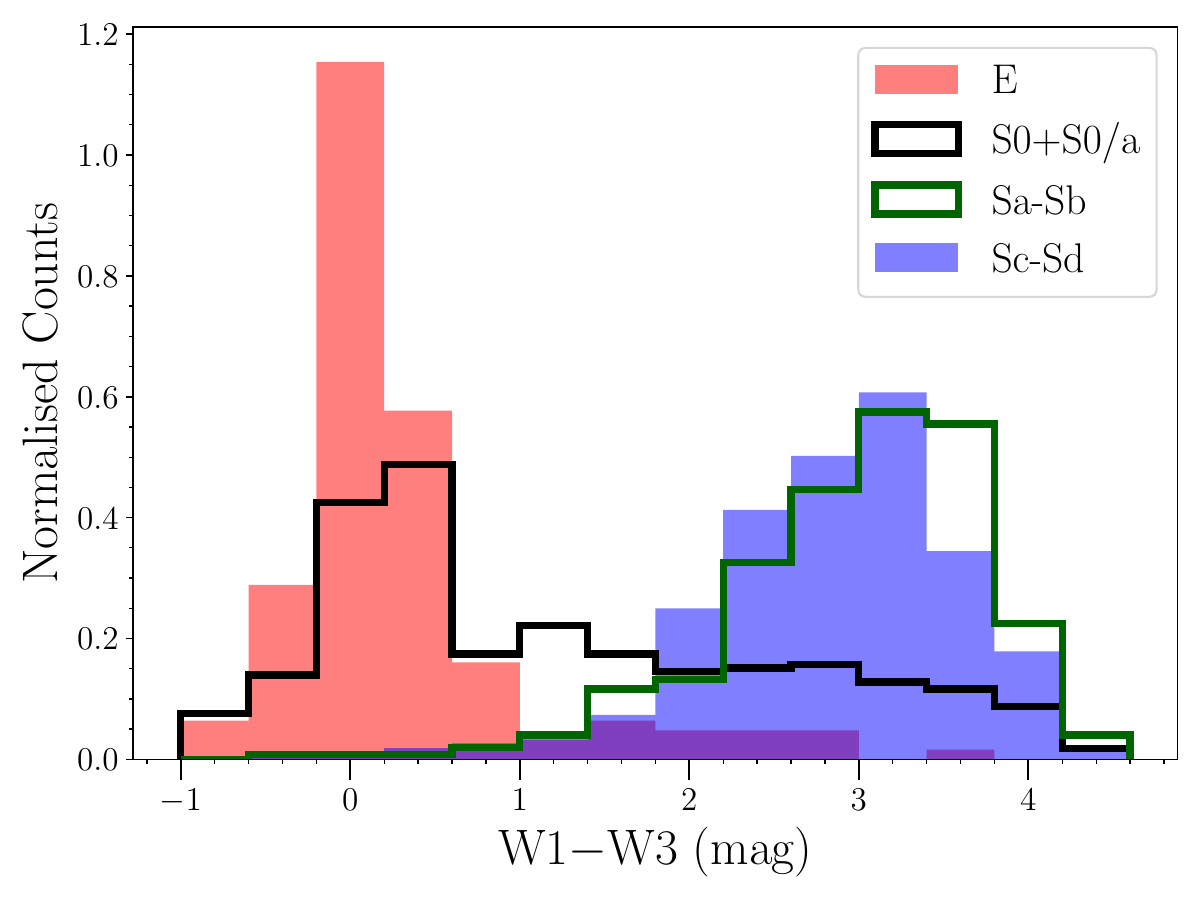}
\caption{Color delineation across morphology: grouping by morphology we see that \WIWIII\ colors separate type E, Sa-Sb and Sc-Sd reasonably well, but S0 and S0/a bulge-dominated disks tend to inhabit a broad range in color space.}
\label{fig:color_type}
\end{center}
\end{figure}

In Figure \ref{fig:Type_Mass_color}a we examine the \WIWIII\ color distribution as a function of morphological type (excluding sources with unreliable colors; \WIWIII\ error $> 0.5$\,mag). This confirms that late-type (disk-dominated) galaxies tend to exhibit redder \WISE\ colors compared to early-type (bulge-dominated) systems, with the clear exception being S0/Sa systems which occupy the full range of \WIWIII\ colors. The blueward shift of the Sd distribution compared to Sc is likely reflecting, on average, lower dust content paired with lower metallicity in these systems. We do not present results for irregular morphological types due to the challenging nature of these measurements.

\begin{figure}[!th]
\begin{center}
\includegraphics[width=8.5cm]{ 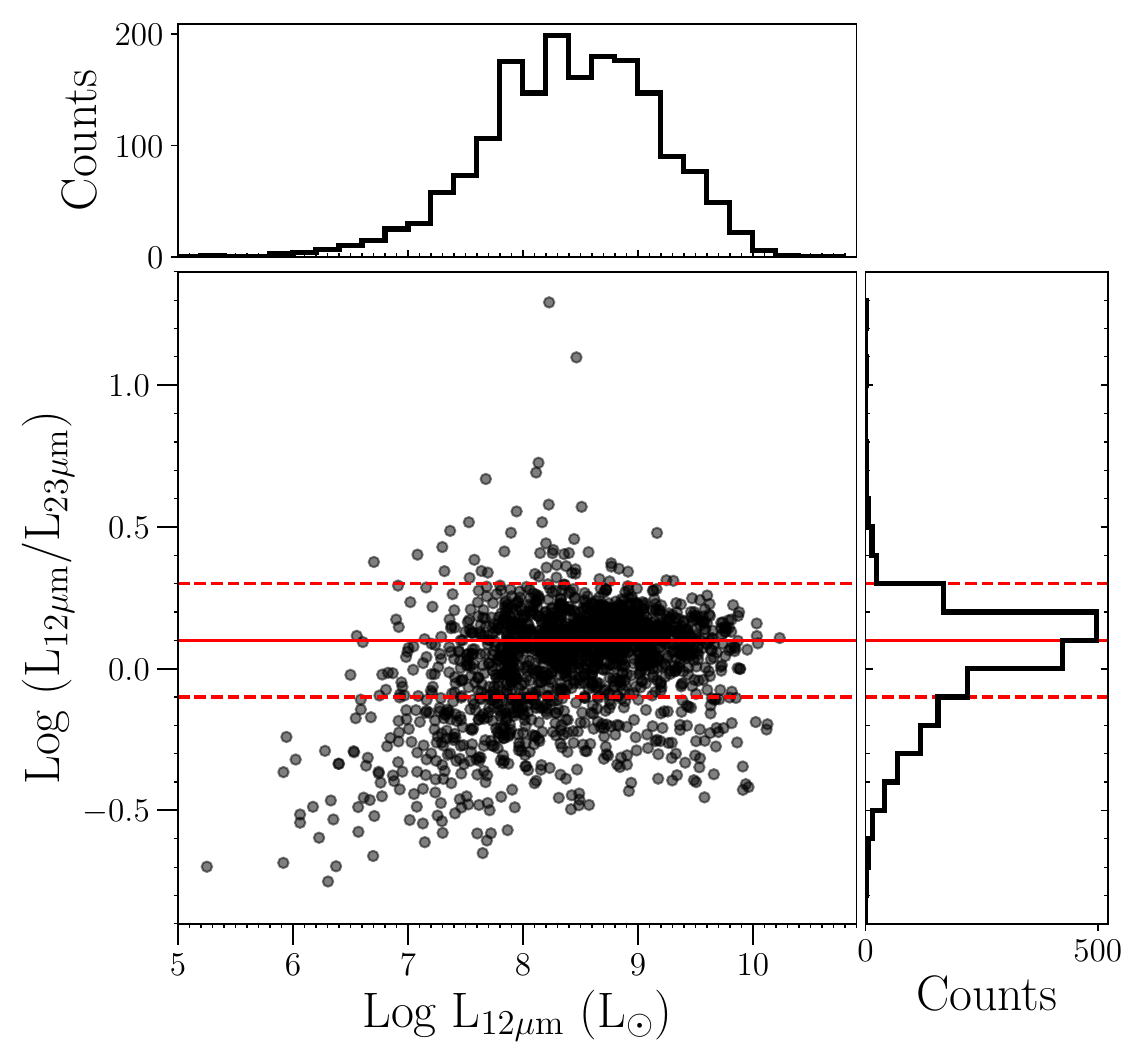}
\caption{W4 ``warm excess" galaxies:  the ratio of $L_{12\mu m}$ to $L_{23\mu m}$ as a function of $L_{12\mu m}$ for the \spitzerg-\wise\ sample. The median log\,(L$_{12\mu m}$/L$_{23\mu m}$) value is shown as a solid line with 0.2 dex on either side as the dashed lines. The histograms of the distributions show the tail that extends to low log\,(L$_{12\mu m}$/L$_{23\mu m}$) values, indicating numerous sources with either a paucity in L$_{12\mu m}$ emission, or a relative excess in L$_{23\mu m}$.}
\label{fig:lumratio}
\end{center}
\end{figure}

The relationship between stellar mass and morphological type is shown in Figure \ref{fig:Type_Mass_color}b; to select the most reliable sources we exclude those with a \WIWII\ error $> 0.2$\,mag \citep[see][]{Jarrett2023}. As expected, we see a trend of disk-dominated galaxies occupying the lower mass regime, compared to high stellar mass systems that are more likely to be bulge-dominated. What is interesting is that early types (E to Sa) occupy a broad range in \Mstar, and hence bulge-dominated systems are on average a poor predictor of stellar mass. Indeed, a number of \spitzerg\ galaxies are classified as dwarf spheroidals, consistent with both their \wise\ color and derived stellar mass.

To better examine the relationship between \WISE\ color and morphology we show the \textit{normalised distribution} of a subset of sources extracted as follows using numerical T-Types: $-5\leq$ E $\leq -3.5$, $-2.5\le$ S0, S0/a $\leq 0.5$, $0.5\le$ Sa-Sb $\leq4.5$, $4.5\le$ Sc-Sd $\leq8.5$. Figure \ref{fig:color_type} underscores the broad range of mid-IR color that S0 and S0/a galaxies occupy, whereas disk galaxies appear to separate relatively cleanly from those classified as pure ellipticals.   

\subsection{The \wise\ Star Formation Bands at $z\sim0$}

We next examine the relationship between the W3 and W4 bands; the \spitzerg-\wise\ sample is particularly powerful in this regard since nearby galaxies provide high quality 12\micron\ and 23\micron\ measurements usually lacking in more distant samples where the lower sensitivity of the W4 band precludes a high signal-to-noise comparison. Although both W3 and W4 trace dust, they measure quite different parts of the spectral energy distribution (see, for example, the SEDs and filter response functions for the \wise\ bands shown in Figure 19 of the Appendix). W3 is a broad filter: in the $z\sim0$ universe, it measures the hot/warm dust continuum including the 7.7 $\mu$m, 8.5 $\mu$m, and 11.3\micron\ PAH (polycyclic aromatic hydrocarbon) families, but also captures nebular emission lines at 12.81 $\mu$m ([Ne {\sc ii}]) and 13.7 $\mu$m ([Ne {\sc iii}]) and the S(2) line of pure rotational hydrogen at 12.3\micron. PAHs are complex tracers; they are excited by starlight photons, but the resulting spectrum is highly dependent on the relative abundance and composition of the PAH population, as well as the source of excitation \citep[see e.g.][]{Draine2021}. 

Although it is often assumed that the \wise\ W3 band is dominated by the 11.3\micron\ PAH at $z\sim0$, as discussed in \citet{Cluver17}, its contribution is on average $<10\%$ and the contribution from all PAHs to W3 is on average $<40\%$ (based on the SINGS sample). This band is hence dominated by non-PAH continuum from stochastically heated small dust grains and the thermal emission from hot/warm dust. But, an additional consideration is the contribution from the 10\micron\ silicate absorption feature which we can expect to be prominent in compact, highly luminous sources. 

In contrast, the W4 band is relatively straightforward: it samples the warm dust grain thermal continuum as well as stochastically heated very small grains. A complication for both, however, is the potential impact of dust geometry.

\begin{figure*}[!tbp]
\begin{center}
\includegraphics[width=14.5cm]{ 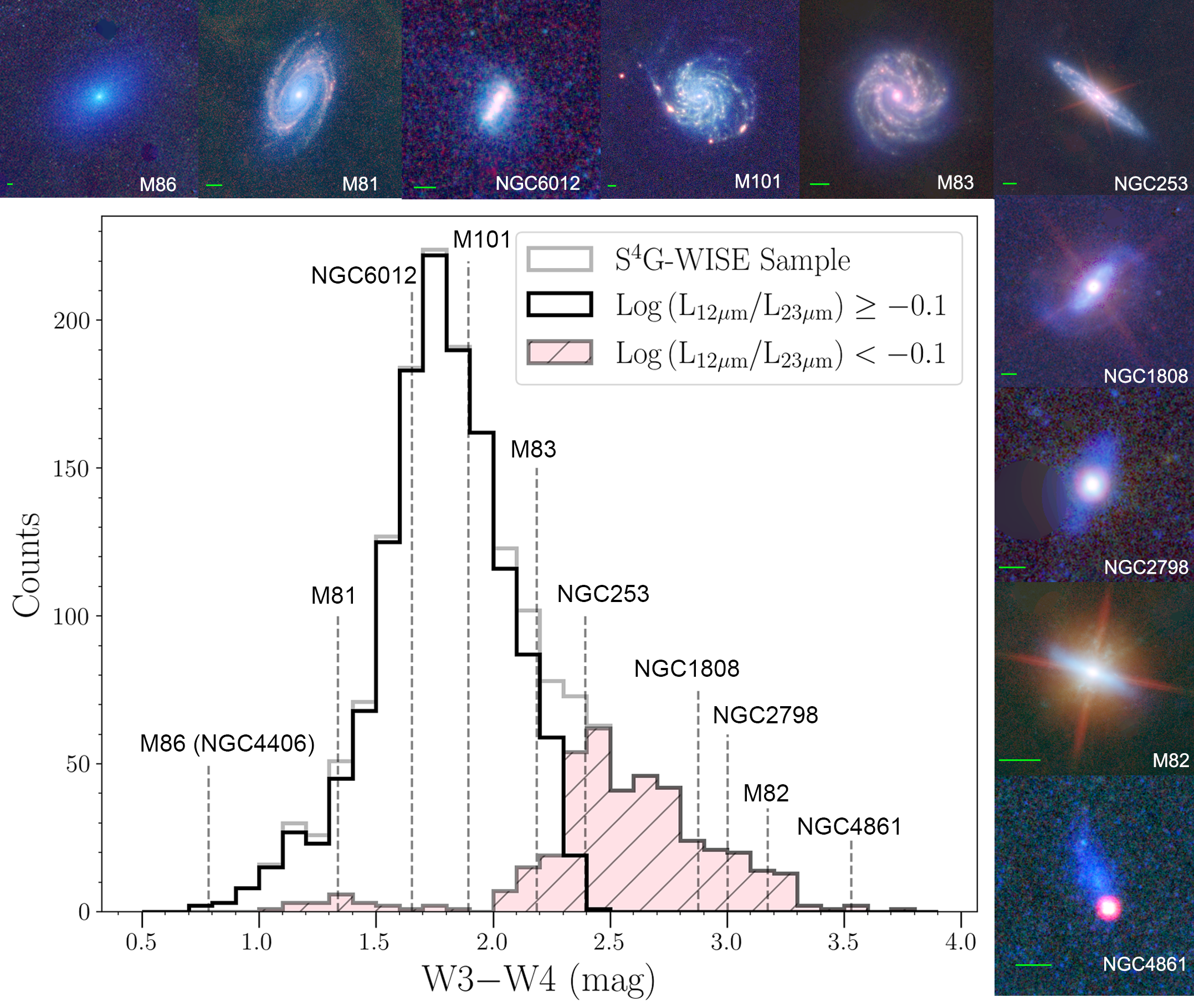}
\caption{Mid-infrared ``warm" galaxies: 
the W3$-$W4 color distribution of \spitzerg-\WISE, separated by log\,(L$_{12\mu m}$/L$_{23\mu m}$) ratio, with selected galaxies shown as four-color composite stamps combining W1 (blue), W2 (cyan), W3 (green), and W4 (red). The green bar in each image is 1\,arcmin.
The steep-spectrum sources, right side (shaded histogram), stand out with large W4 luminosites compared to W3.
}
\label{fig:s4ghist}
\end{center}
\end{figure*}

To investigate the geometric and environmental conditions that may affect W3 and W4 differently, we consider their luminosity ratio.  We expect a range (or intrinsic scatter) that reflects the diverse conditions seen in normal galaxies, but also deviations that arise from extreme conditions (e.g., massive star formation associated with compact starbursts).  In Figure \ref{fig:lumratio} we plot the ratio of L$_{12\mu m}$ to L$_{23\mu m}$ as a function of L$_{12\mu m}$, excluding all galaxies where the log luminosity error is greater than 0.1 dex (i.e. sources with S/N $<5$ after removing the stellar continuum). This shows that the majority of sources cluster within $\pm$0.2 dex (the standard deviation) of the median value of 0.1, with outliers above and below the median distribution. However, we observe an excess in low ratio sources, forming a distinctive tail -- these correspond to systems with a steeper than average mid-IR spectrum (i.e. relatively high L$_{23\mu m}$ emission compared to L$_{12\mu m}$), or alternatively, a paucity of W3 emission compared to W4. These sources are found over roughly 4 orders of magnitude of L$_{12\mu m}$ luminosity. To investigate this further, we consider the W3$-$W4 colour distribution of the sample. This quantity is helpful to examine whether this tail is present before the stellar continuum has been subtracted from the W3 and W4 measurements. 

In Figure \ref{fig:s4ghist} we examine the W3$-$W4 colours of the entire sample, as well as the distribution when the sample is split according to the log\,(L$_{12\mu m}$/L$_{23\mu m}$) ratio using a reference value of $-0.1$ (see Figure 5). We include the locations and images of several galaxies drawn from the \spitzerg\ sample to illustrate the diversity in appearance, among them some of the most famous local galaxies. We see that L$_{23\mu m}$ ``excess" sources do, on average, lie at higher W3$-$W4 color, similarly creating a tail in the distribution. Since we have excluded AGN ``warm" sources from our analysis, a possible driver could be the star formation properties of these galaxies, analogous to the trend seen in the IRAS color of galaxies in which far-IR ``warm" colors (exhibiting strong 60\,$\mu $m emission relative to 100\,$\mu $m) track with massive star formation activity
\citep[see e.g.][]{Helou-1986, Dale_2001}.
M82, for example, with its very red W3$-$W4 colors (i.e, steep mid-IR spectrum), is undergoing an intense nucleated starburst viewed edge-on. 

The impact of including these steep spectrum sources in star formation calibrations is the topic of the next section. A more detailed analysis of the properties of \spitzerg-\wise\ sources based on W3$-$W4 are deferred to a future publication. However, we illustrate the diversity in SED shape corresponding to varying W3$-$W4 color in Figure \ref{fig:seds} of the Appendix using NGC\,1266 (lenticular hosting Seyfert nucleus), NGC\,2798 (SBa undergoing tidal interaction and likely starburst), M101 (Scd and classic face-on grand-design spiral) and M81/NGC\,3031 (Sab).

\subsection{Star Formation Rates}

\begin{figure*}[!thp]
\begin{center}
\gridline{\fig{ 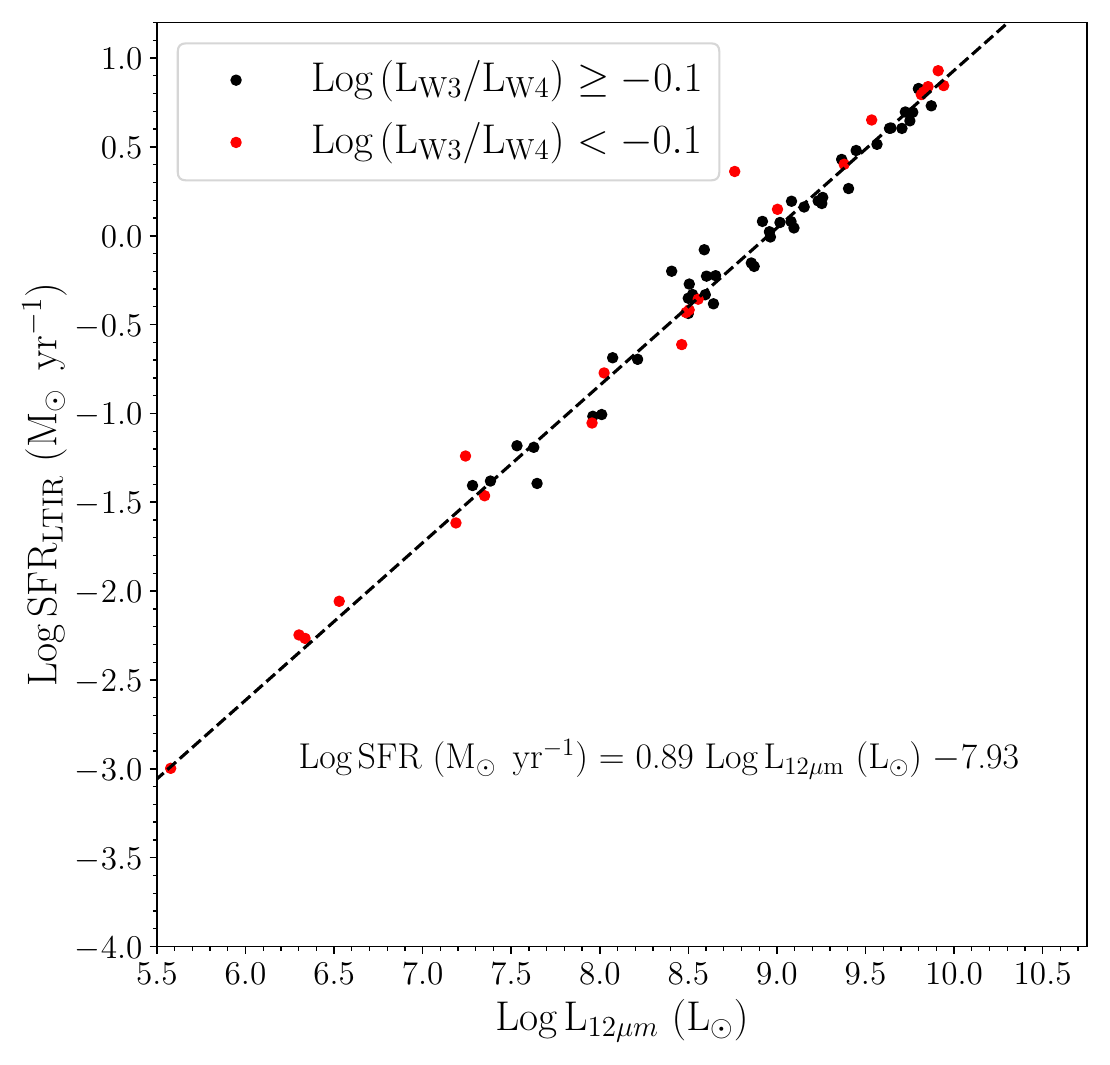}{0.5\textwidth}{(a)}
\fig{ 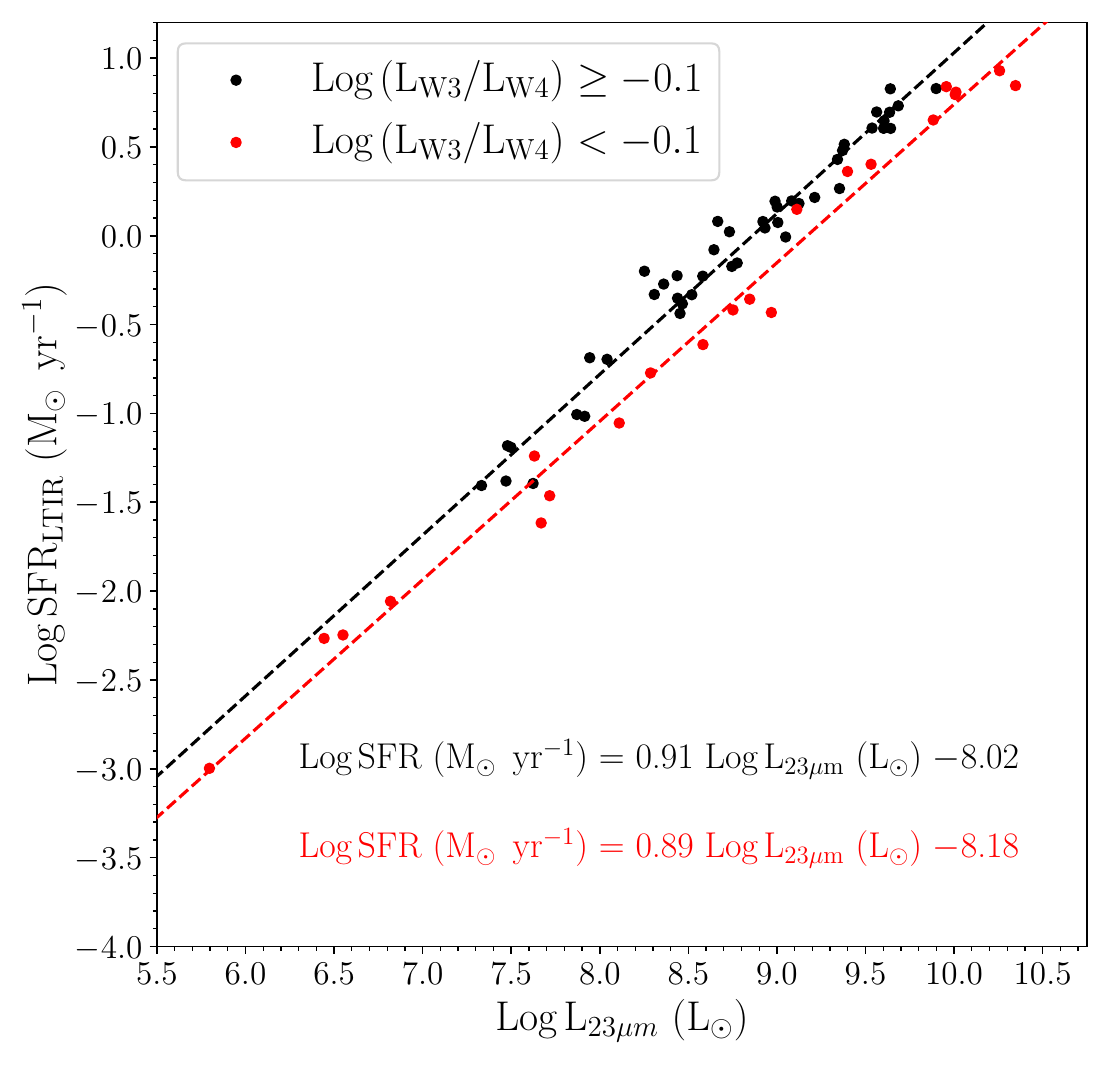}{0.5\textwidth}{(b)}}
\caption{How the mid-IR tracks with the total-infrared SFR:  
(a) The relationship between ${12\mu m}$ luminosity and SFR (calibrated to total infrared luminosity) using the SINGS/KINGFISH sample, with sources color-coded by log\,(L$_{12\mu m}$/L$_{23\mu m}$) ratio. The best fit adequately describes both red (warm) and black (normal) sources. (b) Considering the relationship between SFR$_{\rm TIR}$ and L$_{23\mu m}$ we see that black sources (log\,(L$_{12\mu m}$/L$_{23\mu m}$) $\geq-0.1$) and red sources (log\,(L$_{12\mu m}$/L$_{23\mu m}$) $<-0.1$) can be described by two separate relations thus reducing the overall scatter.  This suggests that steep mid-IR spectrum sources correspond to a lower SFR$_{\rm TIR}$ than would be inferred by the 23$\mu$m emission without taking this into account i.e. they follow a different scaling relation.}
\label{fig:singsfish} 
\end{center}
\end{figure*} 

\subsubsection{Revised WISE Star Formation Rate Calibrations}

In \citet{Cluver17} we derived \WISE\ SFR calibrations tying L$_{12\mu m}$ and L$_{23\mu m}$ to L$_{\rm TIR}$ using the SINGS/KINGFISH sample. Although a relatively small sample of 76 galaxies, by being nearby they offered the best quality mid- and far-infrared measurements necessary to relate \WISE\ fluxes to total infrared luminosity \citep[using the measurements from][]{Dale17}. In this section, we repeat the analysis from \citet{Cluver17} but make a distinction between sources based on their L$_{12\mu m}$/L$_{23\mu m}$ ratio. Revisiting this analysis benefits from improved resolved source photometry of all sources (54 of which are part of \spitzerg-\wise) as the WXSC pipeline has matured, as well as the improved implementation of distances by incorporating more robuste redshift-independent measures. The \wise\ color-color diagram and W3$-$W4 color distribution for the SINGS/KINGFISH sample are included in Section A of the Appendix.

\begin{figure*}[!tbhp]
\begin{center}
\gridline{\fig{ 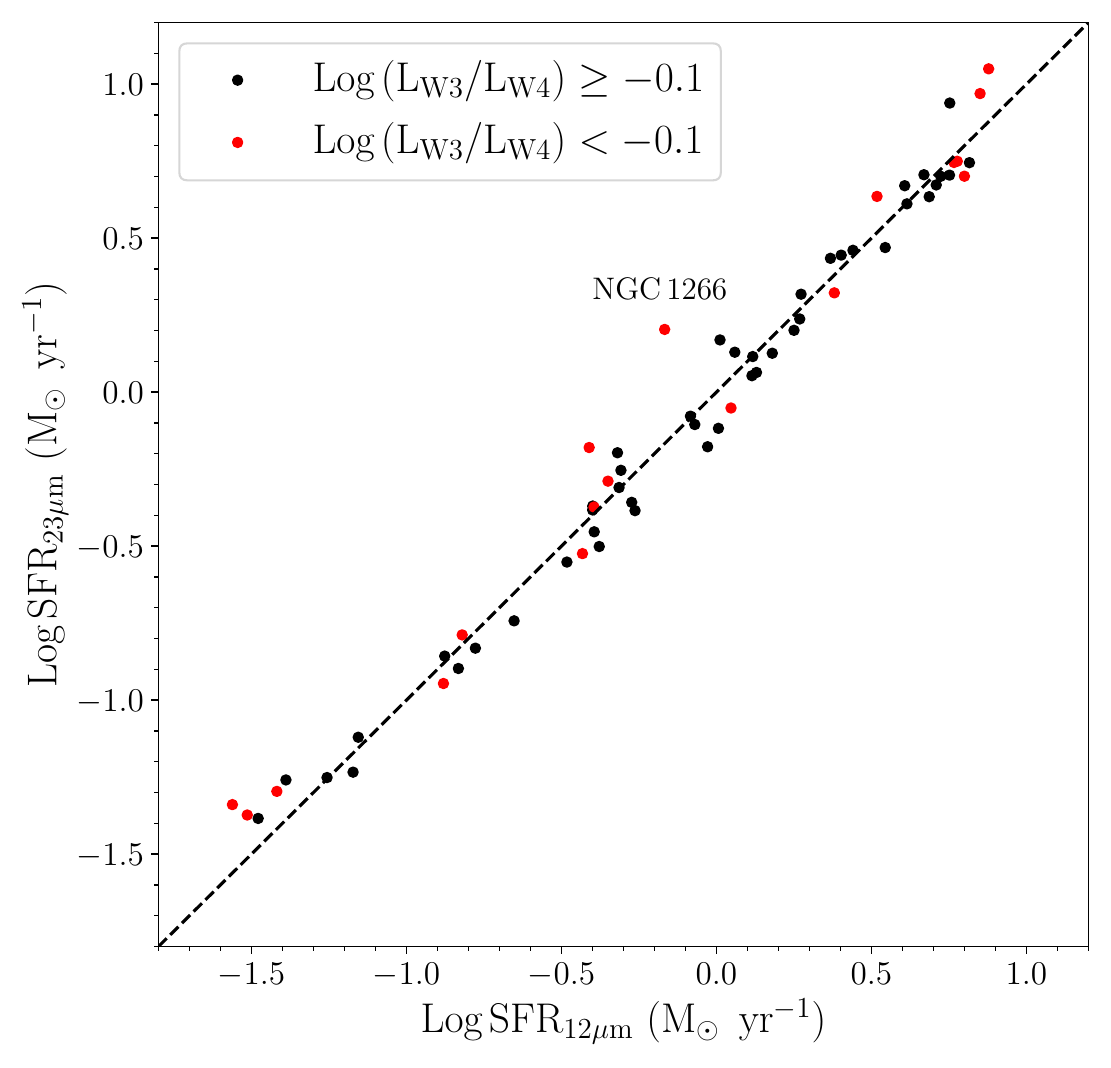}{0.5\textwidth}{(a)}
\fig{ 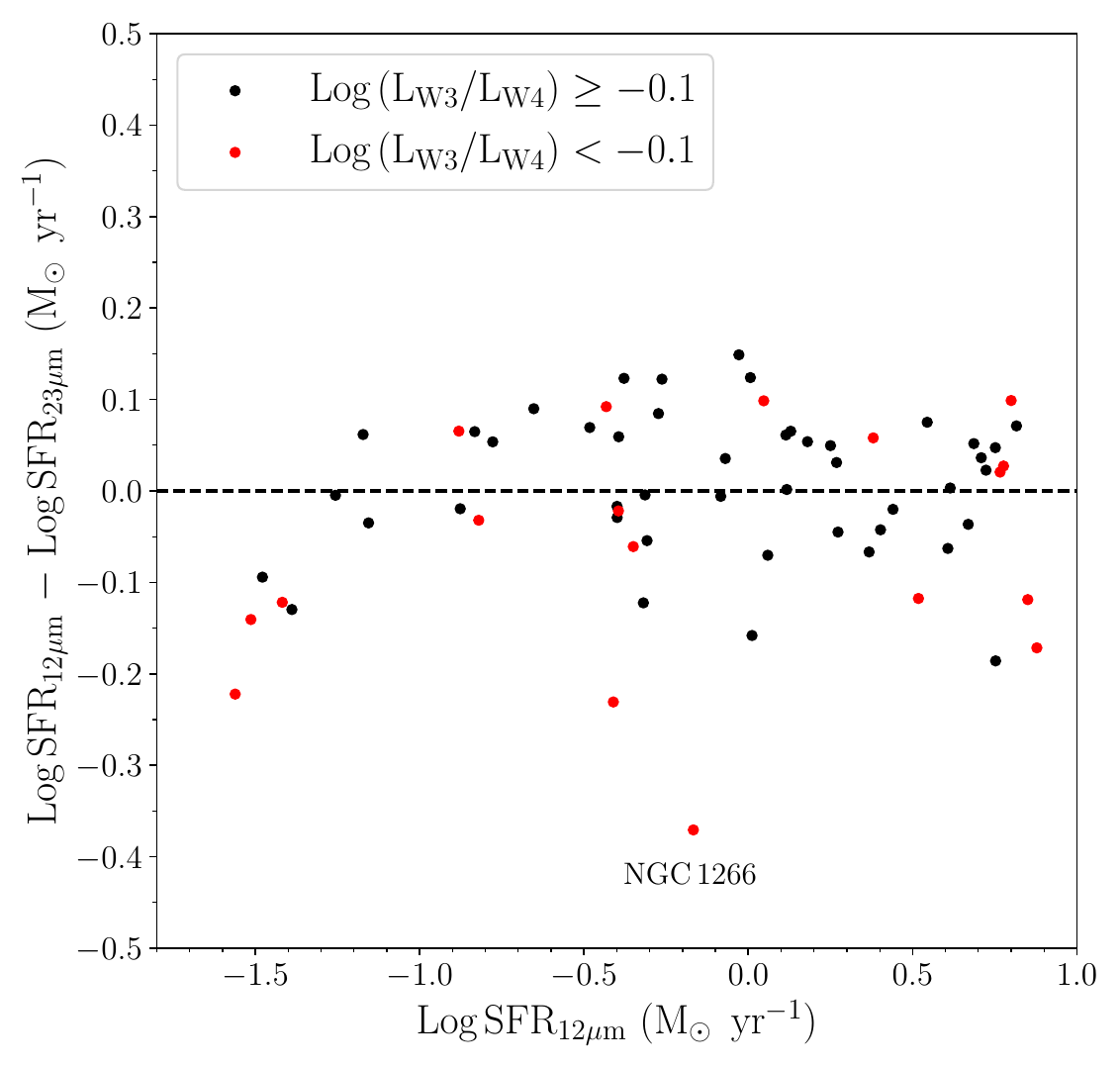}{0.5\textwidth}{(b)}}
\caption{Comparing the L$_{12\mu m}$ and L$_{23\mu m}$ SFRs:  (a) Using the SFR calibrations presented in Figure \ref{fig:singsfish}, i.e. making the distinction based on L$_{12\mu m}$/L$_{23\mu m}$ for SFR$_{23\mu m}$, we find good agreement between SFR$_{12\mu m}$ and SFR$_{23\mu m}$. The dashed line indicates one-to-one correspondance. (b) The comparison between SFR$_{12\mu m}$ and SFR$_{23\mu m}$ indicates that for log\,(L$_{12\mu m}$/L$_{23\mu m}$) $\geq-0.1$ sources the agreement is within $\pm 0.15$ dex.
Galaxy NGC\,1266 is the low-value outlier.
}
\label{fig:sfr_recalib_fin}
\end{center}
\end{figure*}

As in \citet{Cluver17}, we derive L$_{\rm TIR}$ using Equation 5 of \citet{Dale2014} which combines 8\micron, 24\micron, 70\micron, and 160\micron\ using the SINGS/KINGFISH photometry presented in \citet{Dale17}. We convert this to a SFR using the relation (Equation 3) of \citet[][]{Cal2013}, which uses the Starburst99 models \citep{Leith99} assuming Solar metallicity, constant star formation over t = 100 Myr, and a \citet{Kroupa2001} IMF:
\begin{equation}
\textrm{SFR} (M_\odot\, \textrm{yr}^{-1})= 2.8 \times 10^{-44} L_\textrm{TIR} \textrm{ (erg.s}^{-1}).   
\end{equation}

Figure \ref{fig:singsfish}a plots the relationship between log SFR$_{\textrm L_{\rm TIR}}$ and log\,L$_{12\mu m}$ color-coded by L$_{12\mu m}$/L$_{23\mu m}$ ratio. Somewhat surprisingly, the relationship between L$_{12\mu m}$ and SFR$_{\textrm L_{\rm TIR}}$ appears largely agnostic to this quantity. The one significant outlier is NGC\,1266, likely harboring an AGN \citep[e.g.][]{Alatalo2011}. The best-fit line and SFR relation is given by:

\begin{equation}
\begin{split}
\textrm{Log SFR}\,(M_\odot\, \textrm{yr}^{-1})= 0.89(\pm0.02) \, \textrm{Log}\, L_{12\mu m} (L_{\odot})\\ - 7.93(\pm0.20)  
\end{split}
\end{equation}
with a 1$\sigma$ scatter of 0.10 dex.

By contrast, Figure \ref{fig:singsfish}b reveals that what was thought to be increased scatter in \citet{Cluver17} can be attributed to two ``flavours" of sources, with log\,(L$_{12\mu m}$/L$_{23\mu m}$) $<-0.1$ galaxies being systematically lower in the log\,SFR$_\textrm{LTIR}$ versus log\,L$_{23\mu m}$ diagram. In reality, we imagine a continuum of values, but as a first step we fit each population separately resulting in updated SFR$_{23\mu m}$ relations as follows:

\begin{equation}
\begin{split}
\textrm{Log SFR}\,(M_\odot\, \textrm{yr}^{-1})= 0.91(\pm0.03)\, \textrm{Log}\, L_{23\mu m} (L_{\odot}) \\
- 8.02(\pm0.23)  
\end{split}
\end{equation}
with a 1$\sigma$ scatter of 0.12 dex for ``nominal" mid-infrared sources, i.e. log\,(L$_{12\mu m}$/L$_{23\mu m}$) $\geq-0.1$, and,

\begin{equation}
\begin{split}
\textrm{Log SFR}\,(M_\odot\, \textrm{yr}^{-1})= 0.89(\pm0.02)\, \textrm{Log}\, L_{23\mu m} (L_{\odot}) \\
- 8.18(\pm0.19)  
\end{split}
\end{equation}
with a 1$\sigma$ scatter of 0.14 dex for ``warm" mid-infrared sources, i.e. log\,(L$_{12\mu m}$/L$_{23\mu m}$) $<-0.1$.

The intermediate relationship between L$_{\rm TIR}$ and L$_{12\mu m}$ and L$_{23\mu m}$, respectively, is shown in Figure \ref{fig:sfr_recalib} and provided as Equations A1 for L$_{12\mu m}$ and A2 and A3 for L$_{23\mu m}$ in the Appendix.

Next, in Figure \ref{fig:sfr_recalib_fin}, we compare the L$_{12\mu m}$ and L$_{23\mu m}$ derived SFRs and find much better agreement than in \citet{Cluver17}. We note the outlier in Figure \ref{fig:sfr_recalib_fin}b with a log\,SFR$_{\rm 12 \mu m}$ -  log\,SFR$_{\rm 23 \mu m}$ value of $-0.38$, NGC\,1266. Although this galaxy has a warm \WIWII\ color, it does not exceed our threshold (see Figure \ref{fig:singshist}a) but is known to harbor an obscured AGN \citep{Alatalo2011}. The steep spectrum of NGC\,1266 means that its 23$\mu$m derived SFR is 1.55\,M$_\odot$\,yr$^{-1}$, compared to the 12$\mu$m derived value is 0.65\,M$_\odot$\,yr$^{-1}$. We include an SED of this source in Figure \ref{fig:seds} of the Appendix, clearly showing it to be one of the steepest sources in the sample \citep[see also Figure 6 in][]{Dale07}.

The improved calibration of SFR using L$_{12\mu m}$ and L$_{23\mu m}$ allows us to combine these using inverse-variance weighting (photometric + fitting uncertainties), 
since they are calibrated to the same system and do not appear to vary systematically; we designate this combined SFR as SFR$_{\rm MIR}$. Of course, this all hinges on how well L$_{\rm TIR}$ captures star formation and we remind the reader that the use of \wise\ as a SFR indicator should be an informed choice. We investigate this further in the following section.

\begin{figure*}[!thb]
\begin{center}
\gridline{\fig{ 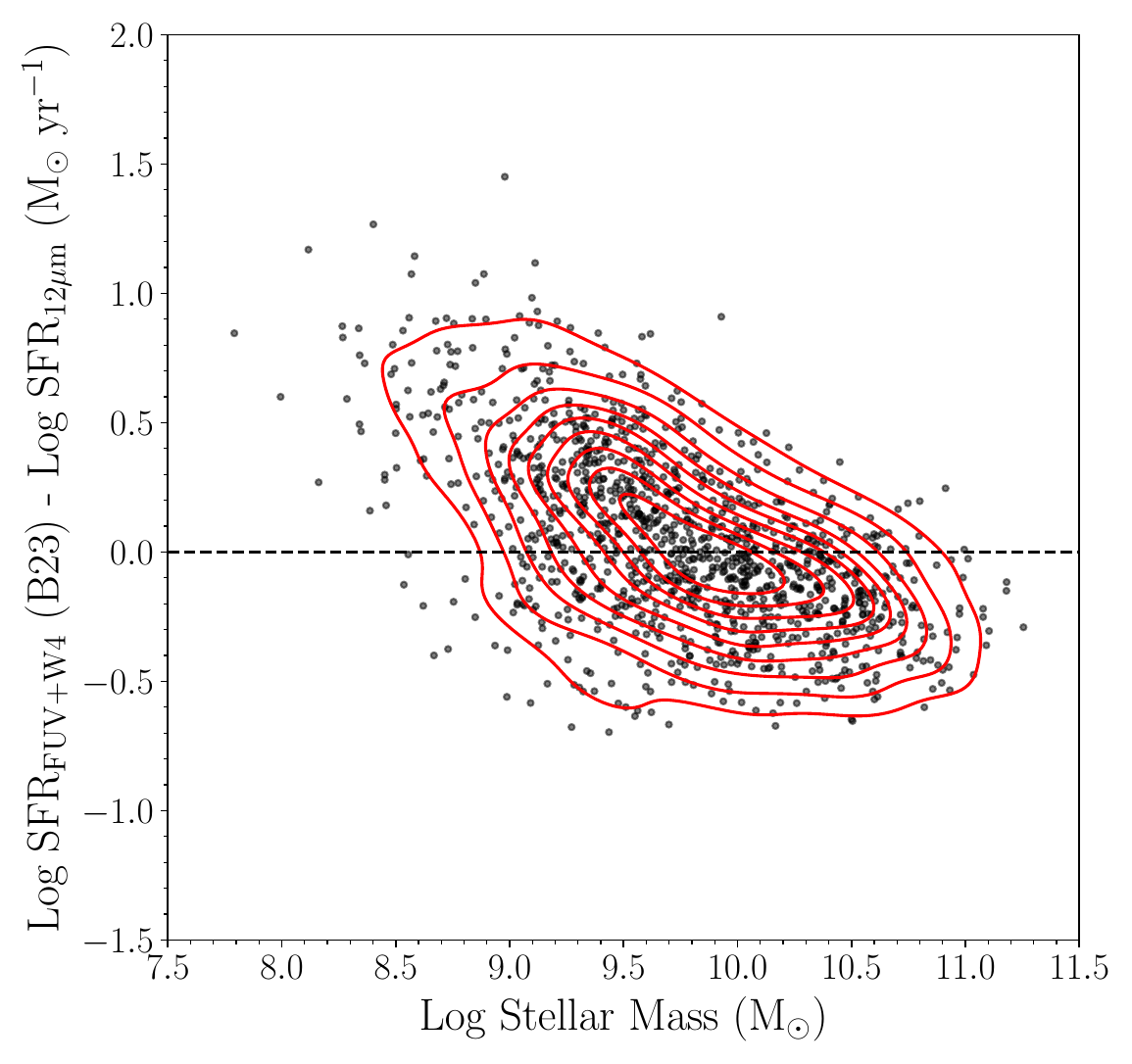}
{0.45\textwidth}{(a)}
\fig{ 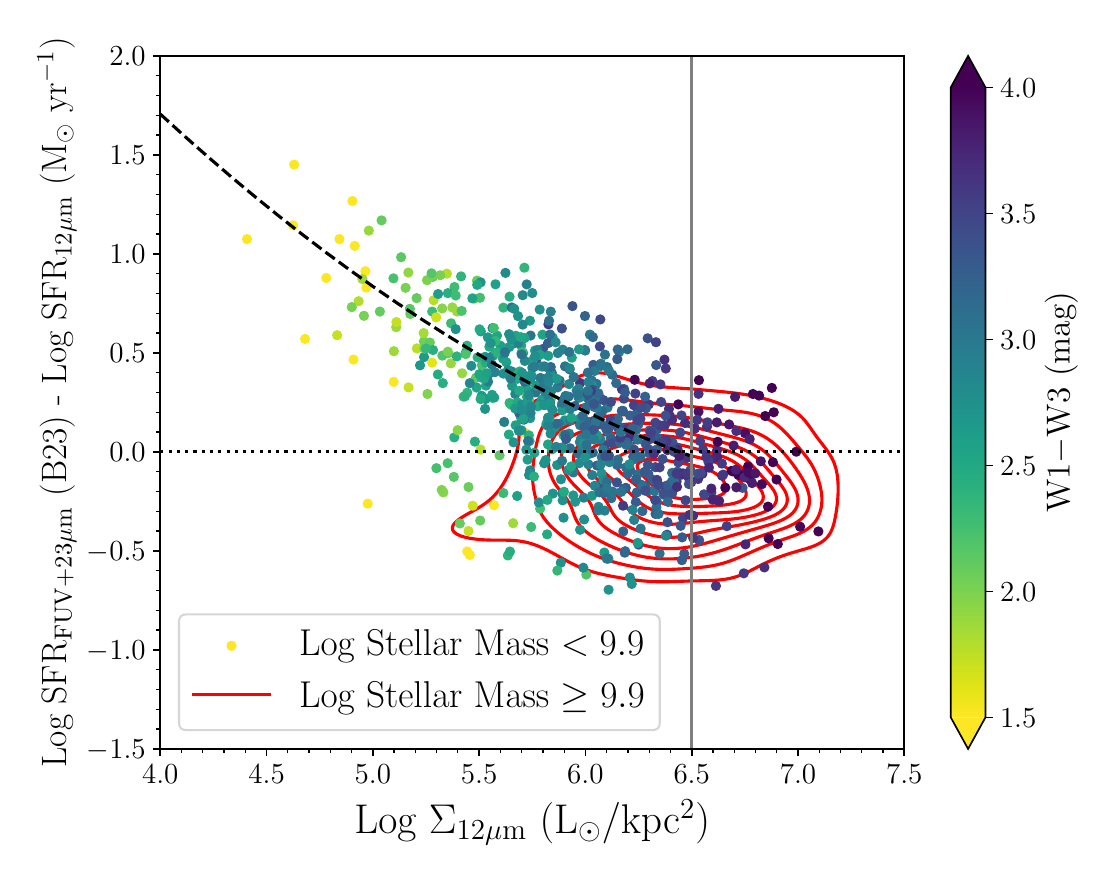}{0.55\textwidth}{(b)}
}
\gridline{\fig{ 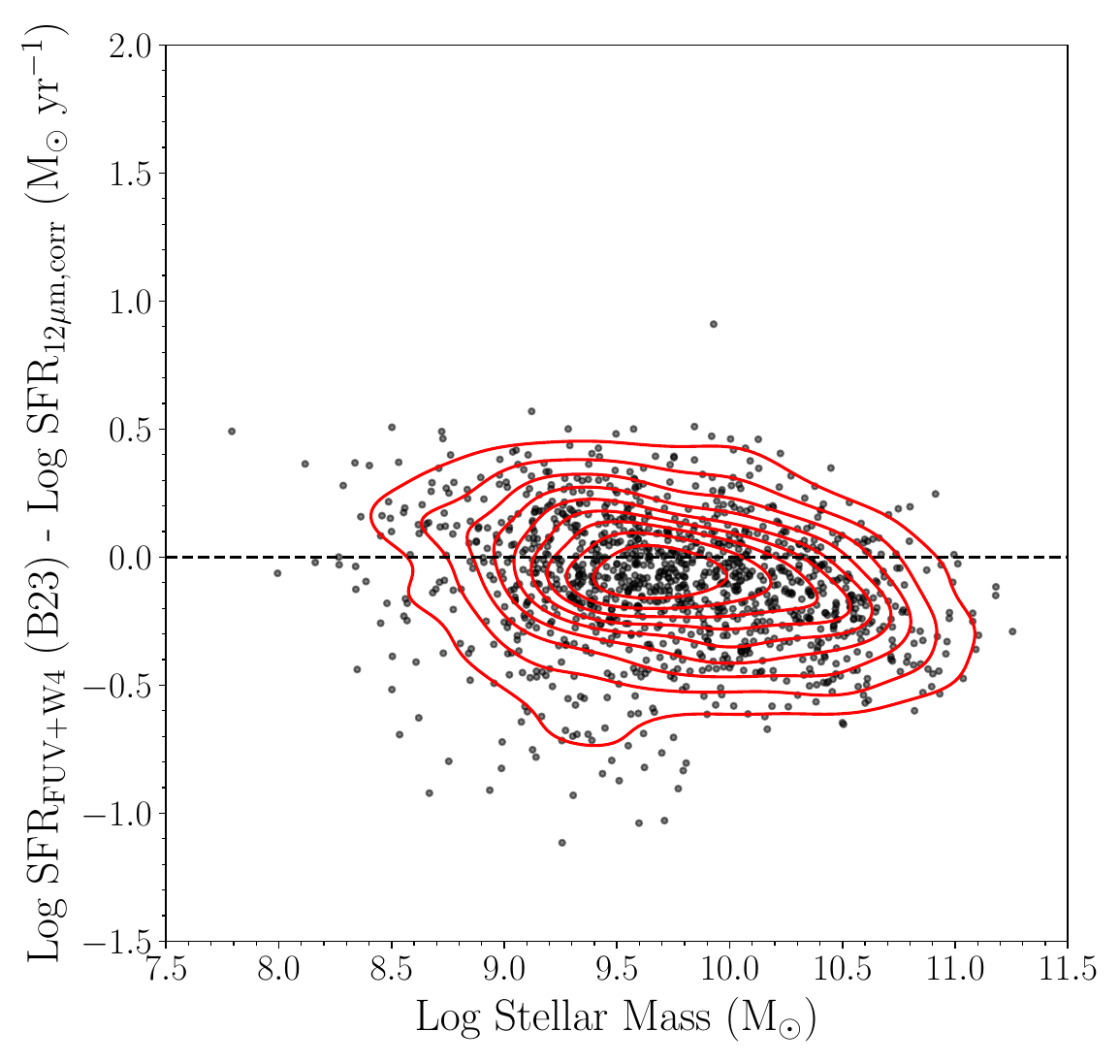}{0.45\textwidth}{(c)}
\fig{ 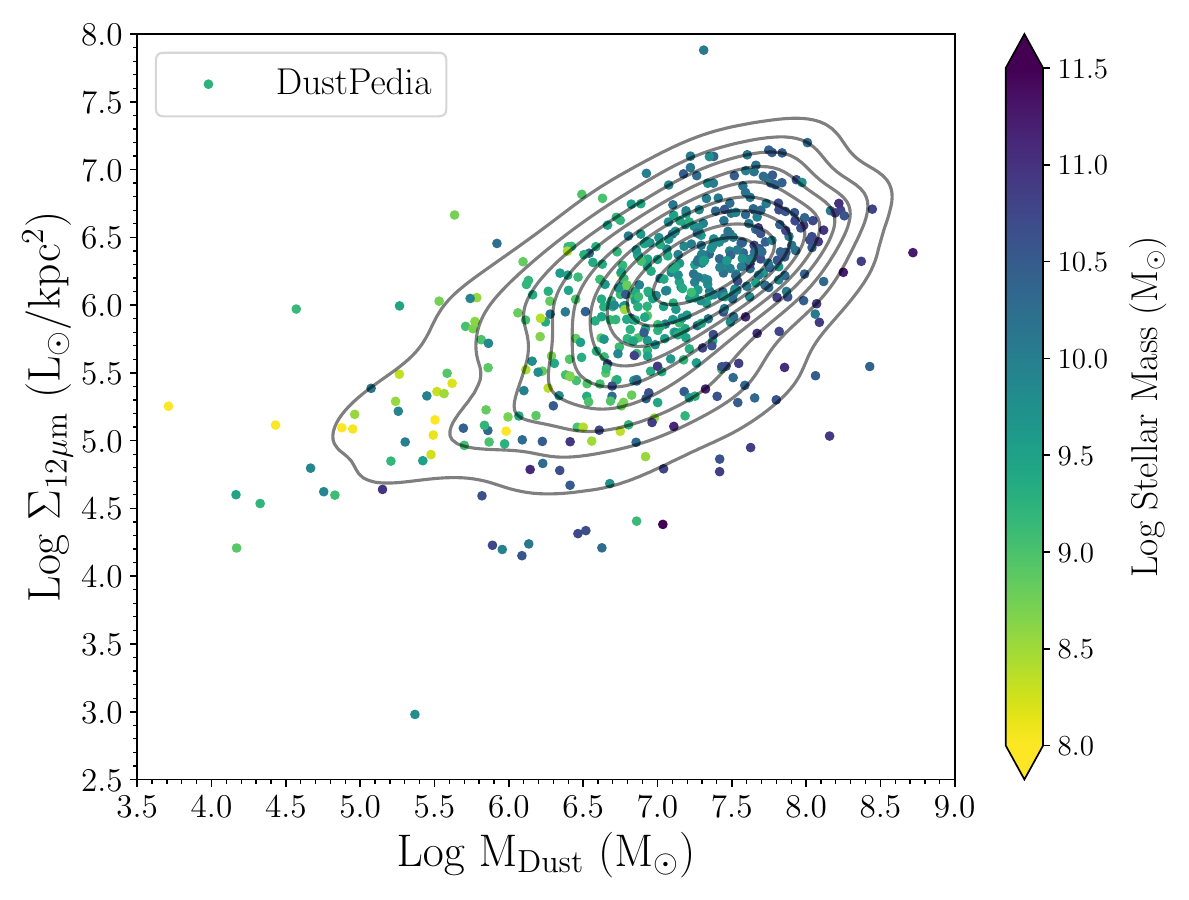}{0.55\textwidth}{(d)}
}
\caption{Assessing the relative difference between the UV and infrared SF tracers: 
(a) The difference between SFR$_{\rm FUV+W4}$ using the relation from \citet{Belf23} and SFR$_{\rm 12 \mu m}$ as function of stellar mass. (b) The difference between SFR$_{\rm FUV+W4}$ using the relation from \citet{Belf23} and SFR$_{\rm 12 \mu m}$ as function of 12\micron\ dust density ($\Sigma_{\rm 12\mu m}$ showing that sources with log \Mstar $<9.9$ show progressively lower SFR$_{\rm 12 \mu m}$ compared to SFR$_{\rm FUV+W4}$ as a function of $\Sigma_{\rm 12\mu m}$. (c) After applying a correction derived from Figure \ref{fig:sf_12_mass}b, the correspondence between SFR$_{\rm 12 \mu m}$ compared to SFR$_{\rm FUV+W4}$ as a function of stellar mass is vastly improved. (d) Using sources in common with the DustPedia project we find a correspondence between $\Sigma_{\rm 12\mu m}$ and dust mass as derived from SED-fitting using CIGALE. }
\label{fig:sf_12_mass}
\end{center}
\end{figure*}

\begin{figure}[!ht]
\begin{center}
\includegraphics[width=8.5cm]{ 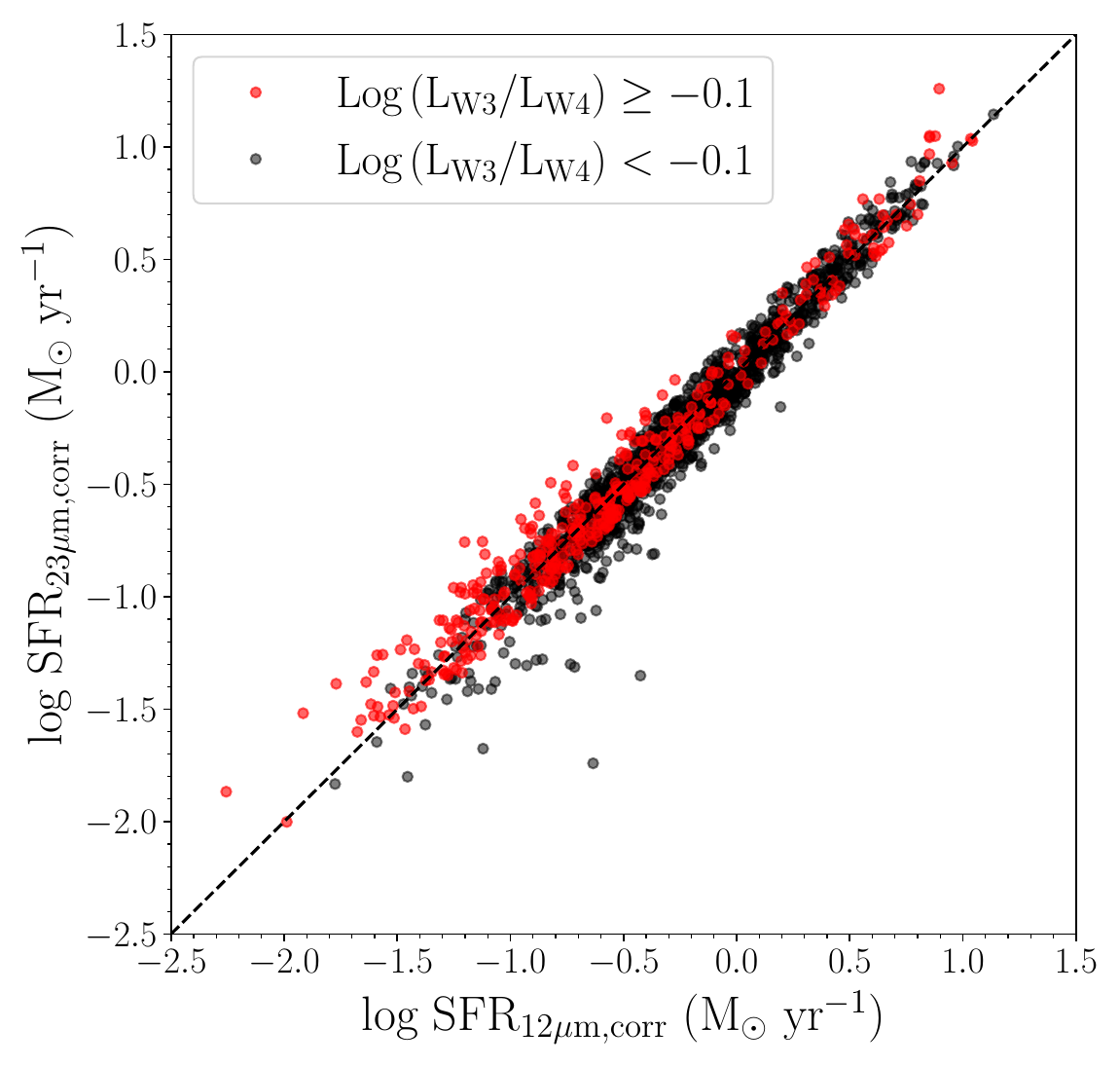}
\caption{Similar to Figure 8a, but now with the entire \spitzerg-\wise\ sample, and using the SFR$_\textrm{deficit}$ correction, we observe the relatively close correspondence between SFR$_{\rm 12\mu m, corr}$ and SFR$_{\rm 23\mu m, corr}$.}
\label{fig:mid_sfr_comp}
\end{center}
\end{figure}

\begin{figure*}[!tbp]
\begin{center}
\gridline{\fig{ 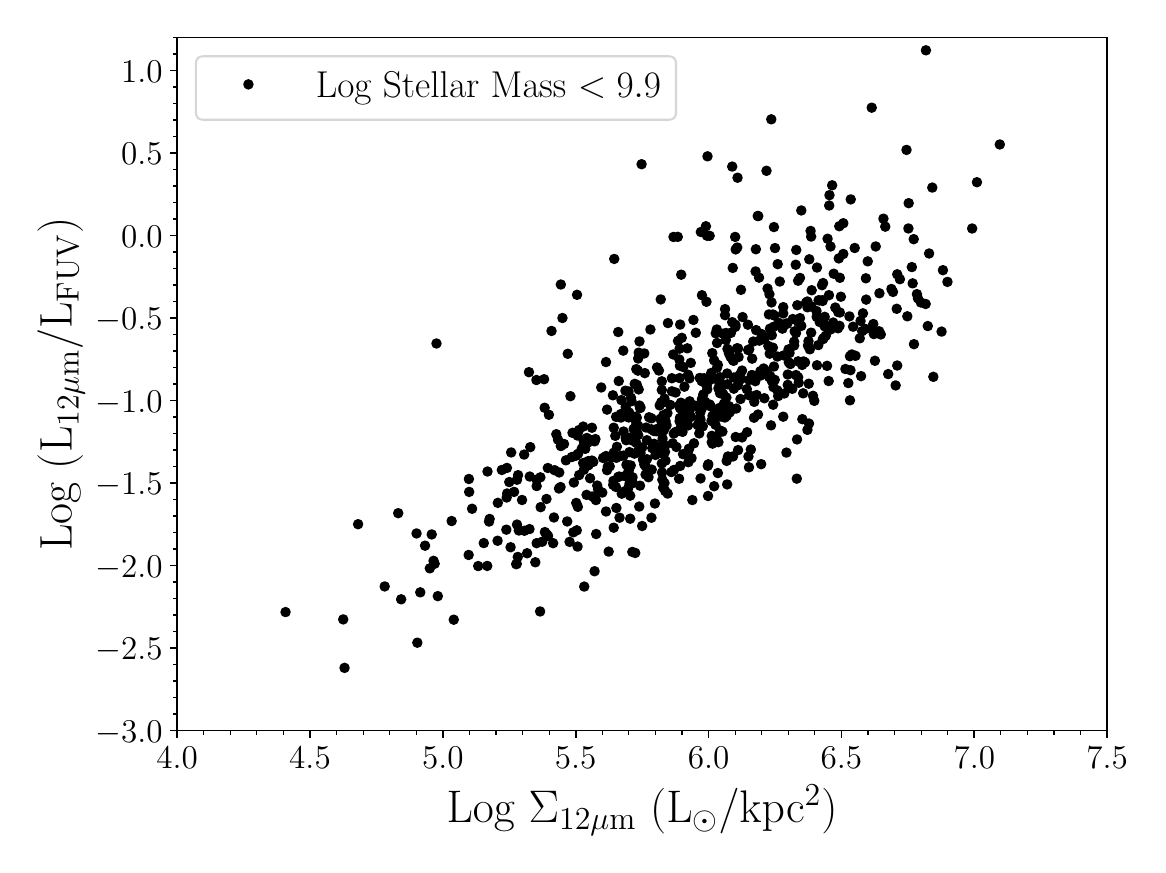}{0.5\textwidth}{(a)}
\fig{ 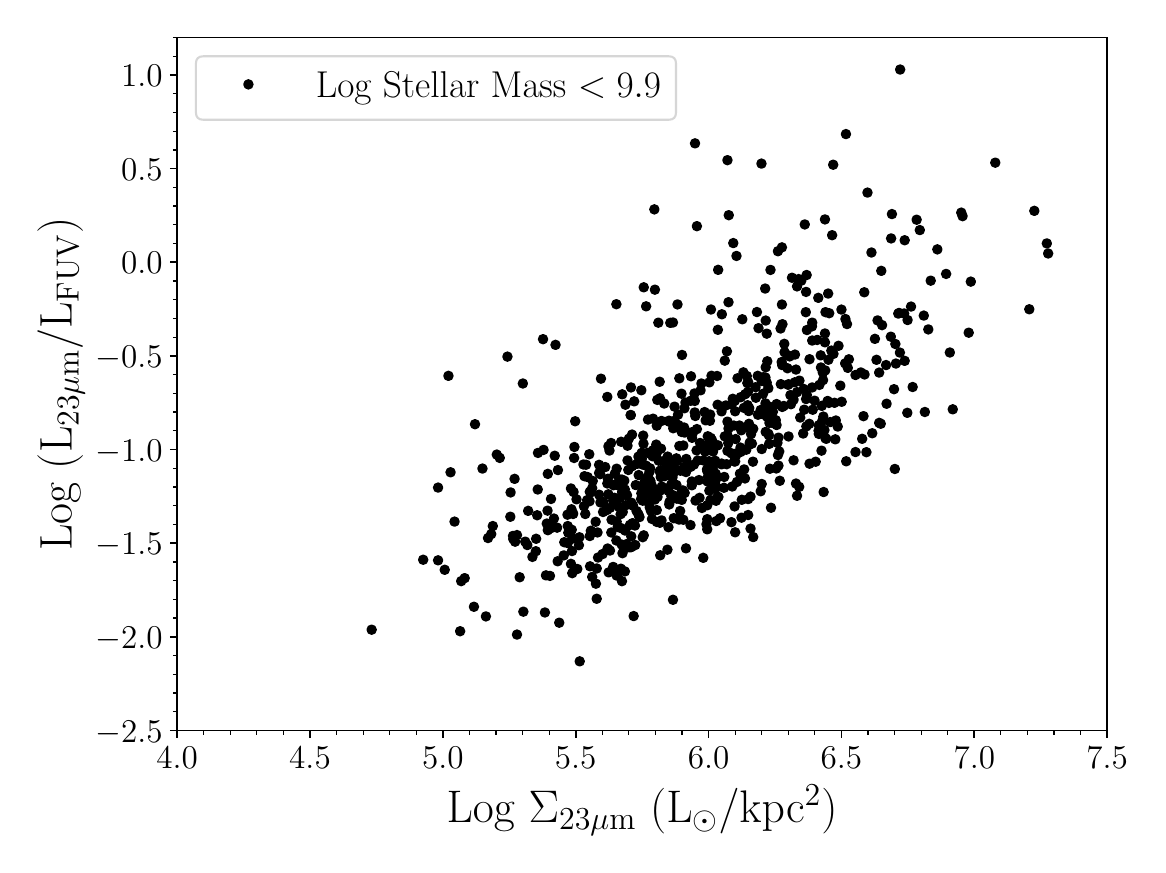}{0.5\textwidth}{(b)}}
\caption{The effect of dust density on the infrared-to-FUV correlation: 
for sources with log\,\Mstar$<9.9$, we show in (a) the broad correspondence between L$_{12\mu m}$/L$_{\rm FUV}$ and $\Sigma_{\rm 12\mu m}$, and in (b) the relationship between L$_{23\mu m}$/L$_{\rm FUV}$ and $\Sigma_{\rm 23\mu m}$. The latter appears to shows more scatter, likely driven by the poorer sensitivity of the W4 band.}
\label{fig:lumden}
\end{center}
\end{figure*}

\begin{figure*}[!thp]
\begin{center}
\includegraphics[width=16cm]{ 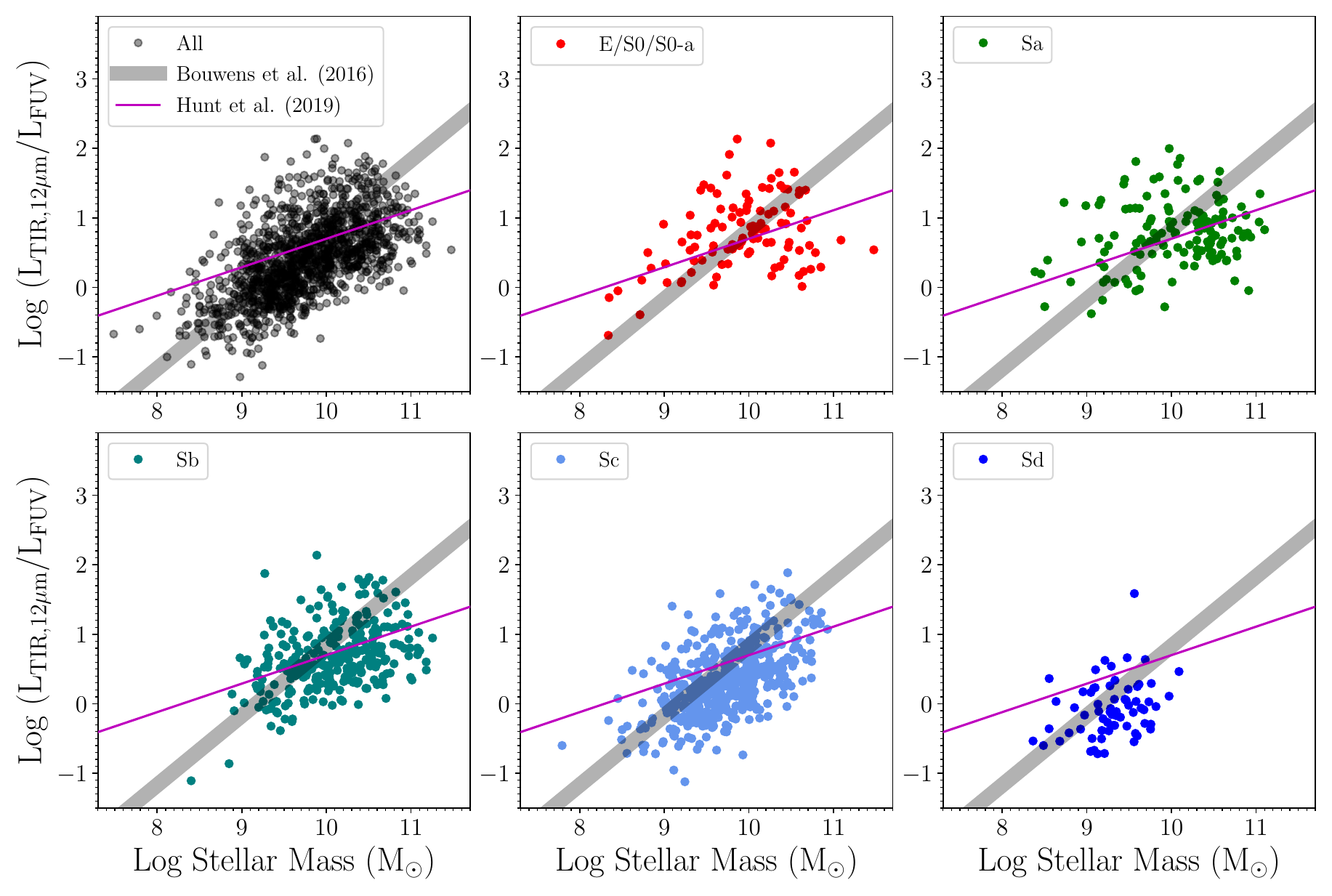}
\caption{Morphology and the infrared-to-UV correlation:  using the relationship between L$_{\rm 12 \mu m}$ and  L$_{\rm TIR}$ (see Section \ref{sings} of the Appendix), we calculate the proxy  L$_{\rm TIR, 12 \mu m}$/L$_{\rm FUV}$ ratio for our sample (top left panel) and then separated by morphological type. The relations from \citet{Bouw2016} and \citet{Hunt2019}  are shown for comparison. See also the similar diagrams in the Appendix, Figures~\ref{fig:IRX12} and \ref{fig:IRX23}, showing this ratio as a function of the dust density.}
\label{fig:lumdenmorph}
\end{center}
\end{figure*}

\subsubsection{Calibration to Hybrid UV$+$IR SFRs}

Although the mid-infrared can be considered a reasonable predictor of SFR for most star-forming galaxies, it will underestimate the SFR of low stellar mass galaxies with relatively low metallicity and dust content. At the other end of the mass spectrum, it will overestimate the current SFR where dust from past star formation is being preferentially heated by old stars (in this case an overly delayed measure of star formation). To investigate this further we exploit the UV data available for \spitzerg. 

\citet{bouquin+2018} present the GALEX-\spitzerg\ sample which furnishes GALEX FUV and NUV photometry for 1931 galaxies in \spitzerg. We make use of these measurements to explore hybrid calibrations that leverage tracers of both dust unobscured and obscured star formation. 

The FUV and NUV photometry is taken from \citet{bouquin+2018} and reddening-corrected using the recommended values from \citet{Wyder2007} derived using a \citet{Card89} extinction law. Galactic reddening is corrected using the relations from \citet{Peek2013} following \citet{Salim+2016}.

In Figure \ref{fig:sf_12_mass}a we use the hybrid FUV$+$W4 SFR indicator from \citet{Belf23} to examine the difference in SFR rate compared to SFR$_{12\mu m}$, derived in the previous section, as a function of \Mstar. There is a pronounced deviation for masses Log\,\Mstar\, $<9.9$, where SFR$_{12\mu m}$ is progressively underestimating the SFR. The question arises as to whether this paucity in dust emission can be quantified. In Figure \ref{fig:sf_12_mass}b we plot the SFR difference as a function of L$_{\rm W3}$ luminosity density, or $\Sigma_{\rm 12\mu m}$ (L$_\odot$/kpc$^2$). This corresponds to L$_{12\mu m}$, i.e. W3 emission with the stellar contribution removed, normalised by the area of the galaxy (in kpc$^2$) as measured in W1 (which traces the entire stellar disk). We see that for sources with Log\,\Mstar\, $\geq 9.9$ the points cluster around 0, but for Log\,\Mstar\ $< 9.9$ the offset increases, also reflected by the decreasing \WIWIII\ color indicating more stellar-dominated emission. This deviation can be described by the following function:
\begin{equation}
{\rm Log\, SFR}_{\rm deficit} = {\rm a} + {\rm b}\,{\rm Log}\Sigma_{\rm 12\mu m} + {\rm c}\, ({\rm Log}\Sigma_{\rm 12\mu m})^2,
\end{equation}\\
where a$ =7.56$, b$=-1.94$, and c$ =0.119$.\\

If we apply this as:
\begin{equation}
{\rm Log\, SFR}_{\rm 12\mu m, corr}  = {\rm Log\, SFR}_{\rm 12\mu m} + {\rm Log\, SFR}_{\rm deficit}
\end{equation}

for sources with Log \Mstar\,$<9.9$ M$_\odot$ and where Log\,$\Sigma_{\rm 12\mu m}$ $<6.5$ L$_\odot$/kpc$^2$ (see Figure \ref{fig:sf_12_mass}b), the result is shown in Figure \ref{fig:sf_12_mass}c. 

Similarly we can define:
\begin{equation}
{\rm Log\, SFR}_{\rm 23\mu m, corr}  = {\rm Log\, SFR}_{\rm 23\mu m} + {\rm Log\, SFR}_{\rm deficit}
\end{equation}

In the absence of UV information, one can therefore estimate the degree to which the \WISE-derived SFRs are underestimating the total SFR. We caution that for any one source, there would be a high degree of uncertainty, but for studying the behaviour of large populations it provides a more accurate reflection of the SFR for the low mass population in the absence of UV photometry. We also note that Log\,$\Sigma_{\rm 12\mu m}$ is by its nature a global measurement, essentially scaling with the dust mass (see below), and is designed as a broad indicator of the dust deficit, and hence SFR deficit in the low dust regime. 

We note that there are sources in Figure \ref{fig:sf_12_mass}c where SFR$_{\rm 12\mu m, corr}$ appears to be an overestimate compared to SFR$_{\rm FUV + W4}$; these are analysed further in Section \ref{hybrid} of the Appendix, which suggests that the UV emission is low compared to both 12$\mu$m and 23$\mu$m and (importantly) there is no correlation with stellar mass.

To further investigate the $\Sigma_{\rm 12\mu m}$ quantity, we make use of the DustPedia project \citep{Davies2017}, which has 725 galaxies in common with \spitzerg, and has SED-derived properties \citep{Dust19} modelled using CIGALE \citep{Boq2019}; we use the outputs derived using the \citet{Drain14} dust model. 

Figure \ref{fig:sf_12_mass}d makes a comparison of Log\,$\Sigma_{\rm 12\mu m}$ and SED-derived dust mass, which we would expect to be correlated. Indeed we see that the quantities do track each other despite the relatively simple nature of the Log\,$\Sigma_{\rm 12\mu m}$ metric. Of course, the latter is only feasible in the low-redshift Universe where the isophotal size of the galaxy has been carefully measured. 

In Figure \ref{fig:mid_sfr_comp} we show the comparison of SFR$_{\rm 12\mu m, corr}$ and SFR$_{\rm 23\mu m, corr}$ for the \spitzerg-\wise\ sample, excluding sources with low S/N measurements (S/N $<5$). Further comparisons of SFR$_{\rm 12\mu m, corr}$ with several hybrid UV$+$IR from the literature can be found in Figure \ref{fig:sfrcomp} of the Appendix.

\begin{figure*}[!thp]
\begin{center}
\gridline{\fig{ 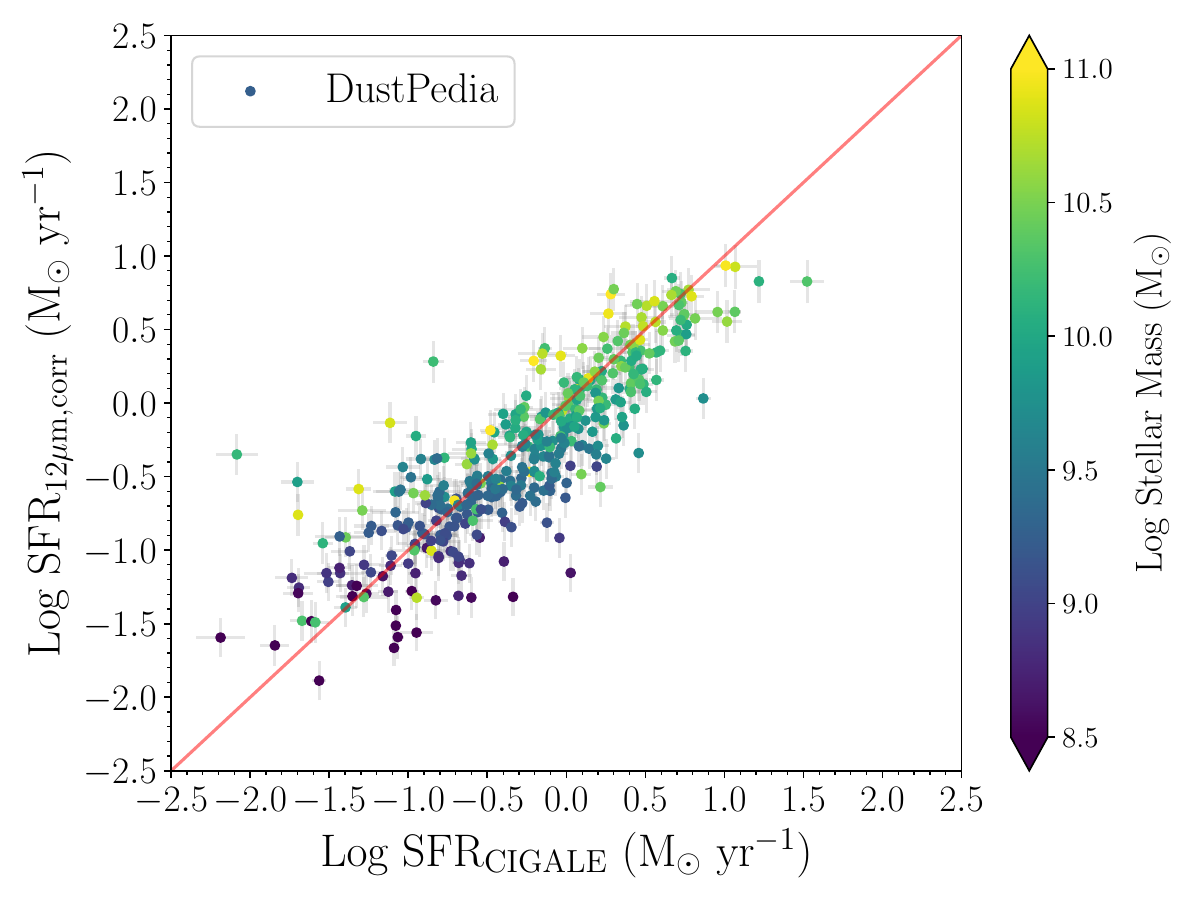}{0.5\textwidth}{(a)}
\fig{ 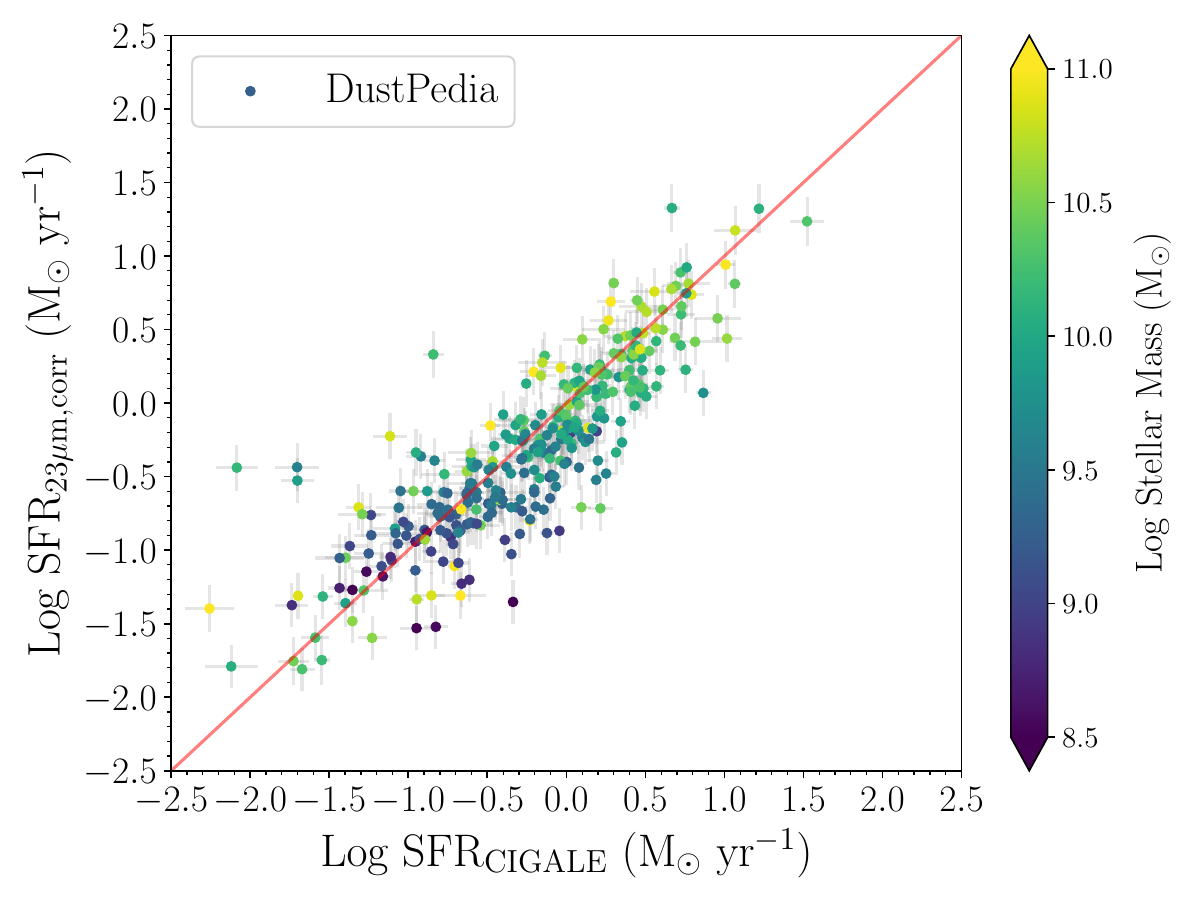}{0.5\textwidth}{(b)}}
\gridline{\fig{ 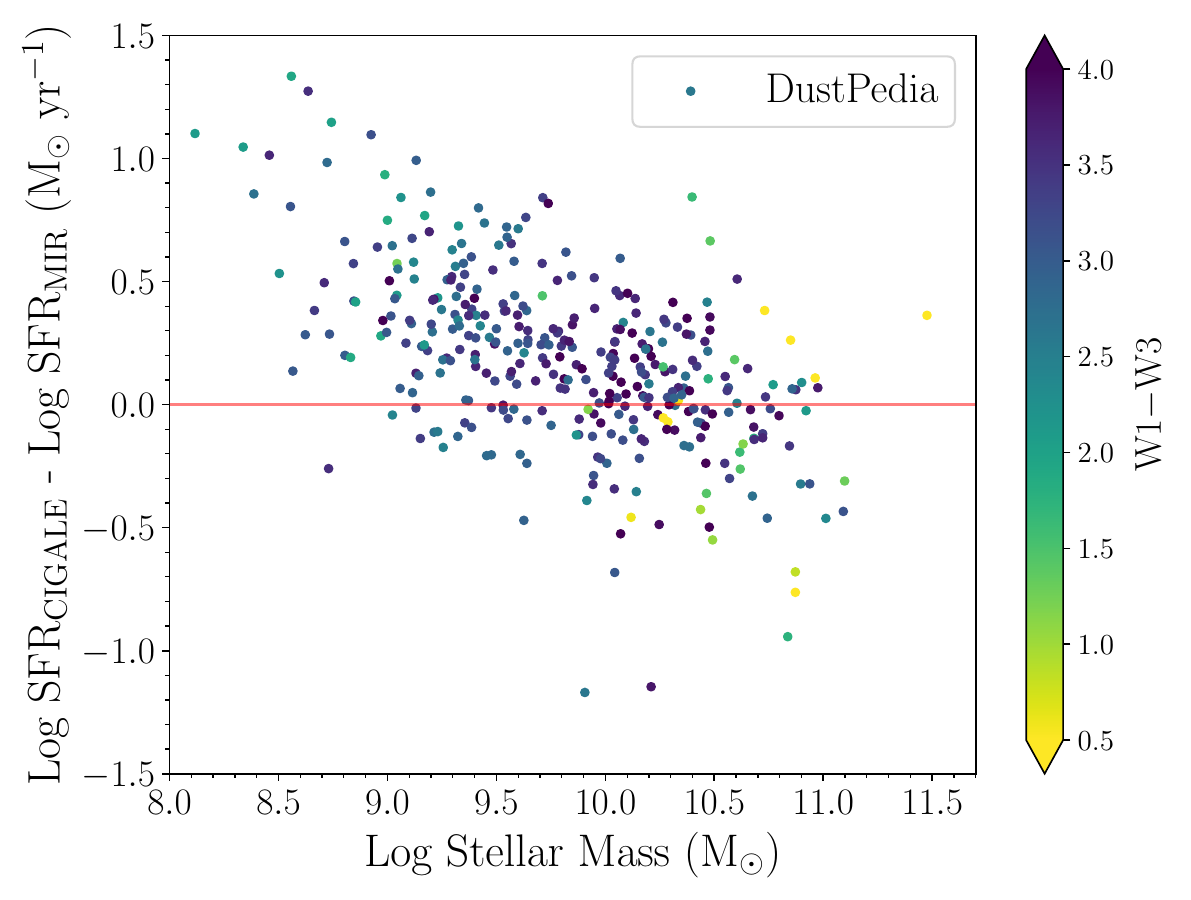}{0.5\textwidth}{(c)}
 \fig{ 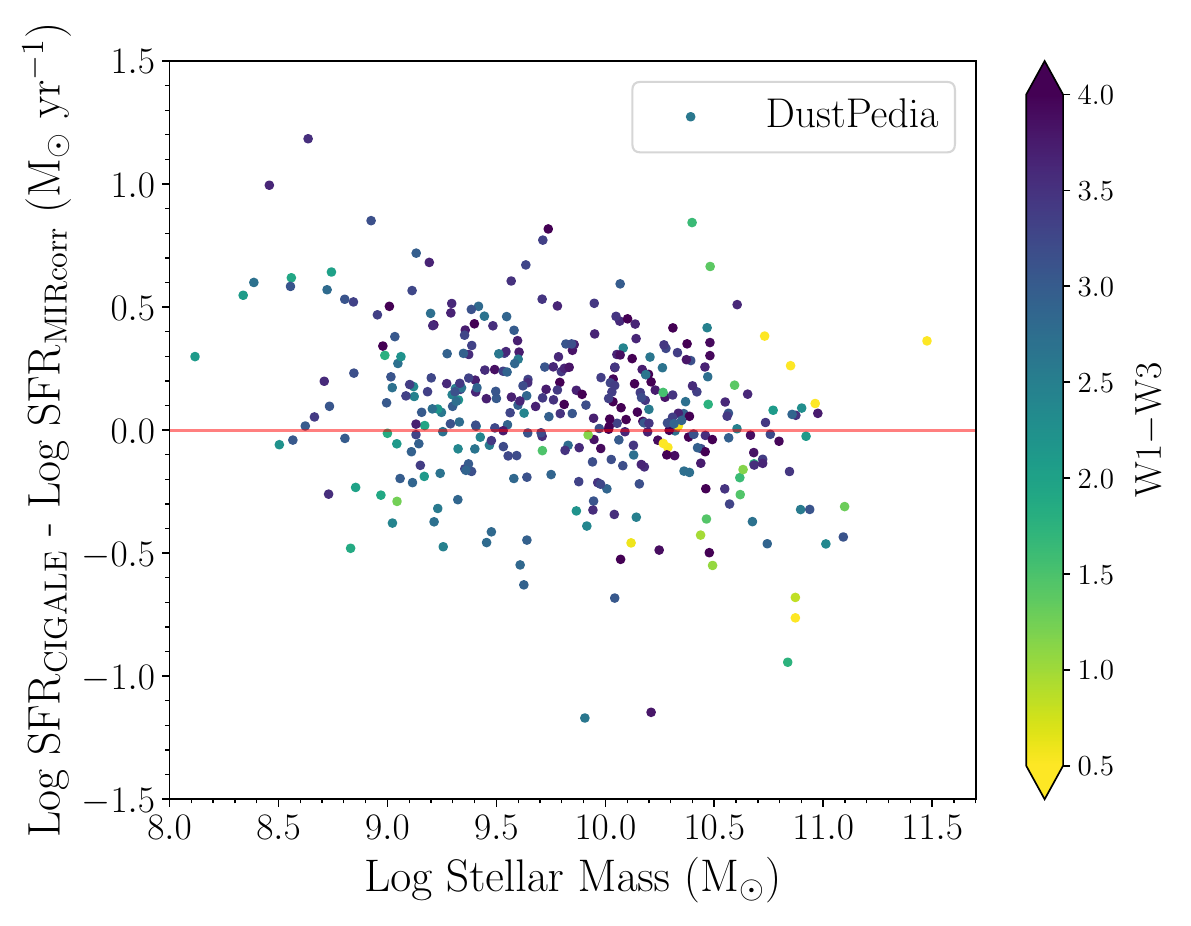}{0.5\textwidth}{(d)}}
\caption{Comparison between CIGALE-derived SFRs from DustPedia  with (a) SFR$_{\rm 12\mu m, corr}$ and (b) SFR$_{\rm 23\mu m, corr}$. In (c) we compare SFR$_{\rm CIGALE}$ to the inverse-variance weighted combination of SFR$_{\rm 12\mu m}$ and SFR$_{\rm 23\mu m}$ (SFR$_{\rm MIR}$), i.e. with no correction applied at low mass, as a function of log\,\Mstar. We see that the clear deficit observed is largely ameliorated in (d) which instead compares SFR$_{\rm CIGALE}$ to the inverse-variance weighted combination of SFR$_{\rm 12\mu m, corr}$ and SFR$_{\rm 23\mu m, corr}$ (SFR$_{\rm MIRcorr}$) as a function of log\,\Mstar.}
\label{fig:dustpedia}
\end{center}
\end{figure*}

We next investigate 
$\Sigma_{\rm 12\mu m}$, and similarly, $\Sigma_{\rm 23\mu m}$ as an ISM  activity metric; in Figure \ref{fig:lumden}a and b we see that these quantities broadly track the ratio of dust to FUV luminosity as traced by L$_{\rm 12\mu m}$ and L$_{\rm 23\mu m}$, for galaxies with Log\,\Mstar\,$<9.9$ (where we have also applied a T-type $>0.5$ requirement to remove low mass spheroidals). The L$_{\rm 12\mu m}$/L$_{\rm FUV}$ and L$_{23\mu m}$/L$_{\rm FUV}$ ratios for the full \spitzerg-\wise\ UV sample (split by morphology) are included in Section \ref{IRX} of the Appendix; this shows that the dust to UV ratios are well-correlated to dust content for late-type galaxies, but show progressively more scatter with earlier types.

The relationship between L$_{\rm TIR}$ and L$_{12\mu m}$ (Equation A1 of the Appendix) can be used to approximate the infrared excess (IRX$\equiv$ L$_{\rm TIR}$/L$_{\rm FUV}$); this is shown in Figure \ref{fig:lumdenmorph} as a function of \Mstar\ separated by morphological type, where we include the relations from \citet{Hunt2019}, derived using a $z\sim0$ sample, and from \citet{Bouw2016} derived for $z=2-10$, for comparison. The agreement appears closest with the \citet{Bouw2016} relation, and perhaps unsurprisingly, for the Sc morphological types, but we note that our sample lacks power at the high mass end (log\,\Mstar $>10.5$). We see that some bulge-dominated (E/S0/S0-a and Sa) galaxies are represented, albeit with larger scatter in the distribution compared to disk-dominated morphologies (Sb to Sd). Noting these are predominately not high mass systems (log\,\Mstar $>11$), the dust and UV properties of these ``early types" are not consistent with being passive systems. This is discussed further in Section \ref{IRX} of the Appendix, but a more complete analysis of these systems is deferred to a future publication.

\begin{figure*}[!thp]
\begin{center}
\includegraphics[width=10.3cm]{ 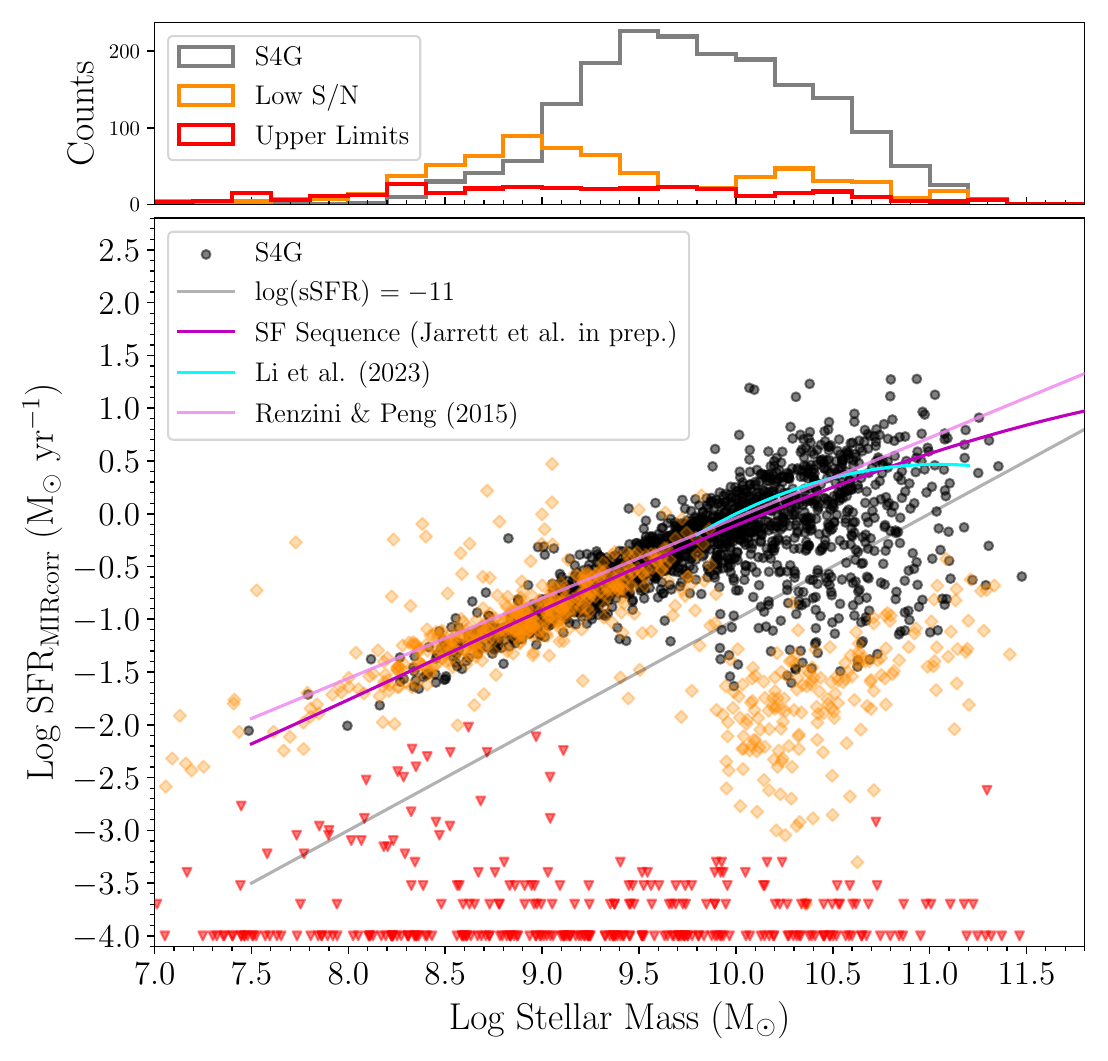}
\caption{The Star-forming ``Main Sequence", showing star formation relative to star formation history, for the \spitzerg-\wise\ sample. The fit derived using a large sample from the 2MRS (Jarrett et al., in prep.) is shown in magenta, the relation of \citet{RP2015} in pink and \citet{LiHo23} in cyan. Low S/N sources (where either L$_{12\mu m}$ or L$_{23\mu m}$ have a S/N $<5$) are shown in orange, and upper limits as red triangles. 
}
\label{fig:sfr}
\end{center}
\end{figure*}

\subsubsection{Comparison to DustPedia CIGALE-derived SFRs}

As mentioned in the previous section, \spitzerg-\wise\ has 725 galaxies in common with the nearby galaxy DustPedia project \citep{Davies2017} where SED-fitting of multiband photometry \citep{Dust19} using CIGALE \citep{Boq2019} can provide an additional test of the \wise-derived SFRs. 

In order to compare SFRs we select sources with well-measured L$_{12\mu m}$ and L$_{23\mu m}$ (S/N $>5$) and with DustPedia SFRs with errors $<40\%$. In Figure \ref{fig:dustpedia}a and b we compare the deficit-corrected \wise-derived SFRs to the CIGALE-derived values, color-coded by Log\,\Mstar. Both SFR$_{\rm 12\mu m, corr}$ and SFR$_{\rm 23\mu m, corr}$ show good agreement with CIGALE-derived star formation rates. It is encouraging that we do not see clear deviation at low levels of SFR, a regime where SFR indicators can produce significantly different values \citep[see][]{LiHo23}. However, our relatively small sample size is a clear limitation.

In Figure \ref{fig:dustpedia}c and d we consider the comparison of DustPedia-derived SFRs to SFR$_{\rm MIR}$, which is the inverse-variance weighted combination of SFR$_{\rm 12\mu m}$ and SFR$_{\rm 23\mu m}$, and SFR$_{\rm MIRcorr}$ which is the inverse-variance weighted combination of SFR$_{\rm 12\mu m, corr}$ and SFR$_{\rm 23\mu m, corr}$. As a function of \Mstar, we see that SFR$_{\rm MIR}$ clearly underestimates the SFR at low mass (Figure \ref{fig:dustpedia}c); this effect is largely ameliorated in Figure \ref{fig:dustpedia}d after applying the SFR$_{\rm deficit}$ correction (as presented in the previous section). In this figure, galaxies have been color-coded by \WIWIII\ color which indicates good agreement between the two SFR methods, irrespective of star formation activity (where \WIWIII\ acts a proxy for specific star formation). In particular, we do not see a clear overestimation of SFR determined using the mid-IR at high \Mstar, which could be ascribed to dust heated by old stars, as opposed to tracing recent star formation. However, we note that the \spitzerg\ sample does not contain many high \Mstar\ galaxies and the limited sample overlap of DustPedia and \spitzerg\ in this regime does not provide a fully reliable test. The accuracy of \wise-derived SFRs in the high \Mstar\ regime will be investigated in a future study using a larger sample drawn from the WXSC \citep[notably including galaxies from the 2MRS;][]{Huchra12}.

\subsection{Stellar Mass and Star Formation in the Nearby Universe}

In this section we examine the relationship between star formation and stellar mass in the \spitzerg\ sample. We derive \wise\ stellar masses using the prescription of \citet{Jarrett2023}, but since the \wise-derived SFRs assume a \citet{Kroupa2001} IMF, we apply a conversion to a \citet{chabrier+2003} IMF using the offsets from \citet{Zahid2012} to be fully consistent (corresponding to a near negligible shift of 0.03 dex). 

\begin{figure*}[!thp]
\begin{center}
\includegraphics[width=16cm]{ 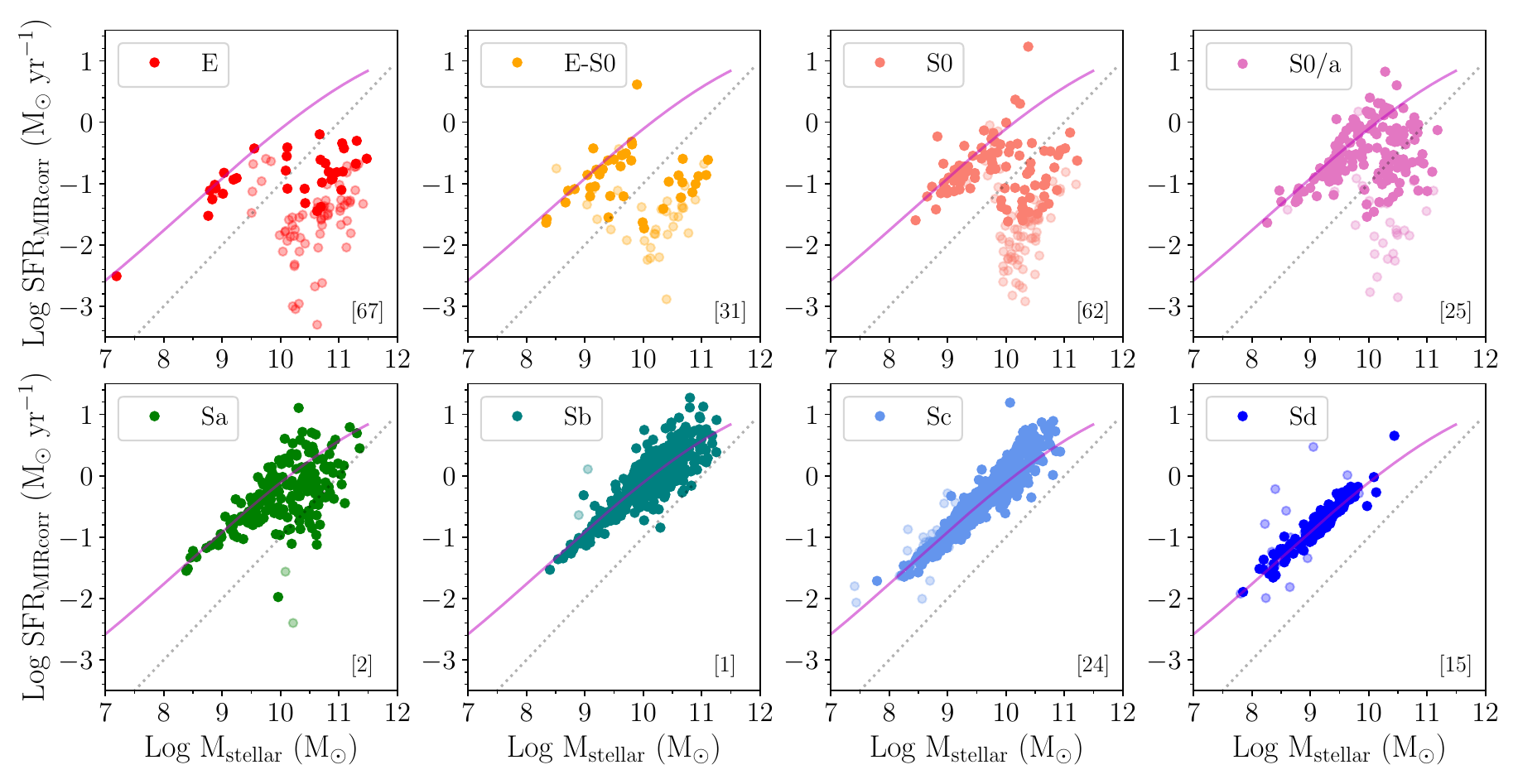}
\caption{Examining the location of different morphological types in the SFR-\Mstar\ plane, the SFMS line from the previous figure (Jarrett et al., in prep.) is shown as the magenta line, and log\,SFR/\Mstar $=-11$ as the dotted grey line. Low S/N sources (where either L$_{12\mu m}$ or L$_{23\mu m}$ has a S/N $<5$) are shown as transparent points, indicating that even given the proximity of the \spitzerg\ sample, low levels of star formation are largely associated with low confidence measurements. The number in the bottom right of each panel reflects the number of galaxies not shown since they are star formation non-detections.}
\label{fig:MS}
\end{center}
\end{figure*}

\begin{figure*}[!thp]
\begin{center}
\includegraphics[width=16cm]{ 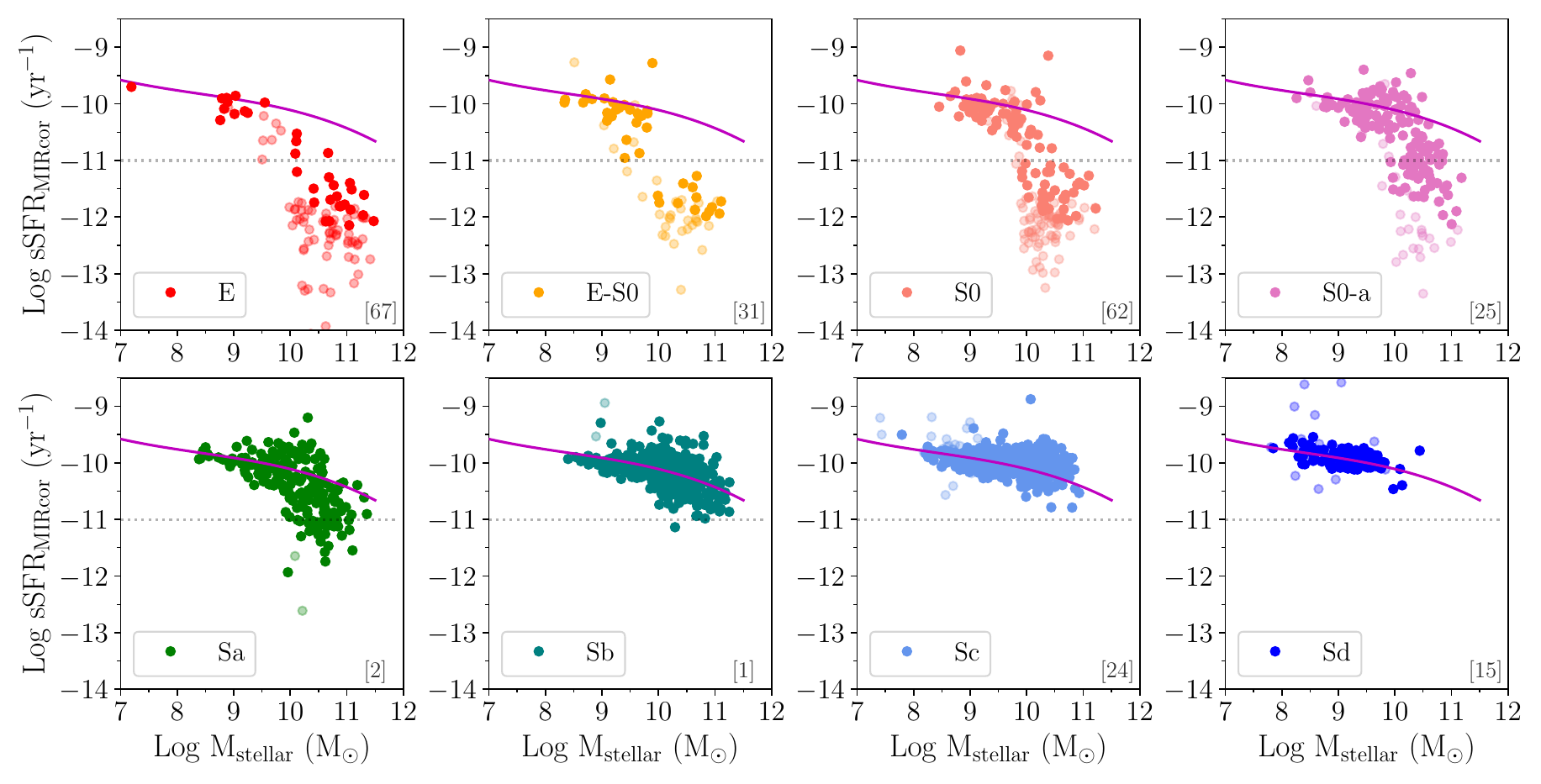}
\caption{Same as the previous figure, but now showing specific SFR, i.e. log\,SFR/\Mstar\ on the y-axis. This more clearly shows the correlation between morphology and galaxies moving off the so-called SFMS.}
\label{fig:ssfr}
\end{center}
\end{figure*}

In Figure \ref{fig:sfr} we plot the \spitzerg\ sample in the log\,SFR$_{\rm MIRcorr}$-log\,\Mstar\ plane. We make the distinction between low S/N sources (where either L$_{12\mu m}$ or L$_{23\mu m}$ have a S/N $<5$) and well-determined SFRs in the sample, and also indicate sources where only an upper limit for the SFR is achievable. The latter does not necessarily mean the original detection W3 detection is low S/N, but could instead arise where the stellar continuum was the dominant source of flux in the band. Low S/N sources appear at both low and high mass reflecting two different low star formation regimes, where the former may be the result of low dust content in low metallicity, but actively star-forming low mass galaxies. We also show the so-called Star forming Main Sequence (SFMS) fit from Jarrett et al. (in prep.) which is derived by combining the \spitzerg\ sample with $\sim$14\,000 sources from the 2MASS Redshift Survey \citep{Huchra12}; this fit shows excellent agreement with the star-forming galaxies in \spitzerg. 

The division between efficiently star-forming galaxies and those that have moved off the SFMS is often delineated using specific star formation rate (sSFR; log\, SFR/\Mstar), for example, log\, SFR/\Mstar $=-11$ \citep[e.g.][]{Wetzel2012, Houston2023}; this is indicated on Figure \ref{fig:sfr} as the grey line. 

Examining the location of sources in Figure \ref{fig:sfr}, we see that even given the proximity of the \spitzerg\ sample, low levels of measured star formation for Log\,\Mstar $>10$ are predominantly associated with low S/N sources. We therefore caution users from interpreting very low \wise-derived SFRs (log\,SFR $<-2$) as these may be spurious measurements or reflect low levels of dust emission that may not be associated with recent star formation. In any event, these galaxies have star formation activity well below the line used to delineate galaxies that have moved off the SFMS. The number of galaxies without detectable star formation (and which are therefore not shown) are indicated by the number in the bottom right corner of each plot. 

We next sub-divide the \spitzerg-\wise\ sample by morphological type, showing the Log\,SFR--Log\,\Mstar\ plane in Figure \ref{fig:MS} and Log\,sSFR--Log\,\Mstar\ plane in Figure \ref{fig:ssfr}. This shows that disk-dominated galaxies (Sc and Sd) exhibit a tight locus around the SFMS. The addition of a more prominent bulge (Sa and Sb) appears to effectively broaden the distribution of the star-forming population, with galaxies increasingly leaving the SFMS as they become bulge-dominated (from S0/a to E), reflecting their decreased likelihood of exhibiting active star formation. The nature of the dust emission in particularly the E and S0 populations, and for example what an appropriate \WIWIII\ color cut would be to distinguish passively heated dust from active star formation, will be the subject of a future publication.

As noted in section 3.1, the S0 and S0/a population occupies a broad range of \WIWIII\ color; we see that this behaviour is consistent with what is observed in the Log\,SFR--Log\,\Mstar\ plane, i.e. showing widely distributed star-forming properties relative to stellar mass content. 

Figure \ref{fig:ssfr}, employing sSFR in place of SFR, more clearly shows the transition from the star-forming sequence to being ``quenched". In Table \ref{tab:MorphQfrac} we summarise the total number of galaxies in each morphological type, the percentage that lie within the star-forming sequence (i.e. above log\, SFR/\Mstar $=-11$), those below, and galaxies where no star formation was detected. Given the proximity of the sample, galaxies classified as Sc and Sd with non-detectable star formation is likely due to relatively low dust content, and not necessarily the absence of star formation (as would be more likely true of the early-type systems). 

\begin{table*}[!htp]
\centering
    \begin{tabular}{lccccc}
       \multicolumn{6}{c}{Star Formation (SF) Classification by Morphological Type}\\
       \hline
       \hline
Class & T-Type & Total & SF & Below SF & SF\\
 &    & Galaxies  & Sequence  & Sequence  & Not Detected \\
\hline
\hline
E    &  -5.0\,--\,-3.5 &  169  & 13.0\% & 48.5\% &  38.5\%\\
E/S0  & -3.5\,--\,-2.5 &  102  &  30.4\% & 39.2\% &  30.4\%\\
S0   &  -2.5\,--\,-1.5 &  232 &  28.9\% & 44.4\% &   26.7\%\\
S0/a &  -1.5\,--\,0.5  &  206  &  58.7\% & 25.9\% &  12.1\%\\
Sa   &   0.5\,--\,2.5  &  207  &  84.5\% & 14.5\% &  1.0\%\\
Sb  &    2.5\,--\,4.5  &  412  &  99.5\% & 0.2\% &  0.2\%\\
Sc   &   4.5\,--\,7.5  &  810   & 97.2\% & 0\% & 2.8\%\\
Sd   &   7.5\,--\,8.5  &  171  &  91.2\% & 0\% & 8.8\%\\
\end{tabular}
\caption{The Star Formation Classification per morphological type (T-Type); see Fig.\ref{fig:ssfr} for the \spitzerg\ sample. The number of galaxies where star formation was not detected (L$_{\rm 12\mu m}$ upper limits) is given in the final column.
}
\label{tab:MorphQfrac}
\vspace{-5pt}
\end{table*}

\section{Discussion}

The \spitzerg-\wise\ sample underscores the dichotomy of complexity in the dust properties of galaxies in the nearby Universe; we see that morphology, stellar mass and dust density can produce clear correlations, or conversely, large variations. Central to this work, we have re-evaluated the relationship between L$_{\rm TIR}$ and L$_{\rm 12\mu m}$ and L$_{\rm 23\mu m}$, respectively, and derived updated SFR scaling relations with the option of using dust density as a metric for estimating the deficit due to the contribution of unobscured star formation in low mass (log\,\Mstar\,$<9.9$ M$_\odot$), low dust galaxies. 

Of course, the use of \wise\ as a SFR tracer relies on how well the reprocessing of recent star formation is captured by dust emission, with the contribution of old stars to the heating of dust as an important source of uncertainty \citep[e.g.][]{Boq2016}. This is particularly important when PAHs are concerned \citep[e.g.][]{LuHo23, Cal2024} and although, as discussed earlier, the \wise\ W3 band is not dominated by PAH emission, it is certainly a significant contributor.

The \spitzerg-\wise\ sample does not have sufficient numbers of high mass galaxies (log\,\Mstar$>10.5$ M$_\odot$) to explore this fully, but examining Figures \ref{fig:IRX12} and \ref{fig:IRX23} of the Appendix, we see that earlier morphological types (irrespective of total stellar mass) show distributions with, on average, elevated log\,(L$_{\rm 12\mu m}$/L$_{\rm FUV}$) and log\,(L$_{\rm 23\mu m}$/L$_{\rm FUV}$) ratios as a function of dust density ($\Sigma_{\rm 12 \mu m}$ and $\Sigma_{\rm 23 \mu m}$, respectively). This may be indicative of dust heated by evolved stellar populations (in this case, correlated with increasing bulge prominence) which would lead to an overestimation of mid-infrared derived star formation rates. Quantification and potential mitigation of such an effect is beyond the scope of this work; this would require detailed radiative transfer modelling \citep[e.g.][]{Baes2011} to disentangle the effects of dust attenuation, dust geometry and the heating from different generations of stellar populations.

What we have shown is that the low scatter in the correlation of L$_{\rm TIR}$ and L$_{\rm 12\mu m}$ can be replicated in L$_{\rm 23\mu m}$ and L$_{\rm TIR}$ by taking into account the steepness of the warm dust continuum (or, alternatively, the paucity of 12$\mu$m emission) as traced by W3$-$W4, or alternately, L$_{\rm 12\mu m}$/L$_{\rm 23\mu m}$. Given the better sensitivity of W3, having an estimate of L$_{\rm TIR}$ in support of large galaxy surveys is arguably more powerful that merely being a star formation rate indicator (with all of its associated uncertainties). In addition, recent work finds a close relationship between L$_{\rm 12\mu m}$ and CO emission \citep[e.g.][]{Leroy2021, Gao2022} which further suggests that the \wise\ W3 band can be used as a powerful tracer of the ISM for galaxies distributed across the whole sky.

\section{Summary and Conclusions}

We have carefully measured the \wise\ properties of galaxies in the extended \spitzerg\ sample, which constitutes a powerful laboratory with which to examine the dust properties of nearby galaxies. We summarise our key results here:

\begin{enumerate}
    
    \item Examining the relationship between \wise\ color and morphology we find that spheroids largely separate from disk galaxies in \WIWIII\ color, except for S0 and S0/a galaxies which can span the full range of \WIWIII\ color.

    \item The distribution of \spitzerg-\wise\ galaxies in W3$-$W4 color shows a tail towards redder \wise\ color indicative of a steep warm dust spectrum or, alternatively, a relative paucity of 12$\mu$m emission.  

    \item Taking into account these ``warm" sources (using the ratio of L$_{\rm 12\mu m}$ to L$_{\rm 23\mu m}$), we derive new relationships for W3 and W4 with L$_{\rm TIR}$, which results in closer statistical agreement between SFR$_{\rm 12\mu m}$ and SFR$_{\rm 23\mu m}$.
    
    \item Using dust density ($\Sigma_{\rm 12\mu m}$ in units of L$_\odot$/kpc$^2$), we derive estimates of the SFR$_{\rm deficit}$ expected for low mass (log\,\Mstar$<9.9$ M$_\odot$), low dust systems. This can be used to correct \wise-derived SFRs in this regime.

    \item We investigate the ratio of mid-infrared emission to L$_{\rm FUV}$ for different morphological types, finding that L$_{\rm 12\mu m}$/L$_{\rm FUV}$ is well-correlated for Sc galaxies as a function of both \Mstar\ and $\Sigma_{\rm 12\mu m}$, but with increasing scatter for progressively earlier morphological types.

    \item We use the weighted combination of SFR$_{\rm 12\mu m, corr}$ and SFR$_{\rm 23\mu m, corr}$ (SFR$_{\rm MIR_{corr}}$) to investigate the location of different morphological types in the SFR-\Mstar\ plane. This shows that galaxies with more prominent bulges are less likely to fall on the SFMS, and that Sc and Sd galaxies lie almost exclusively on the SFMS (as long as star formation is detected).

\end{enumerate}

In conclusion, nearly 15 years since its launch, the \WISE\ mission continues to build its legacy in the study of galaxies in the local Universe thanks to its combination of mid-infrared bands and all-sky coverage.

\vspace{10mm}

The authors would like to acknowledge the tragic passing of co-author, Dr Thomas Jarrett. We will continue your legacy, Tom, but we do so with heavy hearts. We thank the referee for recommendations that improved the content of this paper. MEC acknowledges the support of an Australian Research Council Future Fellowship (Project No. FT170100273) funded by the Australian Government. T.H.J. acknowledges support from the National Research Foundation (South Africa). 
This publication makes use of data products from the Wide-field Infrared Survey Explorer, which is a joint project of the University of California, Los Angeles, and the Jet Propulsion Laboratory/California Institute of Technology (JPL/Caltech), and NEOWISE, which is a project of the JPL/Caltech. WISE and NEOWISE are funded by the National Aeronautics and Space Administration (NASA).  This research has made use of the NASA/IPAC Extragalactic Database (NED) and the NASA/IPAC Infrared Science Archive (IRSA),which is operated by the JPL, Caltech, under contract with NASA.

\vspace{5mm}
\facilities{WISE, Spitzer, GALEX, NED, IRSA}

\software{\texttt{astropy} \citep{astropy2013, astropy2018}, \texttt{matplotlib} \citep{Hunter2007}, \texttt{NumPy} \citep{Walt2011}, and \texttt{SciPy} \citep{Virt20}. }

\appendix

\section{The SINGS/KINGFISH Sample}\label{sings}

As in \citet{Cluver17}, we make use of the combined SINGS \citep{SINGS2003} and KINGFISH samples \citep{KINGSFISH2011} for our primary relations between L$_{12\mu m}$ and L$_{23\mu m}$, and L$_{\rm TIR}$. In total, we have measurements for 76 galaxies (the full SINGS/KINGFISH sample less HoI, HoII, and HoIX which are too faint and diffuse to be measured in \wise). Of these, 6 have upper limits for L$_{12\mu m}$ and 3 have  upper limits for L$_{23\mu m}$ and are not included in the analysis that follows. Upper limits can result from either emission being too faint for reliable measurement (a non-detection) or, alternatively, due to stellar continuum emission dominating the signal in W3 and W4 which after removal results in a low confidence measurement. Photometry and derived properties for the SINGSFISH sample are included in Tables 3 and 4, respectively. 

\begin{figure}[!ht]
\begin{center}
\gridline{\fig{ 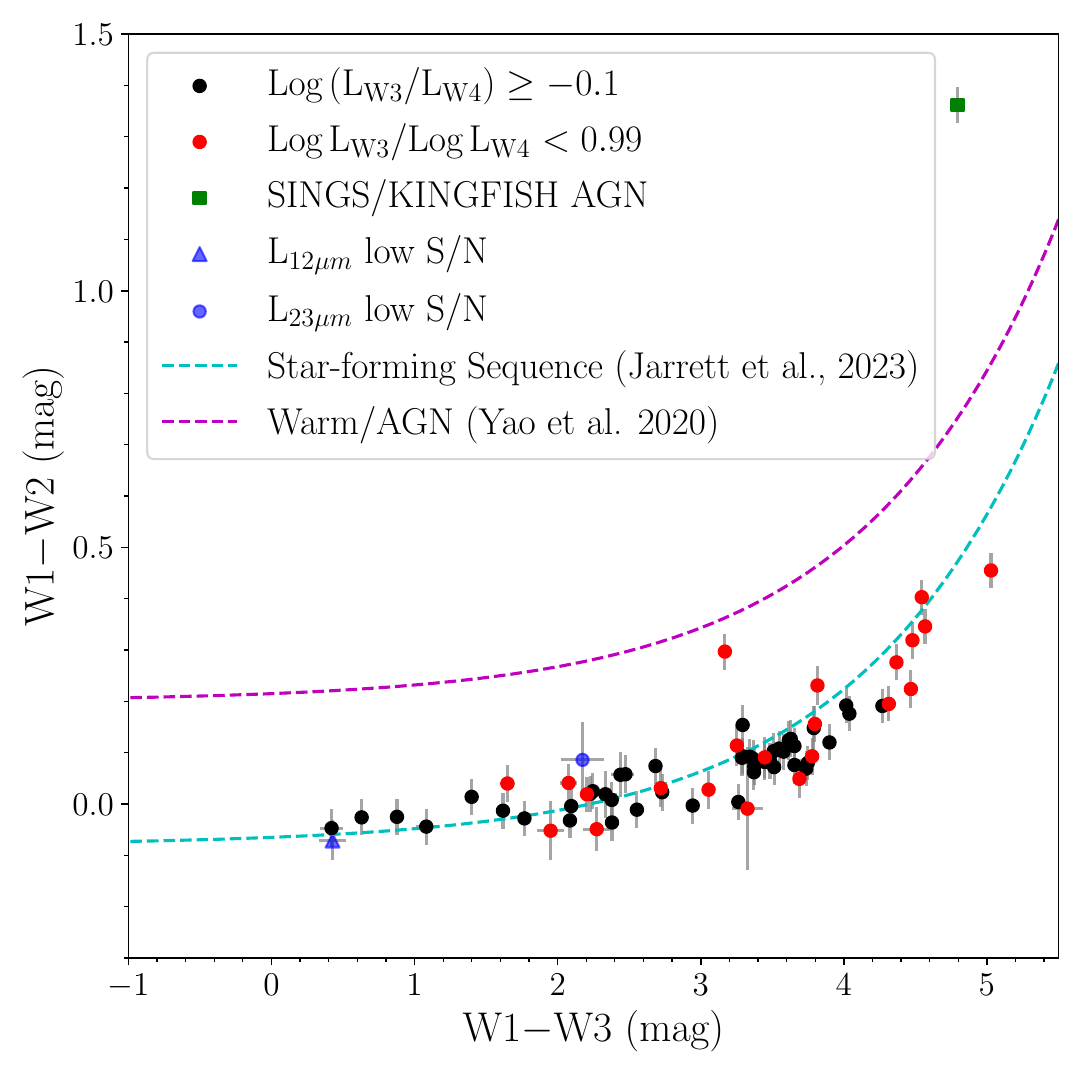}{0.45\textwidth}{(a)}
\fig{ 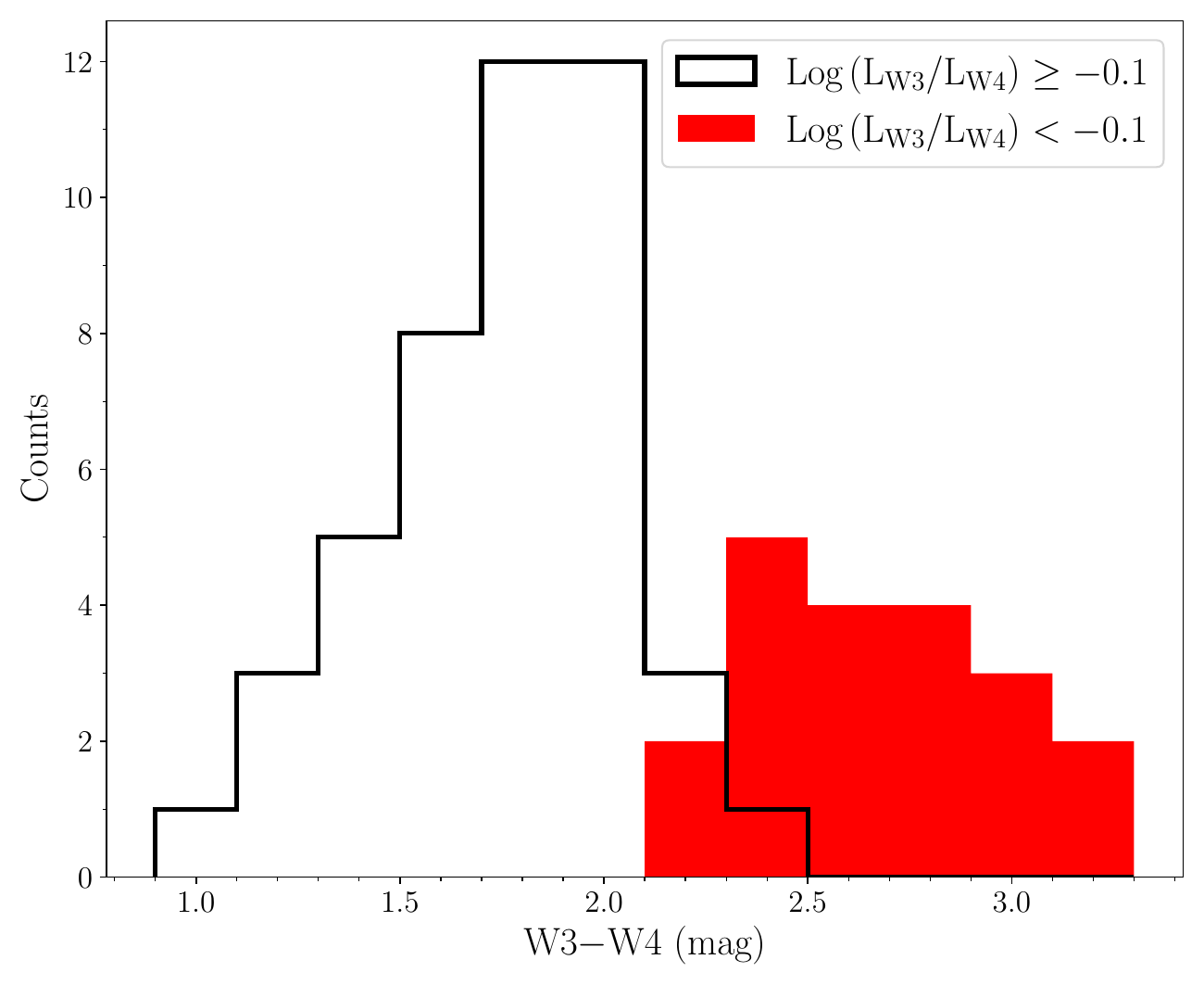}{0.55\textwidth}{(b)}}
\caption{(a) The SINGS/KINGFISH \wise\ color-color diagram showing the color distribution log\,(L$_{12\mu m}$/L$_{23\mu m}$) $\geq-0.1$ sources (black points) and log\,(L$_{12\mu m}$/L$_{23\mu m}$) $<-0.1$ sources (red points). The green point is NGC\,1377 while the red outlier with elevated \WIWII\ is NGC\,1266. (b) Histogram of the W3$-$W4 colors of the SINGS/KINGFISH sources separated by log\,(L$_{12\mu m}$/L$_{23\mu m}$) $\geq-0.1$ or $<-0.1$. The much smaller sample size compared to \spitzerg\ makes it more difficult to see a separation between their distributions (compared to Figure \ref{fig:s4ghist}).}
\label{fig:singshist}
\end{center}
\end{figure}

In Figure \ref{fig:singshist}a we plot the \WIWII\ versus \WIWIII\ colors for the SINGS/KINGFISH sample, separating sources with high 
log\,(L$_{12\mu m}$/L$_{23\mu m}$) ($\geq-0.1$; black points) and low log\,(L$_{12\mu m}$/L$_{23\mu m}$) ($<-0.1$; red points). NGC\,1377 is colored green since it has very warm \WIWII\ colors and would be classified as an AGN by that criterion (this source is therefore not included in the fits that follow). We note that this galaxy is not typical and has been identified as a nascent starburst \citep{Rouss2006}. The galaxy with elevated \WIWII\ color which falls just below our delineation for AGN activity is NGC\,1266, which is thought to harbor an obscured AGN \citep{Alatalo2011}. Galaxies with low S/N measurements (S/N $<5$) are shown in blue and also not included in any fits.

The W3$-$W4 colour distribution for the 
log\,(L$_{12\mu m}$/L$_{23\mu m}$) $\geq-0.1$ sources (black open histogram) and log\,(L$_{12\mu m}$/L$_{23\mu m}$) $<-0.1$ sources (red filled histogram) are shown in Figure \ref{fig:singshist}. Given the small sample size, the division appears contrived, but a significant number of the SINGS/KINGFISH sources fall into this category.

\begin{figure*}[!thb]
\begin{center}
\gridline{\fig{ 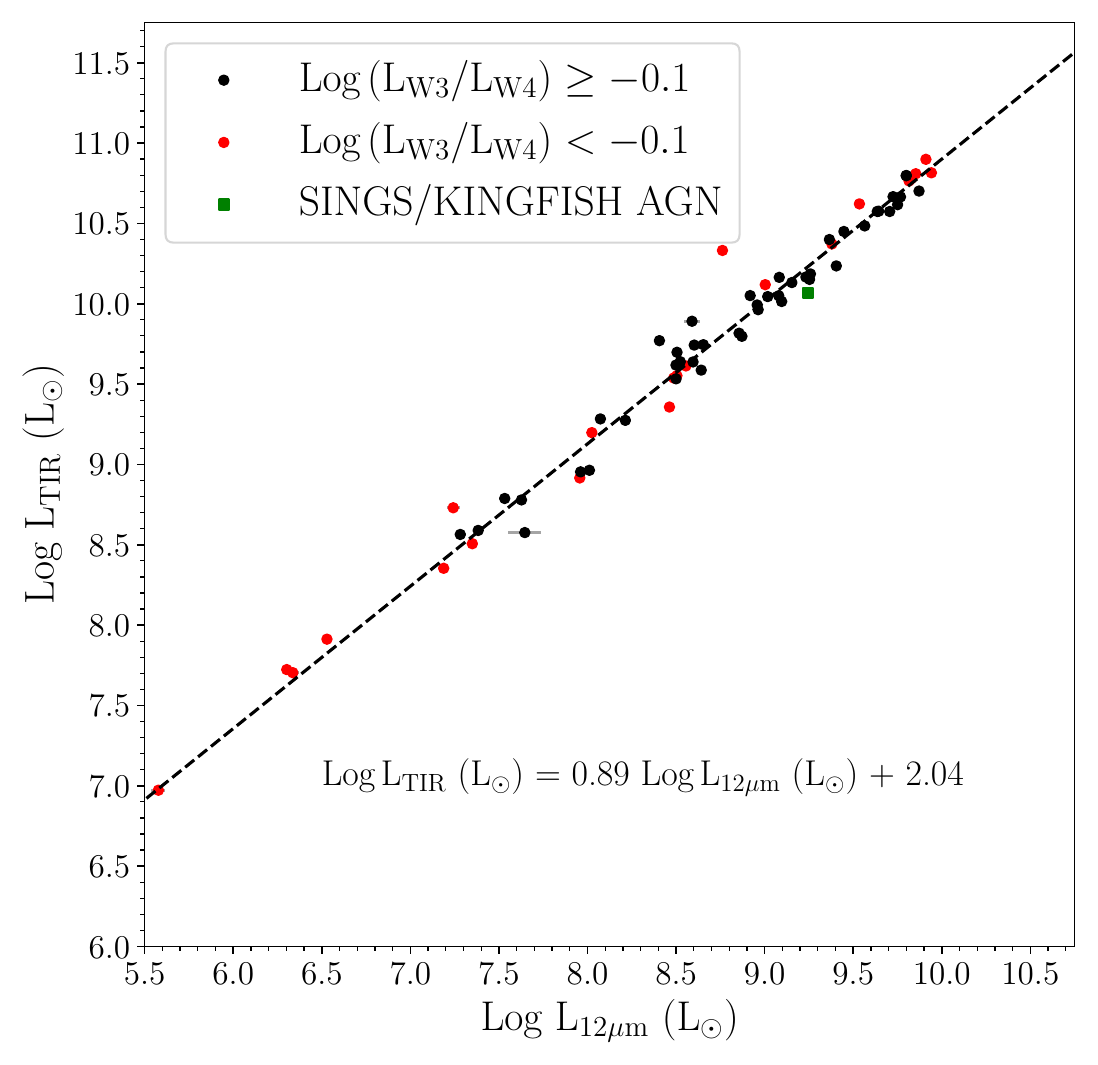}{0.5\textwidth}{(a) Log\,L$_{\rm TIR}$ vs. Log\,L$_{12\mu m}$}
\fig{ 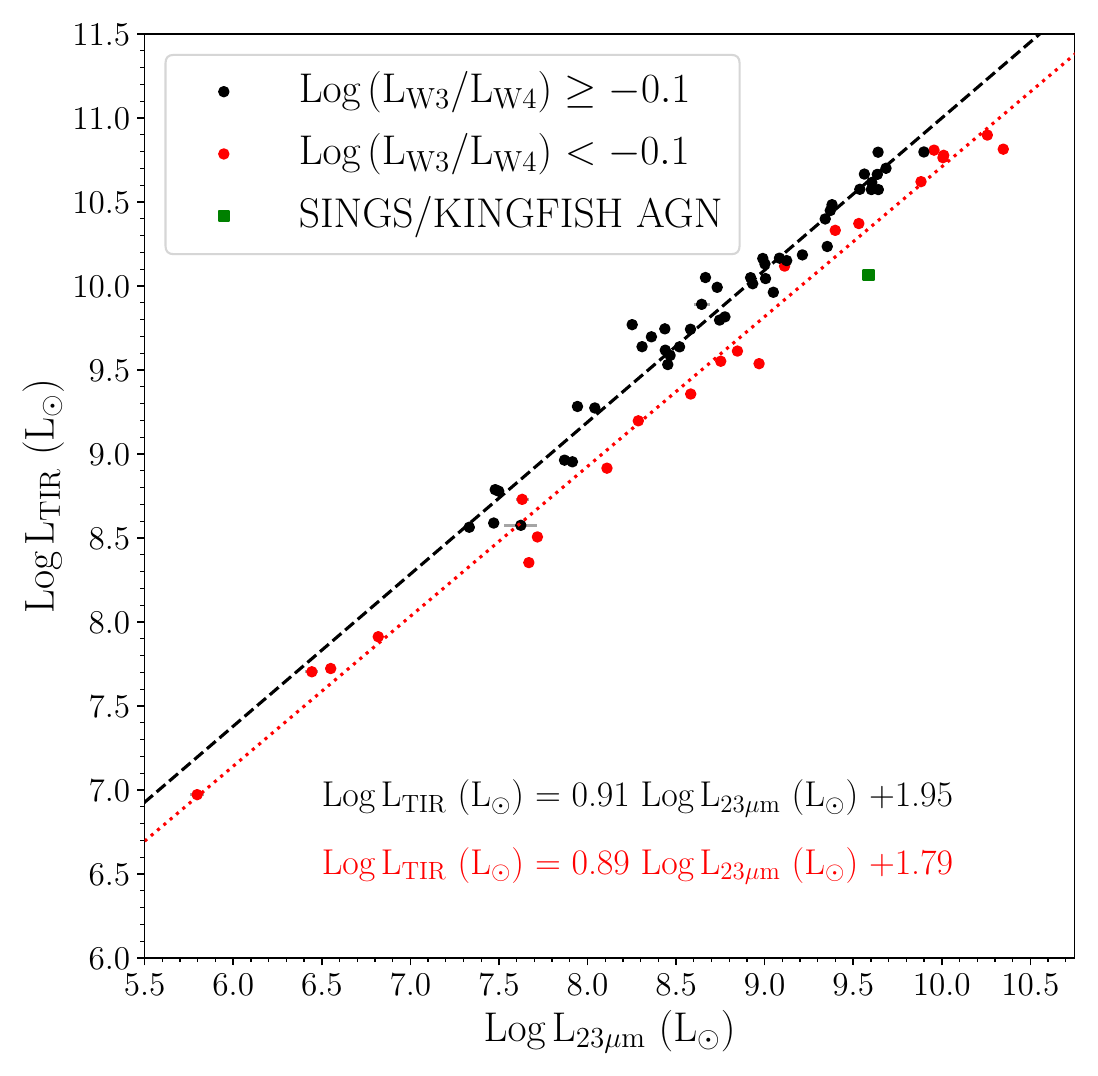}{0.5\textwidth}{(b) Log\,L$_{\rm TIR}$ vs. Log\,L$_{23\mu m}$}}
\gridline{\fig{ 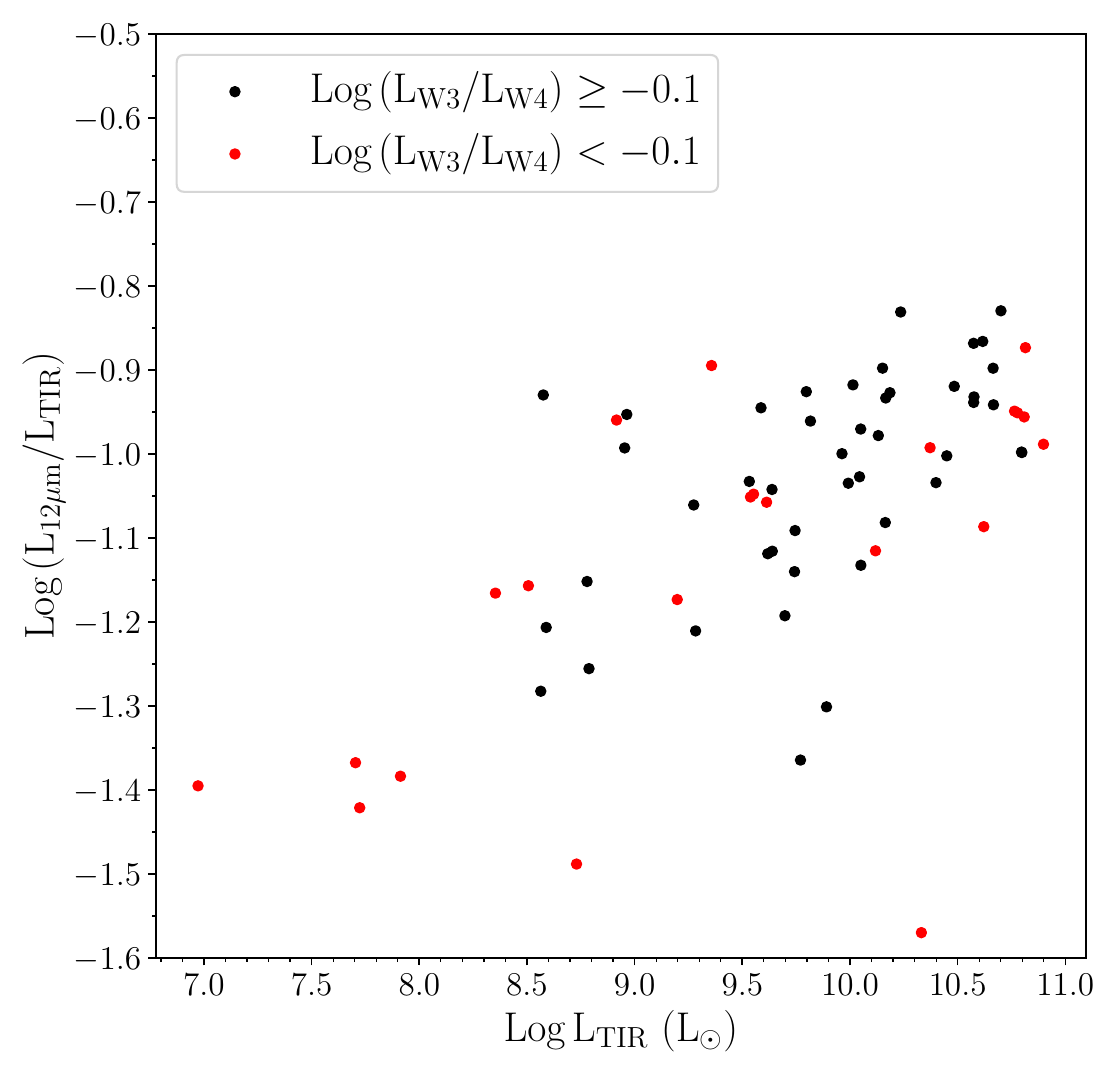}{0.5\textwidth}{(c) Log\,L$_{12\mu m}$/Log\,L$_{\rm TIR}$ vs. Log\,L$_{\rm TIR}$}
\fig{ 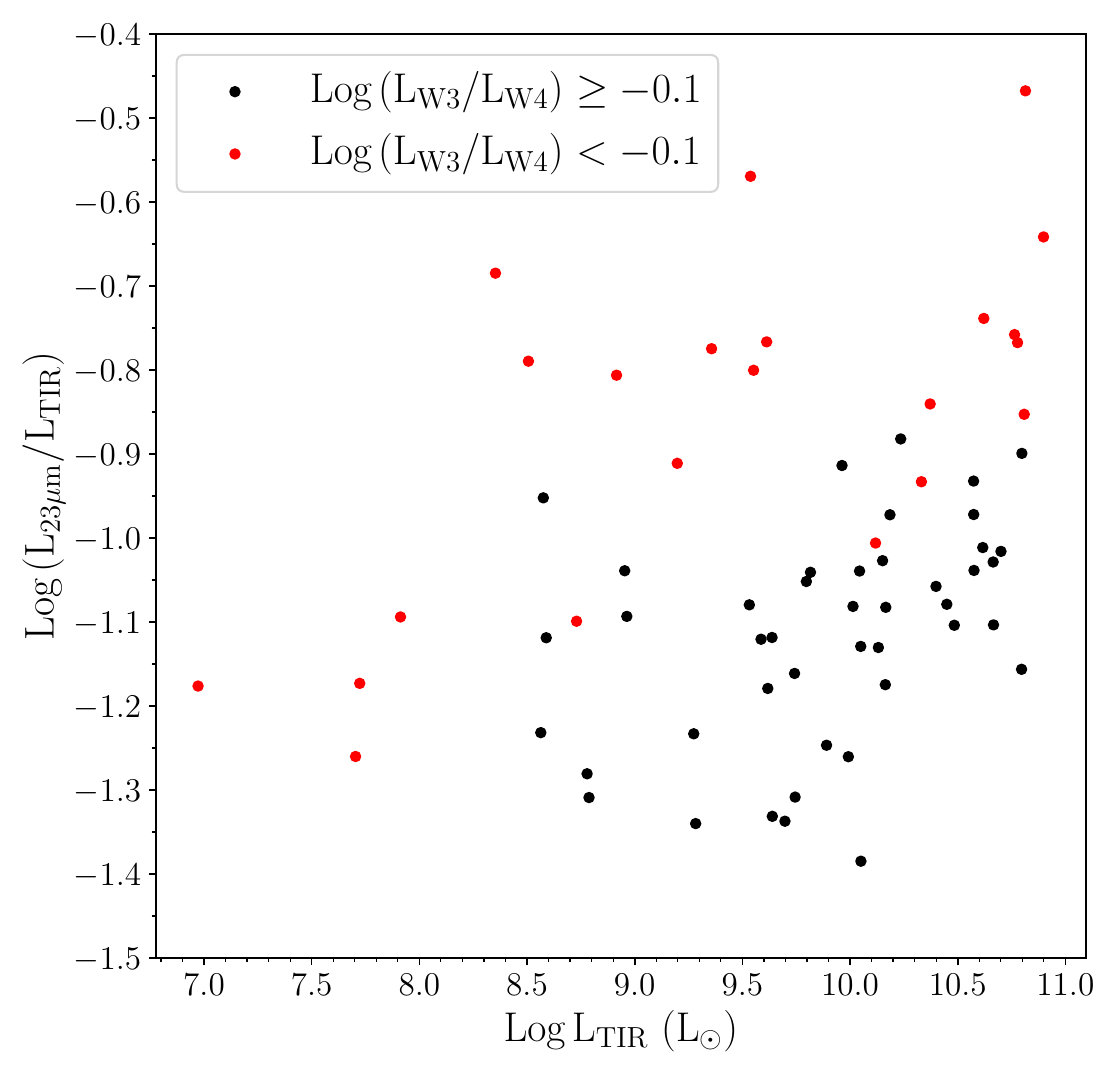}{0.5\textwidth}{(d) Log\,L$_{23\mu m}$/Log\,L$_{\rm TIR}$ vs. Log\,L$_{\rm TIR}$}}
\caption{The relationship between Log\,L$_{12\mu m}$, Log\,L$_{23\mu m}$ and Log\,L$_{\rm TIR}$ with sources separated by L$_{12\mu m}$/L$_{23\mu m}$ ratio.}
\label{fig:sfr_recalib}
\end{center}
\end{figure*}

We next re-examine the relationships derived in \citet{Cluver17} taking advantage of improved measurements and more robust and internally consistent adopted distances. 
The correspondence between Log\,L$_{\rm TIR}$ and Log\,L$_{12\mu m}$ is shown in Figure \ref{fig:sfr_recalib}a; we see little difference in the behaviour of the red and black sources (although most of the low luminosity sources have ``warm" mid-infrared colours relatively high L$_{23\mu m}$ compared to L$_{12\mu m}$). The fit to all sources is given by:

\begin{equation}
\textrm{Log L$_{\rm TIR}$}\,(L_\odot)= 0.886 (\pm0.022) \, \textrm{Log}\, L_{12\mu m} (L_{\odot}) + 2.04 (\pm0.20)  
\end{equation}
with a 1$\sigma$ scatter of 0.10 dex.

However, when we examine the relationship between L$_{\rm TIR}$ and L$_{23\mu m}$ in Figure \ref{fig:sfr_recalib}b, we find the log\,(L$_{12\mu m}$/L$_{23\mu m}$) $<-0.1$ sources (red points) appear systematically offset from the black points. Comparing the slopes of the fits to the black and red points, respectively, we see that they are very similar and we are therefore observing a shift to higher L$_{23\mu m}$ relative to L$_{\rm TIR}$. The best fit relations are given by:

\begin{equation}
\textrm{Log L$_{\rm TIR}$}\,(L_\odot)= 0.91 (\pm0.03)\, \textrm{Log}\, L_{23\mu m} (L_{\odot}) + 1.95 (\pm0.23)  
\end{equation}
with a 1$\sigma$ scatter of 0.12 dex for ``nominal" mid-infrared sources (i.e. log\,(L$_{12\mu m}$/L$_{23\mu m}$) $\geq -0.1$, and,

\begin{equation}
\textrm{Log L$_{\rm TIR}$}\,(L_\odot)= 0.89 (\pm0.02)\, \textrm{Log}\, L_{23\mu m} (L_{\odot}) + 1.79 (\pm0.19)  
\end{equation}
with a 1$\sigma$ scatter of 0.14 dex for ``warm" mid-infrared sources (i.e. log\,(L$_{12\mu m}$/L$_{23\mu m}$) $<-0.1$). We note that in \citet{Cluver17} the 1$\sigma$ scatter of the L$_{23\mu m}$ $-$ L$_{\rm TIR}$ relation (i.e. without making a distinction based on mid-IR color) was 0.18 dex. 

\begin{figure*}[!t]
\begin{center}
\gridline{\fig{ 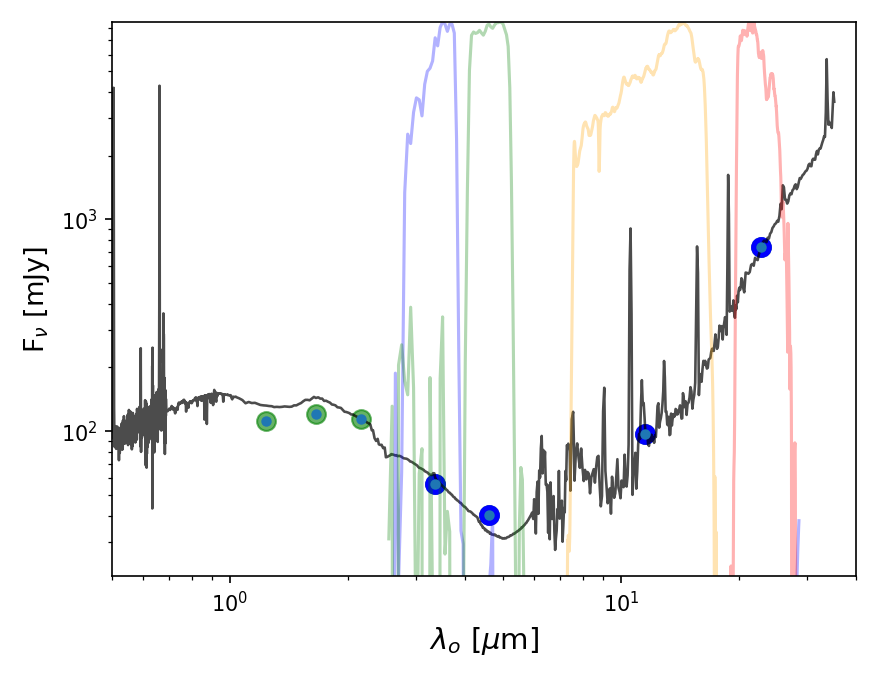}{0.5\textwidth}{(a) NGC\,1266: W3$-$W4 $=3.66$\,mag}
\fig{ 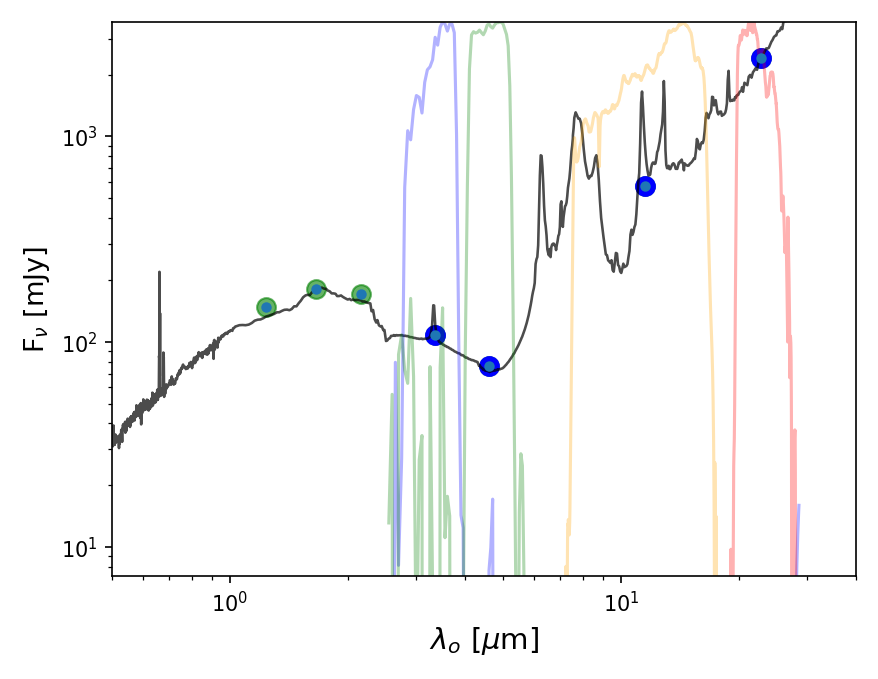}{0.5\textwidth}{(d) NGC\,2798: W3$-$W4 $=3.00$\,mag}}
\gridline{\fig{ 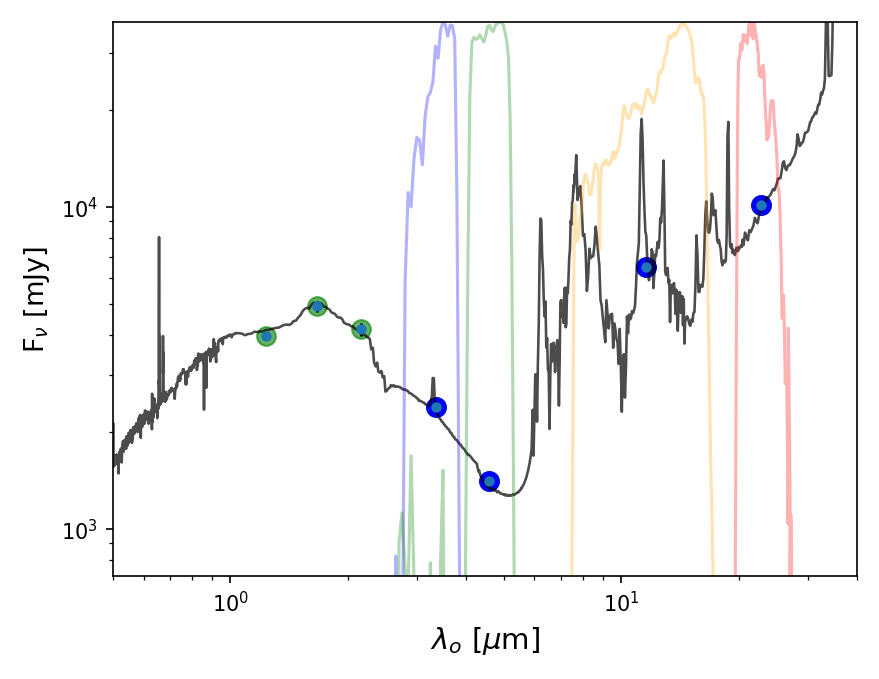}{0.5\textwidth}{(c) M\,101: W3$-$W4 $=1.89$\,mag}
\fig{ 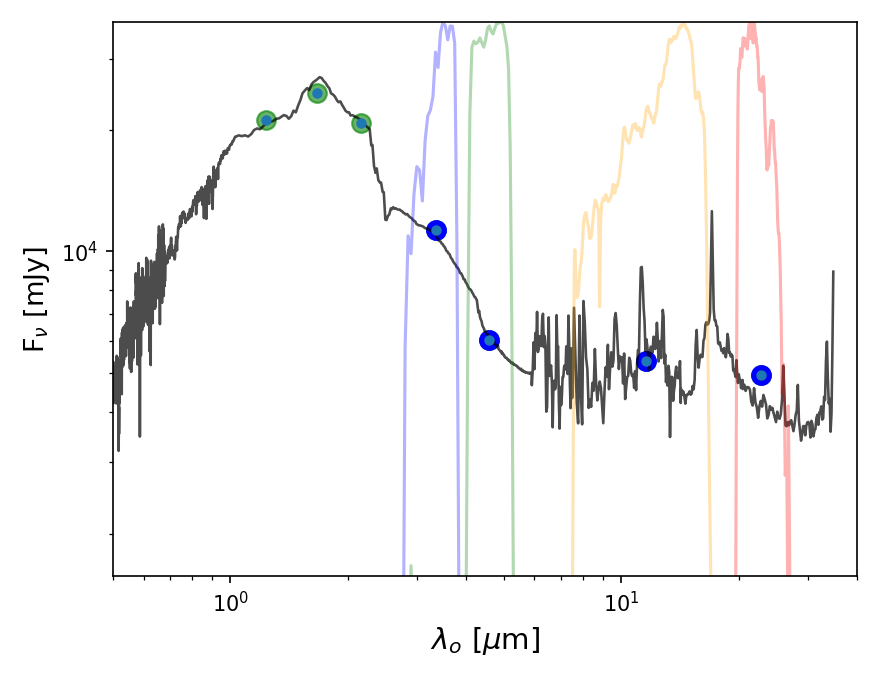}{0.5\textwidth}{(b) NGC\,3031: W3$-$W4 $=1.34$\,mag}}
\label{fig:seds}
\caption{The Spectral Energy Distributions (SEDs) of selected sources illustrating the variation in W3$-$W4 (see also Figure 6); details of the templates used and methodology followed can be found in Appendix A of \citet{Jarrett2023}. The green points correspond to the $J, H,$ and $Ks$ total fluxes from the 2MASS \citep{Skrutskie2006AJ....131.1163S} Extended Source Catalogues \citep[XSC;][]{jarrett-2003}, while the blue points correspond to W1, W2, W3 and W4 bands, respectively. The filter response functions of the four \wise\ bands \citep[]{Jarrett2011} are also shown.}
\end{center}
\end{figure*}

We more closely examine the behaviour of L$_{12\mu m}$ and L$_{23\mu m}$ relative to L$_{\rm TIR}$ in Figure \ref{fig:sfr_recalib}c and d, respectively. For ``nominal" sources (black points) we see that  
L$_{12\mu m}$ and L$_{23\mu m}$ track L$_{\rm TIR}$ very similarly for L$_{\rm TIR} > 8.5\, L_\odot$. However, we see that ``warm" mid-IR sources (red points) at lower L$_{\rm TIR}$ show more of a deficit at L$_{12\mu m}$ compared to L$_{23\mu m}$, but appear well-mixed with the nominal W3$-$W4 color sources (black points) for L$_{\rm TIR} > 8.5\, L_\odot$. On the other hand, the red sources at higher L$_{\rm TIR}$ are systematically higher than the nominal W3$-$W4 color sources. This behaviour is what drives the shift to the right in the relation seen in Figure \ref{fig:sfr_recalib}b. We defer a more detailed examination of the properties of these two populations to a future publication.

\section{Selected SEDs}\label{SEDs}

In Figure 19 we include the spectral energy distributions of four large galaxies to illustrate the variation in W3$-$W4 color due to star formation and nuclear activity. The galaxies are:  NGC\,1266 (a lenticular galaxy hosting a Seyfert nucleus), NGC\,2798 (an SBa type undergoing tidal interaction and likely starburst), M\,101 (an Scd galaxy and classic face-on grand-design spiral) and M\,81/NGC\,3031 (classified as Sab).

\section{Hybrid SFR Comparisons}\label{hybrid}

In this section we compare SFR$_{\rm 12\mu m, corr}$ to selected hybrid FUV$+$W4 SFR indicators from the literature. Since SFR$_{\rm 12\mu m, corr}$ and SFR$_{\rm 23\mu m, corr}$ are calibrated to be consistent with each other, this also reflects the overall performance of SFR$_{\rm 23\mu m, corr}$, but without the complexity of comparing two indicators that both rely on W4. 

\begin{figure*}[!ht]
\begin{center}
\gridline{\fig{ 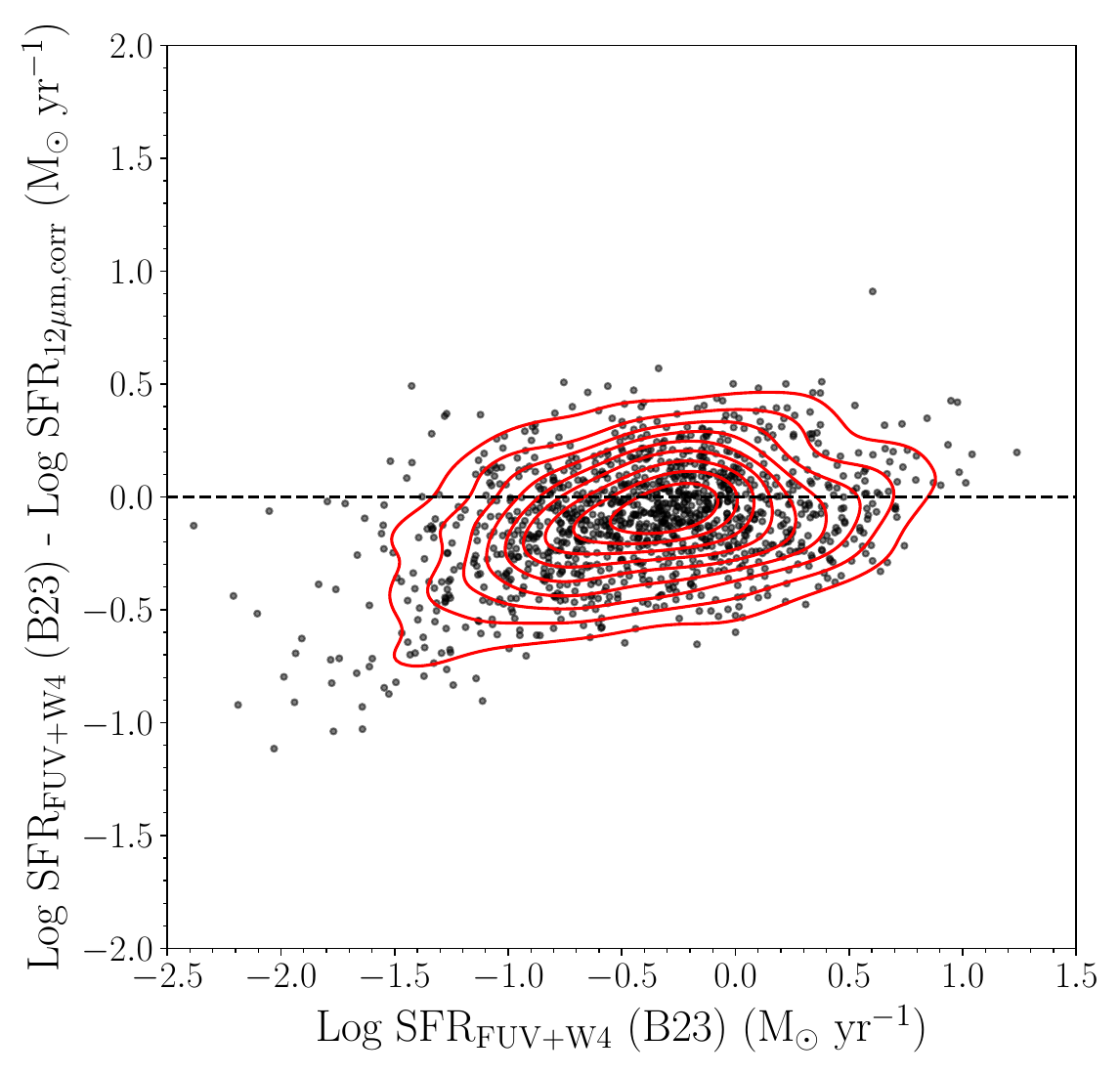}{0.5\textwidth}{(a) Comparison to SFR$_{\rm FUV+W4}$ from \citet{Belf23}.}
\fig{ 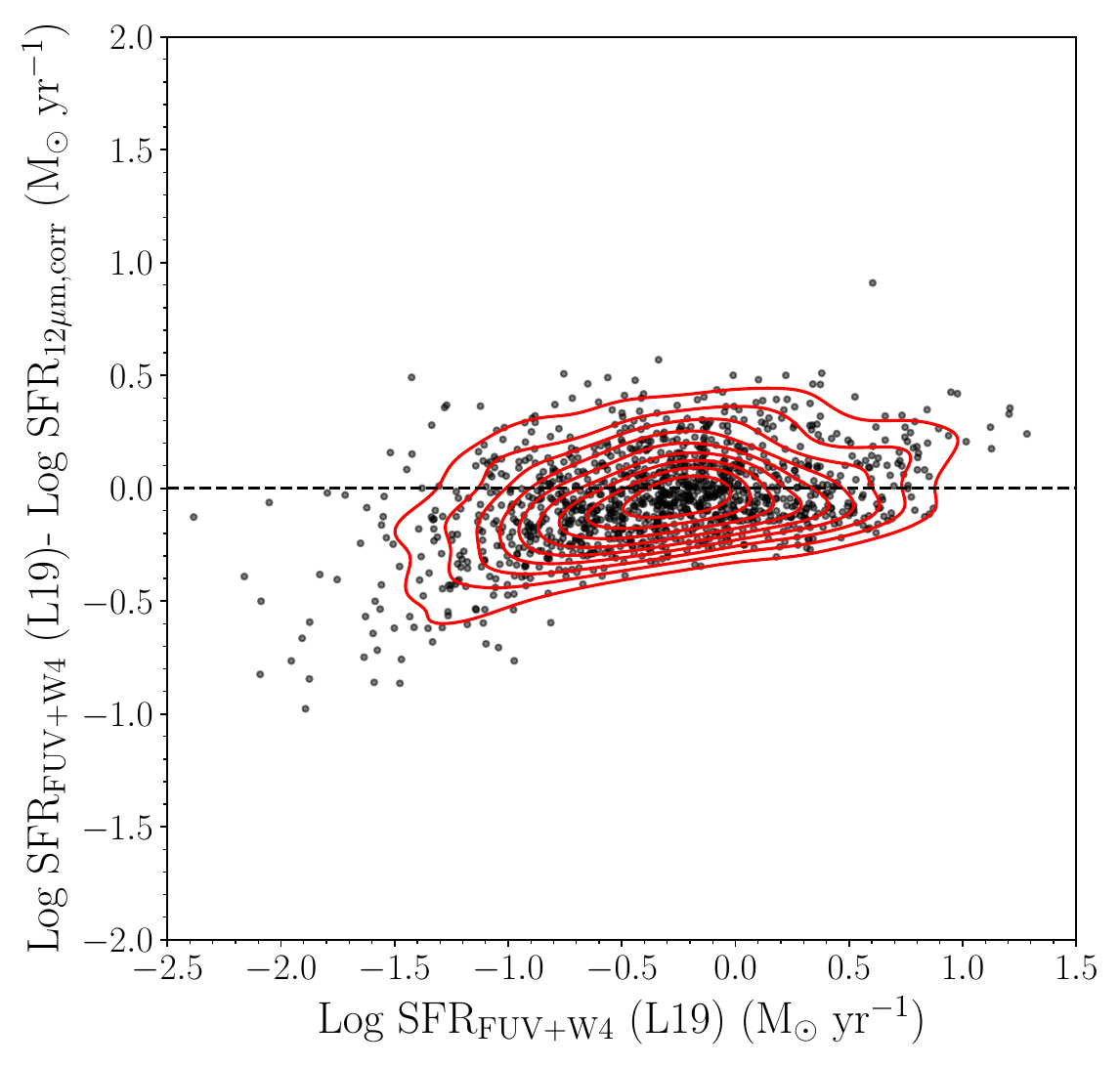}{0.5\textwidth}{(b) Comparison to SFR$_{\rm FUV+W4}$ from \citet{leroy+2019}.}}
\gridline{\fig{ 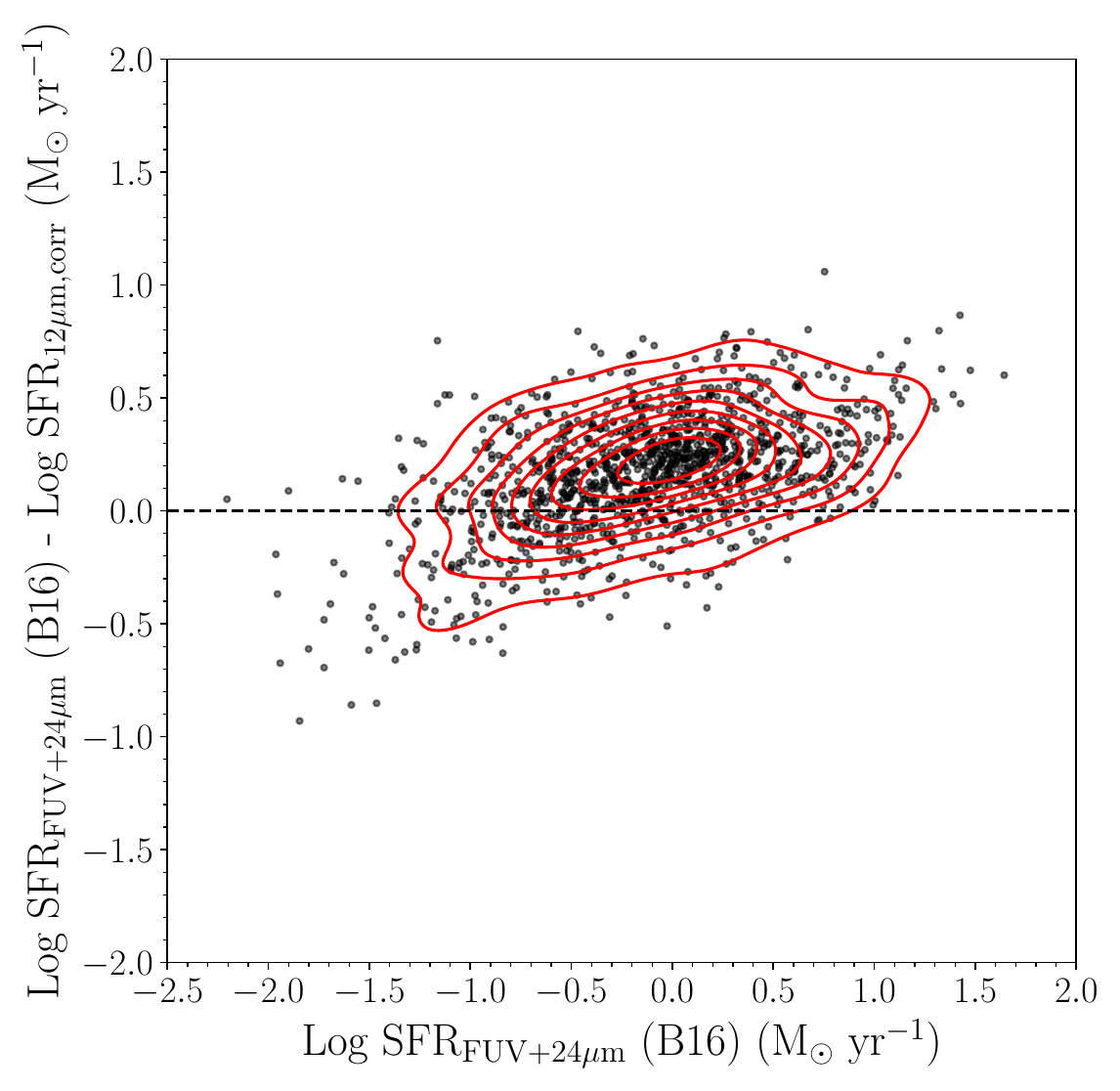}{0.5\textwidth}{(c) Comparison to SFR$_{\rm FUV+24\mu m}$ from \citet{Boq2016}.}
\fig{ 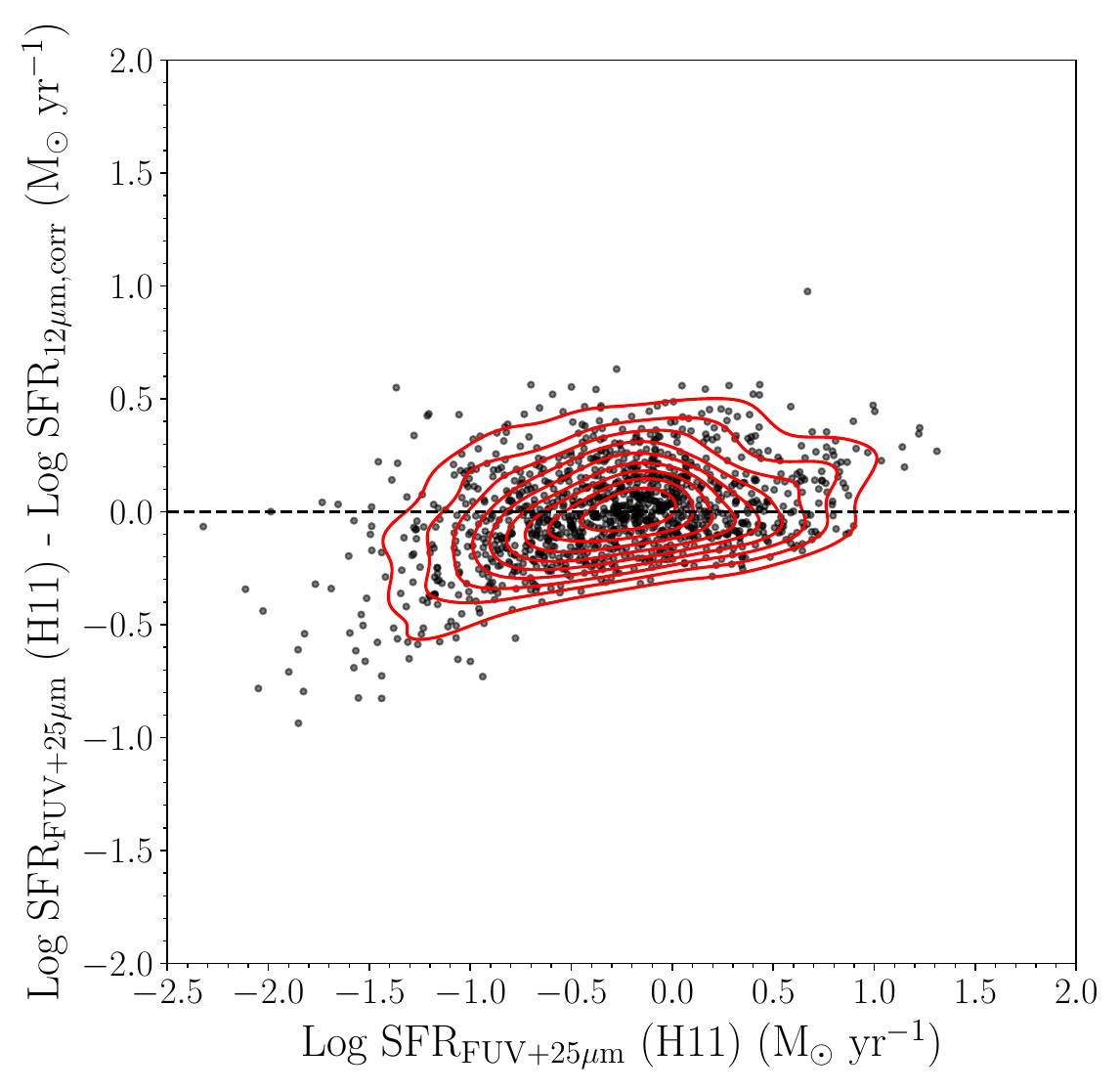}{0.5\textwidth}{(d) Comparison to SFR$_{\rm FUV+25\mu m}$ from \citet{Hao11}.}}
\label{fig:sfrcomp}
\caption{A comparison between corrected 12\micron-derived SFRs (Equation 7) and hybrid UV$+$IR SFR relations from the literature show overall consistency and broad agreement; outliers from panel (a) are examined in the next figure.}
\end{center}
\end{figure*}

In Figure 20a we compare to the FUV$+$W4 relation of \citet{Belf23}, and in panel (b) to the FUV$+$W4 relation \citet{leroy+2019}. In panel (c) we compare to the FUV$+$24\micron\ relation of \citet{Boq2016}, derived using Spitzer MIPS, and in panel (d) to the FUV$+$25\micron\ relation of \citet{Hao11} derived using IRAS 25\micron\ measurements. For the latter two relations we convert our W4 measurement to a proxy MIPS 24\micron\ flux using templates from the GRASIL code \citep{GRASIL} for different morphological types (a factor of 1.1 is appropriate to convert 22.8\micron\ W4 to an approximate 23.675\micron\ MIPS measurement for S0 to Sd galaxy types). We treat the proxy MIPS 24\micron\ measurement as representative of the equivalent IRAS 25\micron\ value for the purposes of this comparison. We note that \citet{Boq2016} provide FUV$-$3.6\micron\ color limits to remove spurious and unphysical measurements; we apply these as FUV$-$W1 color cuts for all sources shown in Figure 20. 

Overall we see good agreement with \citet{Belf23}, albeit with large scatter, and with \citet{leroy+2019} and \citet{Hao11}. We see an offset and SFR dependence when comparing to \citet{Boq2016} (Figure 20c), which is also seen (and discussed) in \citet{Belf23}. 

\begin{figure*}[!ht]
\begin{center}
\gridline{\fig{ 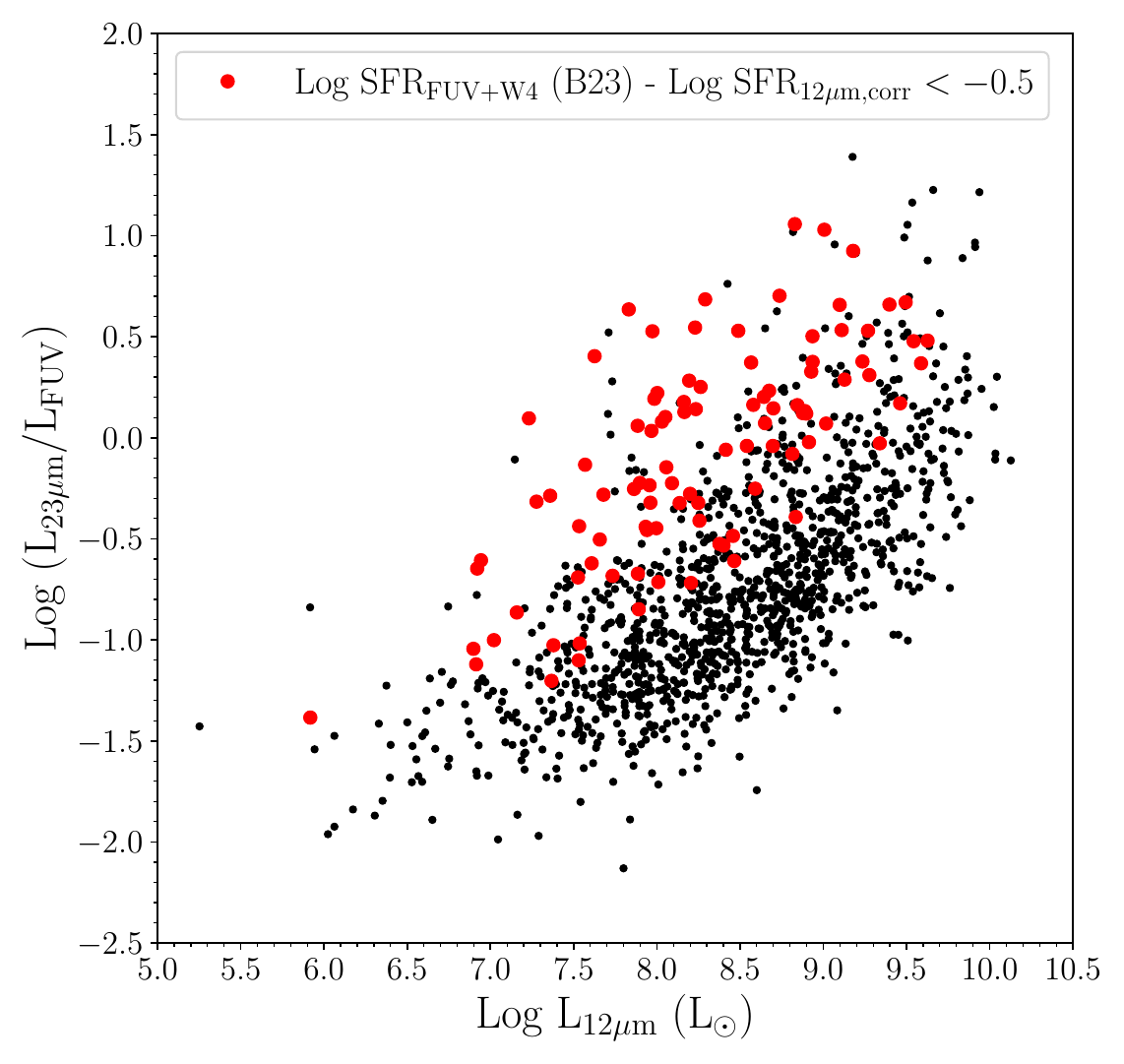}{0.5\textwidth}{(a)}
\fig{ 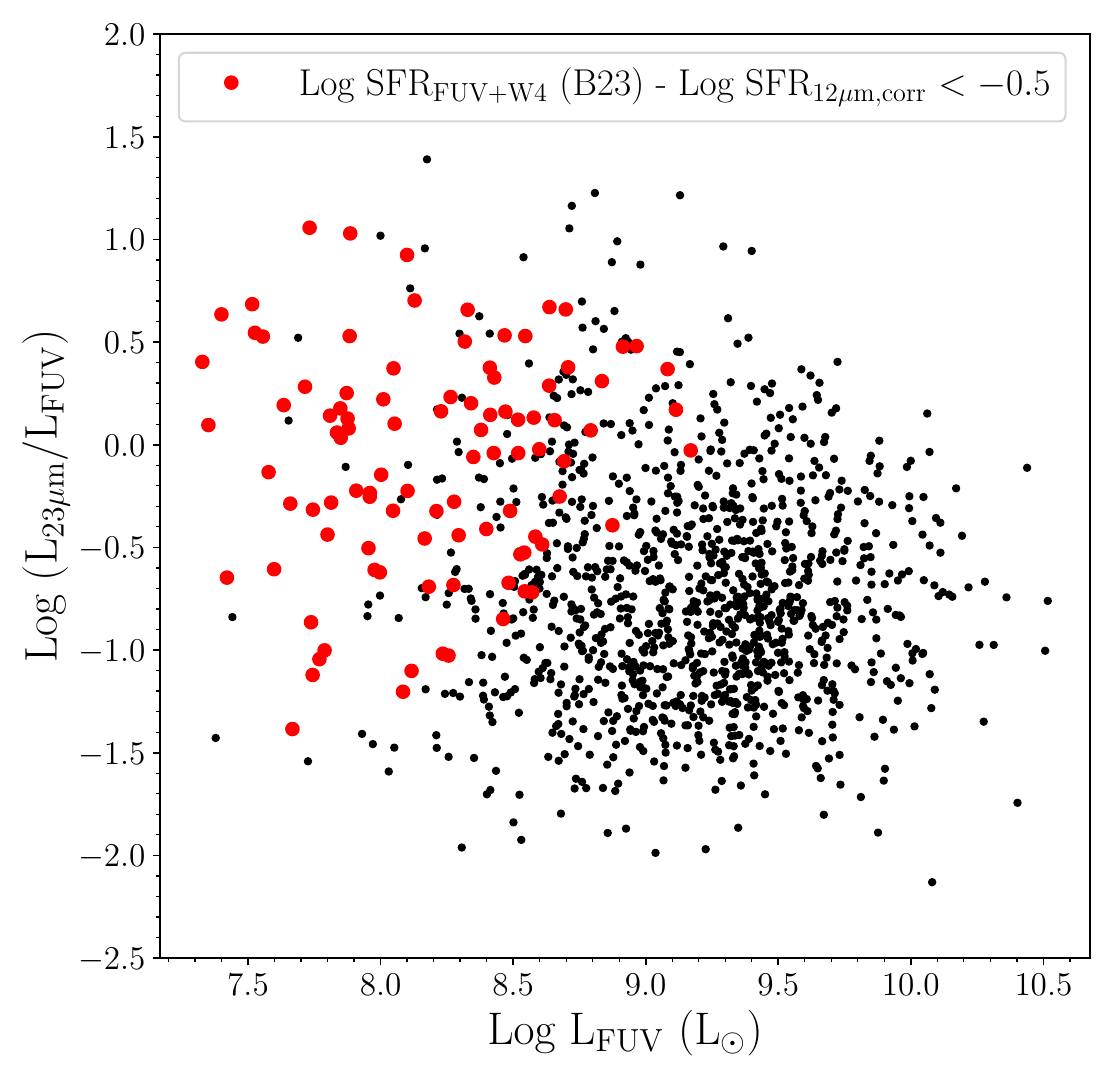}{0.5\textwidth}{(b)}}
\gridline{\fig{ 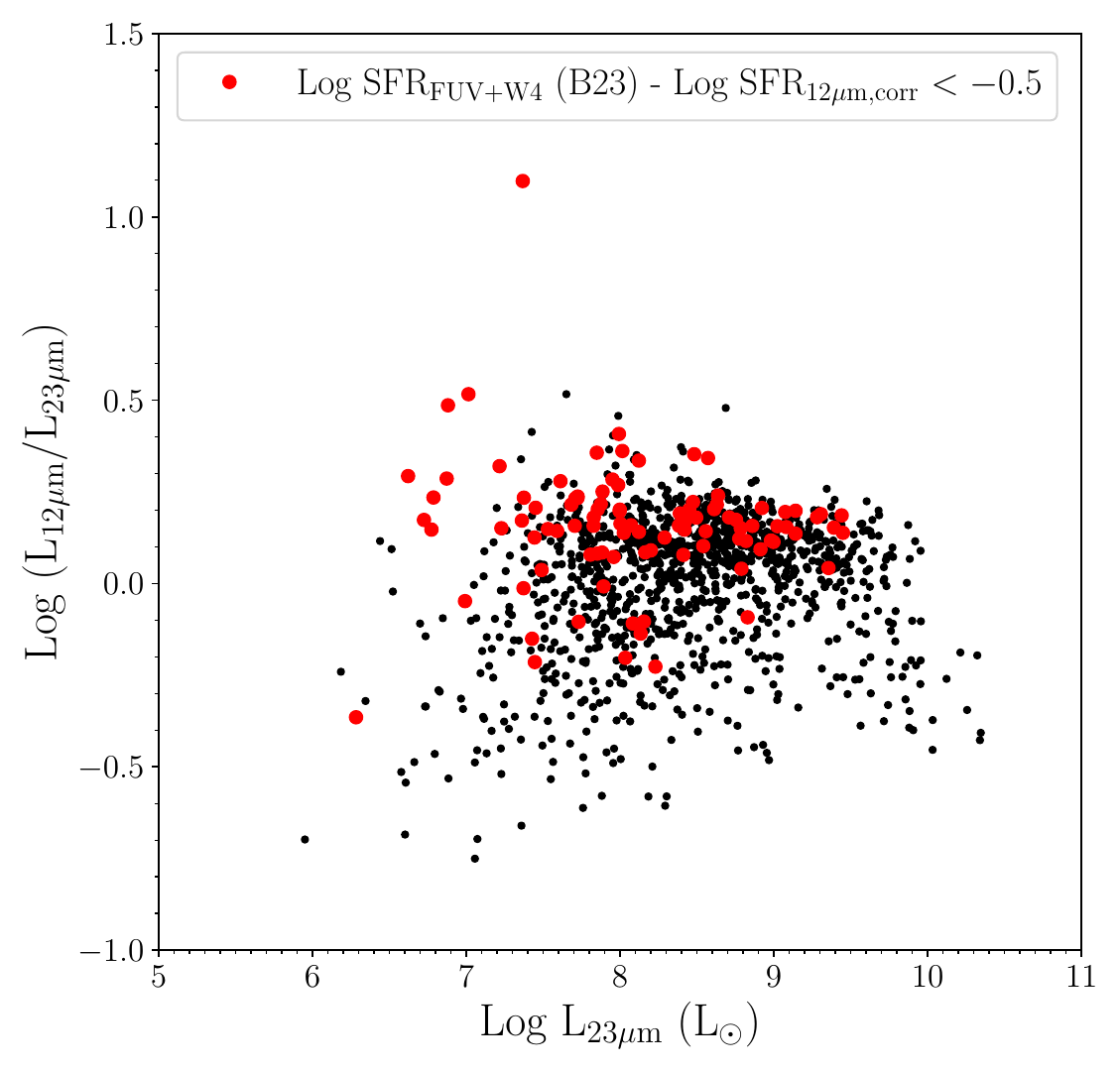}{0.5\textwidth}{(c)}
\fig{ 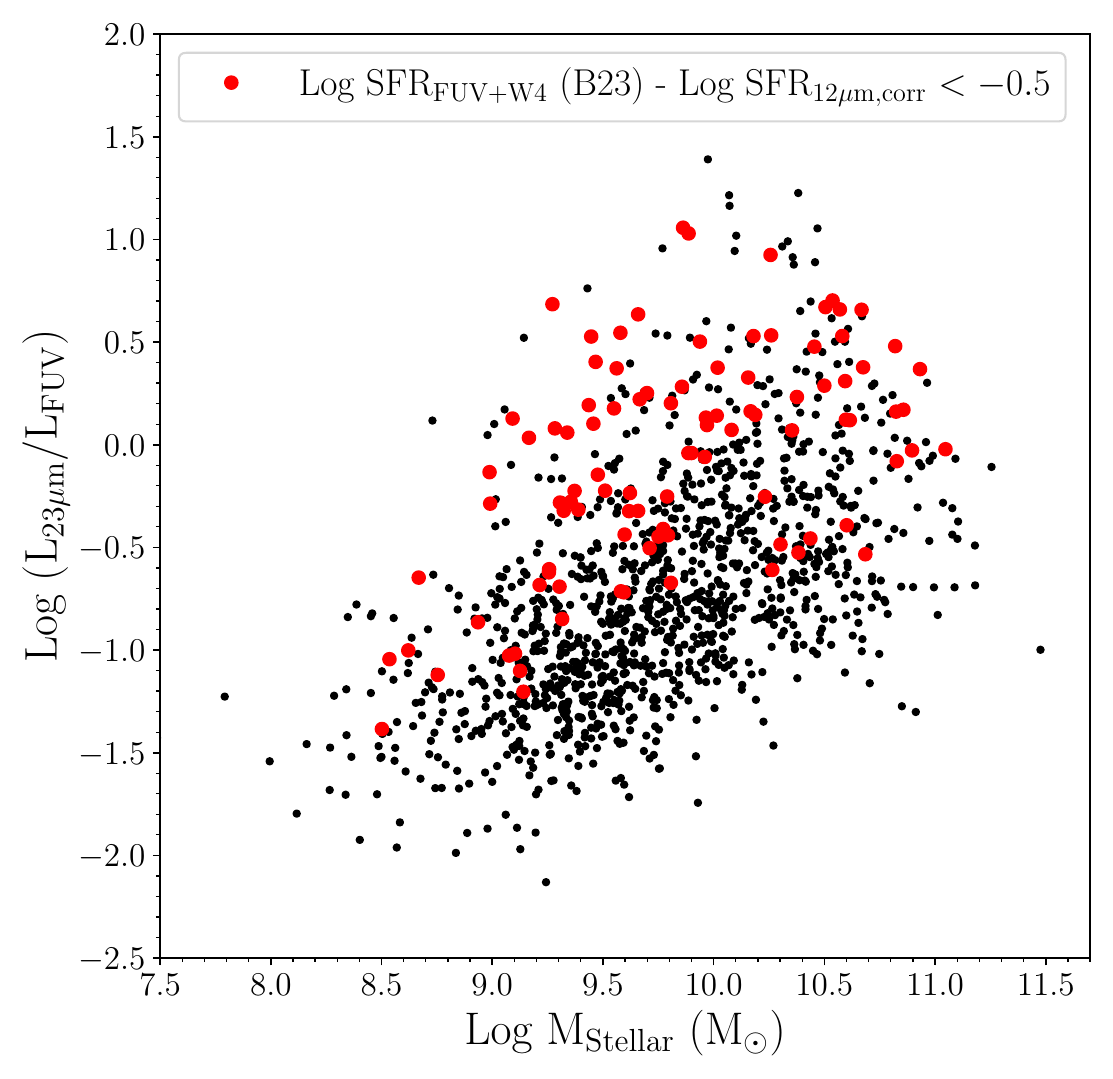}{0.5\textwidth}{(d)}}
\label{fig:sfroutliers}
\caption{Investigating the outliers from Figure \ref{fig:sfrcomp} (red points) compared to the ensemble (black points) by considering the ratio of L$_{23\mu m}$ to L$_{\rm FUV}$ and L$_{12\mu m}$ to L$_{\rm FUV}$ as a function of L$_{12\mu m}$, L$_{23\mu m}$, L$_{\rm FUV}$ and \Mstar. We see that the red points do not show a stellar mass dependence, but do appear to be associated with relatively low levels of UV emission relative to mid-IR dust emission.}
\end{center}
\end{figure*}

All 4 panels of Figure 20 show a clump of sources to the bottom left which imply a large SFR$_{\rm 12\mu m, corr}$ compared to the hybrid FUV$+$MIR derived value. We select all sources with Log SFR$_{\rm FUV + W4}$ (B23) - Log SFR$_{\rm 12 \mu m, corr}$ $<-0.5$, since this captures the largest population that show this offset, and examine their properties further in Figure 21.

Figure 21a shows the ratio of L$_{23\mu m}$ to L$_{\rm FUV}$ as a function of L$_{12\mu m}$, with the outliers from Figure \ref{fig:sfrcomp} shown as red points. We see that these lie above the track inhabited by the other sources in the sample; to further understand this we show the same ratio as a function of L$_{\rm FUV}$ in Figure 21b. This suggests that these sources have a relative paucity of FUV emission which shifts the L$_{23\mu m}$/L$_{\rm FUV}$ ratio to higher values. In order to investigate the possibility that the L$_{23\mu m}$ is elevated compared to L$_{12\mu m}$, thus causing a higher L$_{23\mu m}$/L$_{\rm FUV}$ ratio, we examine the L$_{12\mu m}$/L$_{23\mu m}$ ratio as a function of L$_{23\mu m}$ in Figure 21c. Here we see that the majority of sources have typical L$_{12\mu m}$ to L$_{23\mu m}$ values, which further suggests that the FUV emission for the outliers is low compared to both L$_{12\mu m}$ and L$_{23\mu m}$. Since we are comparing SFR$_{\rm 12\mu m, corr}$ to SFR$_{\rm FUV + W4}$, it is important to check that this is not induced by applying the correction as discussed in Section 3.2.2. In Figure 21d we plot the L$_{23\mu m}$/L$_{\rm FUV}$ ratio as a function of \Mstar, which shows that these sources are found across the stellar mass range of the sample and the majority are found in a mass range where the SFR$_{\rm 12\mu m}$ deficit correction is negligibly small or irrelevant.

\section{IRX Morphology Plots}\label{IRX}

\begin{figure*}[!ht]
\begin{center}
\includegraphics[width=16cm]{ 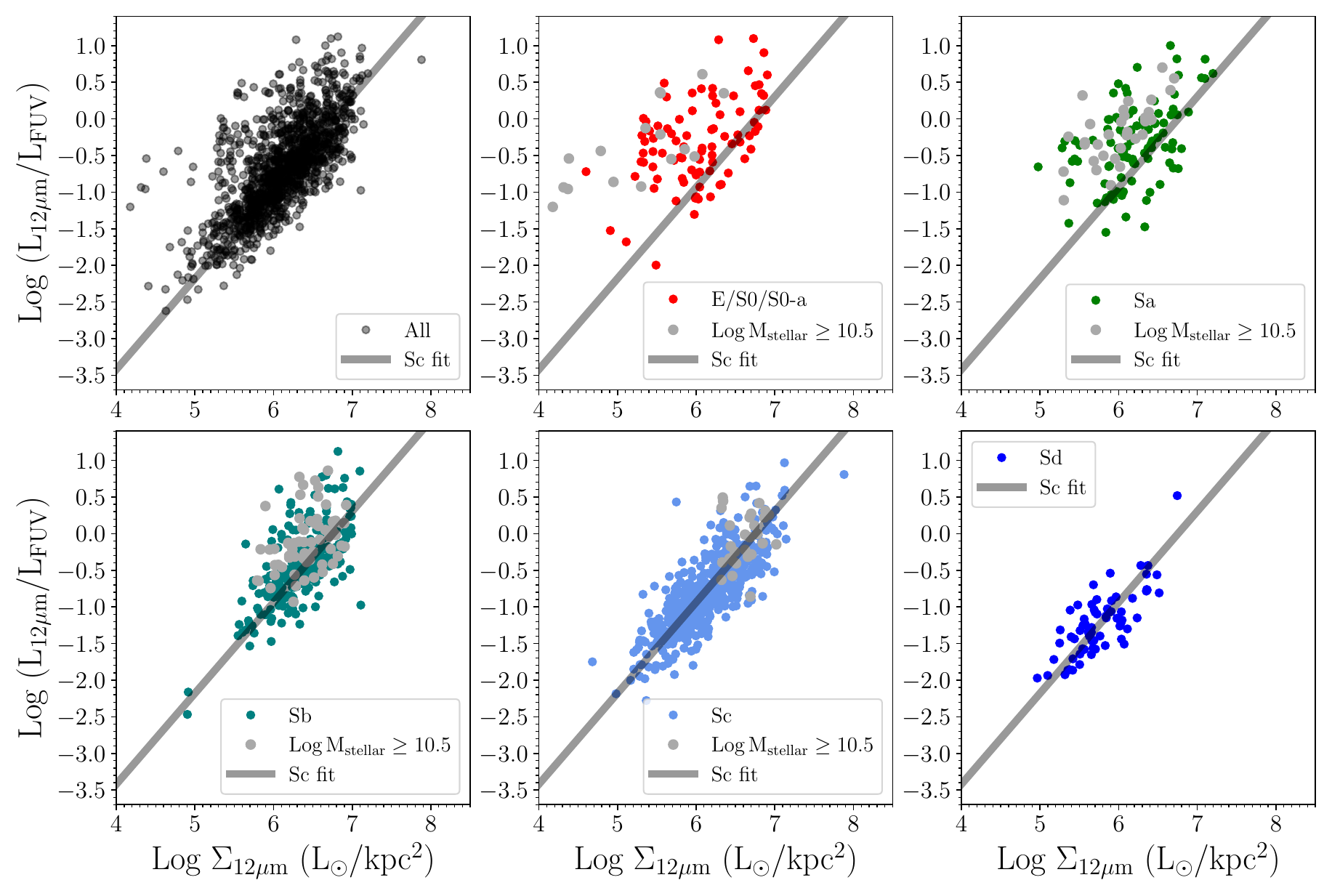}
\caption{The log\,(L$_{\rm 12\mu m}$/L$_{\rm FUV}$) versus log\,$\Sigma_{\rm 12 \mu m}$ for the whole \spitzerg-\wise\ UV sample (first panel) and separated by morphology. Galaxies with log\,\Mstar $>10.5$\,M$_\odot$ are hghlighted as silver points. The fit to sources in the Sc panel (light blue points) is shown in each panel (as the grey line). This figure is complementary to Figure \ref{fig:lumdenmorph} showing the same panels as a function of \Mstar.} 
\label{fig:IRX12}
\end{center}
\end{figure*}

\begin{figure*}[!ht]
\begin{center}
\includegraphics[width=16cm]{ 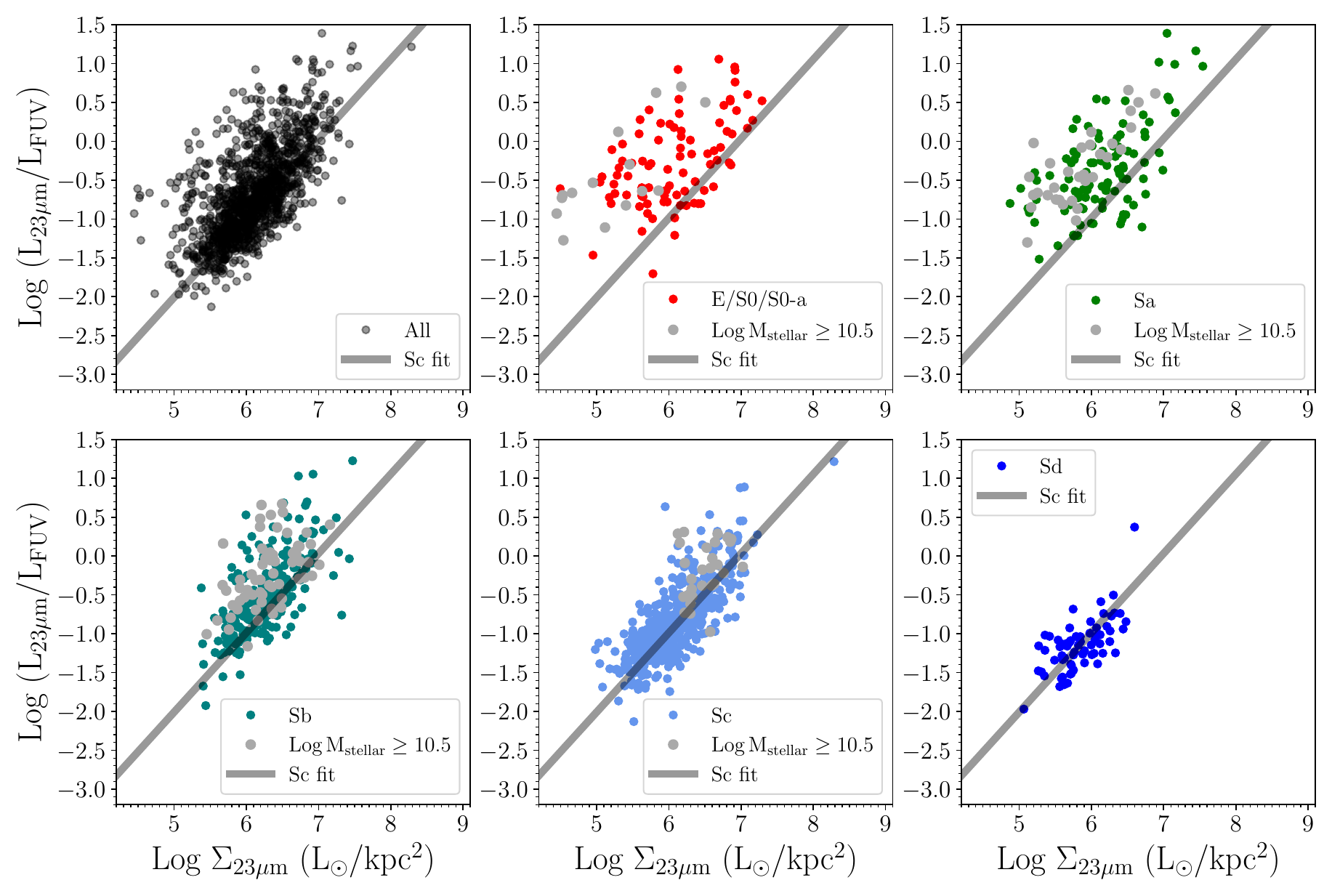}
\caption{Same as in Figure \ref{fig:IRX12}, but for log\,(L$_{\rm 23\mu m}$/L$_{\rm FUV}$ as a function of log\,$\Sigma_{\rm 23 \mu m}$.}
\label{fig:IRX23}
\end{center}
\end{figure*}

In Figure \ref{fig:lumden} we showed the log\,(L$_{\rm 12\mu m}$/L$_{\rm FUV}$) as a function of log\,$\Sigma_{\rm 12 \mu m}$ for galaxies with masses log\,\Mstar\ $<9.9$\,M$_\odot$. We include here the log\,(L$_{\rm 12\mu m}$/L$_{\rm FUV}$) values as a function of log\,$\Sigma_{\rm 12 \mu m}$ (Figure \ref{fig:IRX12}) and log\,(L$_{\rm 23\mu m}$/L$_{\rm FUV}$) values as a function of log\,$\Sigma_{\rm 23 \mu m}$ (Figure \ref{fig:IRX23}) for the full sample (first panel) and separated by morphological type in the subsequent panels. The fit to the Sc galaxies is given by:

\begin{equation}
\textrm{Log\,(L$_{\rm 12\mu m}$}/\textrm{L}_{\rm FUV}) = 1.24\, \textrm{log}\,\Sigma_{\rm 12 \mu m} - 8.39,  
\end{equation}
in Figure \ref{fig:IRX12} and
\begin{equation}
\textrm{Log\,(L$_{\rm 23\mu m}$}/\textrm{L}_{\rm FUV}) = 1.02\, \textrm{log}\,\Sigma_{\rm 23 \mu m} - 7.11,  
\end{equation}
in Figure \ref{fig:IRX23} (reproduced in each panel(as the grey line). Galaxies with log\,\Mstar $>10.5$\,M$_\odot$ are shown as silver points in each panel.

The distribution of points in both figures is similar, albeit with more scatter in Figure \ref{fig:IRX23} (consistent with what is seen in Figure \ref{fig:lumden}). Galaxies with log\,\Mstar $>10.5$\,M$_\odot$ do not appear to be noticeably different in their distribution compared to lower mass galaxies, but there appears to be a trend with morphology where earlier types tend to show elevated L$_{\rm 12\mu m}$/L$_{\rm FUV}$ and L$_{\rm 23\mu m}$/L$_{\rm FUV}$, respectively. This suggests decreased L$_{\rm FUV}$ relative to L$_{\rm 12\mu m}$ and L$_{\rm 23\mu m}$ and may be the result of increased dust not associated with recent star formation; this is particularly noticeable in the E/S0/S0-a panel.

\bibliography{S4G}{}
\bibliographystyle{aasjournal}

\newpage


\begin{splitdeluxetable}{lllrcrcBrcrcrrr}




\tablecaption{Measured WISE Properties of the \spitzerg\ Sample}

\tablenum{1}

\tablehead{\colhead{Name} & \colhead{R.A.} & \colhead{Dec.} & \colhead{W1}  & \colhead{W1 flag} & \colhead{W2} & \colhead{W2 flag} &
\colhead{W3}  & \colhead{W3 flag} & \colhead{W4} & \colhead{W4 flag} & \colhead{Radius$^\tablenotemark{a}$} & \colhead{b/a$^\tablenotemark{a}$} & \colhead{P.A.$^\tablenotemark{a}$} \\ 
\colhead{} & \colhead{(deg)} & \colhead{(deg)} & \colhead{(mag)}  & \colhead{}  & \colhead{(mag)} & \colhead{}  & \colhead{(mag)} & \colhead{} & \colhead{(mag)} & \colhead{} & \colhead{(arcsec)} & \colhead{} & \colhead{(deg)} } 

\startdata
IC5269 & 344.43176 & -36.02611 & 9.402 $\pm$ 0.012 & 10 & 9.425 $\pm$ 0.015 & 10 & 8.717 $\pm$ 0.080 & 10 & 7.987 $\pm$ 0.292 & 2 & 79.82 & 0.556 & 50.1 \\
IC5269A & 343.98248 & -36.34820 & 11.777 $\pm$ 0.015 & 10 & 11.738 $\pm$ 0.038 & 10 & 8.647 $\pm$ 0.087 & 10 & 7.065 $\pm$ 0.274 & 0 & 45.88 & 0.827 & 61.3 \\
IC5269B & 344.15277 & -36.25002 & 10.056 $\pm$ 0.015 & 10 & 9.913 $\pm$ 0.023 & 10 & 7.620 $\pm$ 0.055 & 10 & 5.908 $\pm$ 0.124 & 10 & 149.87 & 0.240 & 97.1 \\
IC5269C & 345.20078 & -35.37063 & 11.228 $\pm$ 0.013 & 10 & 11.230 $\pm$ 0.031 & 10 & 8.764 $\pm$ 0.109 & 10 & 6.588 $\pm$ 0.250 & 0 & 81.89 & 0.451 & 47.6 \\
IC5270 & 344.47852 & -35.85812 & 9.146 $\pm$ 0.011 & 10 & 8.995 $\pm$ 0.013 & 10 & 5.439 $\pm$ 0.017 & 10 & 3.630 $\pm$ 0.023 & 10 & 109.00 & 0.380 & 102.8 \\
IC5271 & 344.50778 & -33.74209 & 8.158 $\pm$ 0.011 & 10 & 8.107 $\pm$ 0.013 & 10 & 5.130 $\pm$ 0.017 & 10 & 3.346 $\pm$ 0.022 & 10 & 110.44 & 0.483 & 138.1 \\
IC5273 & 344.8613 & -37.70278 & 8.511 $\pm$ 0.012 & 10 & 8.465 $\pm$ 0.013 & 10 & 5.126 $\pm$ 0.022 & 10 & 3.048 $\pm$ 0.027 & 10 & 113.38 & 0.762 & 48.3 \\
IC5321 & 351.5834 & -17.95617 & 11.192 $\pm$ 0.013 & 10 & 11.062 $\pm$ 0.022 & 10 & 7.502 $\pm$ 0.033 & 10 & 5.304 $\pm$ 0.047 & 10 & 44.12 & 0.691 & 35.3 \\
IC5325 & 352.18112 & -41.33379 & 8.187 $\pm$ 0.011 & 10 & 8.113 $\pm$ 0.012 & 10 & 4.656 $\pm$ 0.017 & 10 & 2.883 $\pm$ 0.021 & 10 & 112.58 & 0.907 & 12.2 \\
IC5332 & 353.61423 & -36.10089 & 7.961 $\pm$ 0.015 & 10 & 7.910 $\pm$ 0.025 & 10 & 5.246 $\pm$ 0.059 & 10 & 3.571 $\pm$ 0.092 & 10 & 246.26 & 0.968 & 169.5 \\
\enddata
\tablenotetext{a}{Measured in the W1 band}

\tablecomments{Table 1 is published in its entirety in the machine-readable format. A portion is shown here for guidance regarding its form and content.\\ 
A null value in the uncertainty of a quantity when the quantity is non-null indicates an upper limit value for the quantity.\\
The flag for each band indicates if the measurement was from: (0) isophotal aperture of semi-major axis "Radius," axis ratio, and position angle, or (10) the total flux combining the isophotal with the radial surface brightness extrapolation \citep[see][]{jarrett+2019}, or (5) an isophotal aperture with an aperture correction \citep[see][]{Jarrett2023}, or (2) the standard aperture measurement from the ALLWISE measurement, or (1) the point-source measurement from the ALLWISE catalog.}


\end{splitdeluxetable}

\newpage

\begin{splitdeluxetable}{llrrrrBrrrrrrr}




\tablecaption{Derived WISE Properties of the \spitzerg\ Sample}

\tablenum{2}

\tablehead{\colhead{Name} & \colhead{D$_\textrm{lum}$} & \colhead{W1-W2} & \colhead{W2-W3} &  \colhead{W1-W3}  & \colhead{W3-W4}  & \colhead{Log L$_\textrm{W1}$}  & \colhead{Log L$_\textrm{W2}$}  & \colhead{Log L$_\textrm{W3}$$^\tablenotemark{a}$ } &  \colhead{Log L$_\textrm{W4}$$^\tablenotemark{b}$ }   & \colhead{Log L$_{\textrm W1_\textrm{Sun}}$$^\tablenotemark{c}$}  & \colhead{Log M$_\textrm{Stellar}$} & \colhead{SFR} \\ 
\colhead{} & \colhead{(Mpc)} & \colhead{(mag)} &  \colhead{(mag)}  & \colhead{(mag)} & \colhead{(mag)} & \colhead{(L$_\odot$)} & \colhead{(L$_\odot$)}  & \colhead{(L$_\odot$)} & \colhead{(L$_\odot$)}  & \colhead{(L$_{\odot,\textrm W1}$)} & \colhead{(M$_\odot$)} & \colhead{(M$_\odot$/yr)}  } 

\startdata
IC5269 & 27.37 & -0.021 $\pm$ 0.035 & 0.529 $\pm$ 0.063 & 0.508 $\pm$ 0.063 & 0.788 $\pm$ 0.286 & 9.0392 $\pm$ 0.0099 & 8.6336 $\pm$ 0.0103 & 7.2903 $\pm$ 0.0347 & 6.5773 $\pm$ 0.1165 & 10.4013 $\pm$ 0.0139 & 10.035 $\pm$ 0.082 & 0.0241 $\pm$ 0.0059 \\
IC5269A & 37.48 & 0.028 $\pm$ 0.051 & 2.864 $\pm$ 0.100 & 2.891 $\pm$ 0.093 & 1.898 $\pm$ 0.289 & 8.3601 $\pm$ 0.0107 & 7.9785 $\pm$ 0.0178 & 8.005 $\pm$ 0.0365 & 7.8137 $\pm$ 0.1121  & 9.7231 $\pm$ 0.0145 & 9.298 $\pm$ 0.084 & 0.2764 $\pm$ 0.0730 \\
IC5269B & 21.49 & 0.115 $\pm$ 0.041 & 2.368 $\pm$ 0.067 & 2.483 $\pm$ 0.065 & 1.563 $\pm$ 0.139 & 8.5712 $\pm$ 0.0107 & 8.2312 $\pm$ 0.0128 & 7.8952 $\pm$ 0.0241 & 7.7841 $\pm$ 0.0511  & 9.9327 $\pm$ 0.0145 & 9.476 $\pm$ 0.083 & 0.2741 $\pm$ 0.0675 \\
IC5269C & 21.7 & -0.009 $\pm$ 0.045 & 2.491 $\pm$ 0.117 & 2.482 $\pm$ 0.114 & 2.218 $\pm$ 0.274 & 8.1113 $\pm$ 0.0103 & 7.7128 $\pm$ 0.0153 & 7.4509 $\pm$ 0.0451 & 7.5313 $\pm$ 0.1019  & 9.4728 $\pm$ 0.0142 & 9.079 $\pm$ 0.083 & 0.1386 $\pm$ 0.0359 \\
IC5270 & 26.29 & 0.149 $\pm$ 0.035 & 3.547 $\pm$ 0.037 & 3.696 $\pm$ 0.037 & 1.836 $\pm$ 0.042 & 9.1086 $\pm$ 0.0099 & 8.7727 $\pm$ 0.0103 & 9.0021 $\pm$ 0.0112 & 8.8880 $\pm$ 0.0129  & 10.4705 $\pm$ 0.0139 & 9.946 $\pm$ 0.082 & 1.0429 $\pm$ 0.2510 \\
IC5271 & 25.59 & 0.05 $\pm$ 0.035 & 2.970 $\pm$ 0.037 & 3.020 $\pm$ 0.037 & 1.795 $\pm$ 0.041 & 9.4808 $\pm$ 0.0099 & 9.1035 $\pm$ 0.0103 & 9.0819 $\pm$ 0.0112 & 8.9736 $\pm$ 0.0126  & 10.8427 $\pm$ 0.0139 & 10.406 $\pm$ 0.082 & 1.2351 $\pm$ 0.2984 \\
IC5273 & 16.6 & 0.057 $\pm$ 0.035 & 3.316 $\pm$ 0.040 & 3.373 $\pm$ 0.040 & 2.058 $\pm$ 0.046 & 8.9665 $\pm$ 0.0100 & 8.5886 $\pm$ 0.0103 & 8.7228 $\pm$ 0.0126 & 8.7223 $\pm$ 0.0141  & 10.3275 $\pm$ 0.0141 & 9.869 $\pm$ 0.082 & 0.6644 $\pm$ 0.1656 \\
IC5321 & 43.22 & 0.110 $\pm$ 0.040 & 3.516 $\pm$ 0.050 & 3.626 $\pm$ 0.047 & 2.249 $\pm$ 0.065 & 8.7168 $\pm$ 0.0103 & 8.3717 $\pm$ 0.0126 & 8.6041 $\pm$ 0.0161 & 8.6522 $\pm$ 0.0211  & 10.0805 $\pm$ 0.0142 & 9.579 $\pm$ 0.083 & 0.4485 $\pm$ 0.1117 \\
IC5325 & 18.03 & 0.073 $\pm$ 0.035 & 3.450 $\pm$ 0.037 & 3.523 $\pm$ 0.037 & 1.793 $\pm$ 0.041 & 9.1681 $\pm$ 0.0098 & 8.8012 $\pm$ 0.0100 & 8.9870 $\pm$ 0.0112 & 8.8582 $\pm$ 0.0123 & 10.5292 $\pm$ 0.0139 & 10.055 $\pm$ 0.082 & 0.9977 $\pm$ 0.2399 \\
IC5332 & 6.18 & 0.076 $\pm$ 0.042 & 2.660 $\pm$ 0.071 & 2.736 $\pm$ 0.068 & 1.715 $\pm$ 0.113 & 8.3334 $\pm$ 0.0107 & 7.9570 $\pm$ 0.0134 & 7.7888 $\pm$ 0.0255 & 7.6439 $\pm$ 0.0383 & 9.6934 $\pm$ 0.0145 & 9.249 $\pm$ 0.083 & 0.2350 $\pm$ 0.0571 \\
IC5333 & 27.28 & 0.149 $\pm$ 0.036 & 3.326 $\pm$ 0.038 & 3.476 $\pm$ 0.037 & 1.958 $\pm$ 0.047 & 8.6371 $\pm$ 0.0100 & 8.3013 $\pm$ 0.0105 & 8.4319 $\pm$ 0.0115 & 8.3970 $\pm$ 0.0151 & 9.9992 $\pm$ 0.0141 & 9.487 $\pm$ 0.082 & 0.3486 $\pm$ 0.0858 \\
\enddata
\tablenotetext{a}{W3 band spectral luminosity, Log\,$\nu$L$_\nu$(12$\mu$m), after subtraction of the stellar continuum}
\tablenotetext{b}{W4 band spectral luminosity, Log\,$\nu$L$_\nu$(23$\mu$m), after subtraction of the stellar continuum}
\tablenotetext{c}{W1 in-band luminosity}

\tablecomments{Table 2 is published in its entirety in the machine-readable format. A portion is shown here for guidance regarding its form and content.\\
A null value in the uncertainty of a quantity when the quantity is non-null indicates an upper limit value for the quantity.}


\end{splitdeluxetable}

\newpage

\begin{splitdeluxetable}{lrrrcrcBrcrcrrr}


\tablecaption{Measured WISE Properties of the SINGS/KINGFISH Sample}

\tablenum{3}
\tablehead{\colhead{Name} & \colhead{R.A.} & \colhead{Dec.} & \colhead{W1}  & \colhead{W1 flag} & \colhead{W2} & \colhead{W2 flag} &
\colhead{W3}  & \colhead{W3 flag} & \colhead{W4} & \colhead{W4 flag} & \colhead{Radius$^\tablenotemark{a}$} & \colhead{b/a$^\tablenotemark{a}$} & \colhead{P.A.$^\tablenotemark{a}$} \\ 
\colhead{} & \colhead{(deg)} & \colhead{(deg)} & \colhead{(mag)}  & \colhead{}  & \colhead{(mag)} & \colhead{}  & \colhead{(mag)} & \colhead{} & \colhead{(mag)} & \colhead{} & \colhead{(arcsec)} & \colhead{} & \colhead{(deg)} }

\startdata
NGC0337 & 14.95818 & -7.57715 & 8.755 $\pm$ 0.012 & 10 & 8.512 $\pm$ 0.020 & 10 & 4.928 $\pm$ 0.017 & 10 & 2.551 $\pm$ 0.020 & 10 & 101.98 & 0.622 & 131.8 \\
NGC0584 & 22.83642 & -6.86802 & 7.173 $\pm$ 0.012 & 10 & 7.262 $\pm$ 0.013 & 10 & 6.702 $\pm$ 0.046 & 10 & 5.813 $\pm$ 0.093 & 10 & 190.48 & 0.784 & 70.6 \\
NGC0628 & 24.17394 & 15.78366 & 6.416 $\pm$ 0.012 & 10 & 6.298 $\pm$ 0.016 & 10 & 2.736 $\pm$ 0.015 & 10 & 0.943 $\pm$ 0.019 & 10 & 324.47 & 0.819 & 86.4 \\
NGC0855 & 33.51487 & 27.87732 & 9.626 $\pm$ 0.015 & 10 & 9.563 $\pm$ 0.029 & 10 & 7.230 $\pm$ 0.059 & 10 & 5.043 $\pm$ 0.073 & 10 & 103.48 & 0.449 & 65.4 \\
NGC0925 & 36.81964 & 33.57917 & 7.511 $\pm$ 0.015 & 10 & 7.452 $\pm$ 0.030 & 10 & 4.350 $\pm$ 0.034 & 10 & 2.266 $\pm$ 0.058 & 10 & 311.71 & 0.501 & 110.1 \\
NGC1097 & 41.57904 & -30.27462 & 5.995 $\pm$ 0.011 & 10 & 5.890 $\pm$ 0.012 & 10 & 2.487 $\pm$ 0.012 & 10 & 0.152 $\pm$ 0.014 & 10 & 403.14 & 0.793 & 127.4 \\
NGC1266 & 49.00312 & -2.42736 & 9.371 $\pm$ 0.012 & 10 & 9.077 $\pm$ 0.013 & 10 & 6.207 $\pm$ 0.018 & 10 & 2.562 $\pm$ 0.017 & 10 & 65.89 & 0.733 & 115.4 \\
NGC1316 & 50.67380 & -37.20796 & 5.058 $\pm$ 0.011 & 10 & 5.098 $\pm$ 0.014 & 10 & 4.259 $\pm$ 0.037 & 10 & 3.176 $\pm$ 0.046 & 10 & 769.06 & 0.645 & 34.7 \\
NGC1377 & 54.16280 & -20.90195 & 9.483 $\pm$ 0.013 & 10 & 8.142 $\pm$ 0.012 & 10 & 4.689 $\pm$ 0.014 & 10 & 1.704 $\pm$ 0.013 & 10 & 72.06 & 0.636 & 92.0 \\
NGC1404 & 54.71632 & -35.59413 & 6.565 $\pm$ 0.014 & 10 & 6.602 $\pm$ 0.016 & 10 & 5.878 $\pm$ 0.074 & 10 & 5.145 $\pm$ 0.067 & 10 & 251.23 & 0.880 & 160.0 \\
IC0342 & 56.70144 & 68.09635 & 3.935 $\pm$ 0.013 & 10 & 3.880 $\pm$ 0.013 & 10 & 0.225 $\pm$ 0.034 & 10 & -1.815 $\pm$ 0.012 & 10 & 778.51 & 0.946 & 75.5 \\
\enddata
\tablenotetext{a}{Measured in the W1 band}
\tablecomments{Table 3 is published in its entirety in the machine-readable format. A portion is shown here for guidance regarding its form and content. A null value in the uncertainty of a quantity when the quantity is non-null indicates an upper limit value for the quantity.\\
The flag for each band indicates if the measurement was from: (0) isophotal aperture of semi-major axis "Radius," axis ratio, and position angle, or (10) the total flux combining the isophotal with the radial surface brightness extrapolation \citep[see][]{jarrett+2019}, or (5) an isophotal aperture with an aperture correction \citep[see][]{Jarrett2023}, or (2) the standard aperture measurement from the ALLWISE measurement, or (1) the point-source measurement from the ALLWISE catalog.}

\end{splitdeluxetable}

\begin{splitdeluxetable}{lrrrrrBrrlrrrr}




\tablecaption{Derived WISE Properties of the SINGS/KINGFISH Sample}

\tablenum{4}

\tablehead{\colhead{Name} & \colhead{D$_\textrm{lum}$} & \colhead{W1-W2} & \colhead{W2-W3} &  \colhead{W1-W3}  & \colhead{W3-W4}  & \colhead{Log L$_\textrm{W1}$}  & \colhead{Log L$_\textrm{W2}$}  & \colhead{Log L$_\textrm{W3}$$^\tablenotemark{a}$} &  \colhead{Log L$_\textrm{W4}$$^\tablenotemark{b}$} & \colhead{Log L$_{\textrm W1_\textrm{Sun}}$$^\tablenotemark{c}$}  & \colhead{Log M$_\textrm{Stellar}$} & \colhead{SFR} \\ 
\colhead{} & \colhead{(Mpc)} & \colhead{(mag)} &  \colhead{(mag)}  & \colhead{(mag)} & \colhead{(mag)} & \colhead{(L$_\odot$)} & \colhead{(L$_\odot$)}  & \colhead{(L$_\odot$)} & \colhead{(L$_\odot$)} & \colhead{(L$_{\odot,\textrm W1}$)} & \colhead{(M$_\odot$)} & \colhead{(M$_\odot$/yr)}  }

\startdata
NGC0337 & 20.62 & 0.231 $\pm$ 0.038 & 3.586 $\pm$ 0.040 & 3.816 $\pm$ 0.037 & 2.369 $\pm$ 0.040 & 9.0612 $\pm$ 0.0099 & 8.7620 $\pm$ 0.0119 & 9.0033 $\pm$ 0.0111 & 9.1125 $\pm$ 0.0120 & 10.4226 $\pm$ 0.0140 & 9.849 $\pm$ 0.082 & 0.9720 $\pm$ 0.2349 \\
NGC0584 & 21.53 & -0.058 $\pm$ 0.035 & 0.359 $\pm$ 0.057 & 0.301 $\pm$ 0.056 & 0.975 $\pm$ 0.108 & 9.7284 $\pm$ 0.0100 & 9.2946 $\pm$ 0.0102 & 4.5717 $\pm$ ...  & 7.3612 $\pm$ 0.0437 & 11.0898 $\pm$ 0.0140 & 10.749 $\pm$ 0.082 & 0.0431 $\pm$ 0.0151 \\
NGC0628 & 10.05 & 0.069 $\pm$ 0.034 & 3.668 $\pm$ 0.036 & 3.737 $\pm$ 0.035 & 1.797 $\pm$ 0.037 & 9.3512 $\pm$ 0.0098 & 8.9836 $\pm$ 0.0100 & 9.2535 $\pm$ 0.0104 & 9.1243 $\pm$ 0.0109 & 10.7115 $\pm$ 0.0139 & 10.229 $\pm$ 0.082 & 1.7232 $\pm$ 0.4193 \\
NGC0855 & 9.93 & 0.057 $\pm$ 0.044 & 2.382 $\pm$ 0.072 & 2.439 $\pm$ 0.068 & 2.166 $\pm$ 0.098 & 8.0809 $\pm$ 0.0106 & 7.7077 $\pm$ 0.0146 & 7.3828 $\pm$ 0.0254 & 7.4705 $\pm$ 0.0309 & 9.4412 $\pm$ 0.0145 & 9.014 $\pm$ 0.083 & 0.0683 $\pm$ 0.0163 \\
NGC0925 & 10.14 & 0.004 $\pm$ 0.035 & 3.259 $\pm$ 0.043 & 3.263 $\pm$ 0.042 & 1.910 $\pm$ 0.072 & 8.9029 $\pm$ 0.0097 & 8.5073 $\pm$ 0.0102 & 8.6025 $\pm$ 0.014 & 8.5813 $\pm$ 0.0258  & 10.2632 $\pm$ 0.0139 & 9.840 $\pm$ 0.082 & 0.7979 $\pm$ 0.1986 \\
NGC1097 & 16.91 & 0.103 $\pm$ 0.035 & 3.408 $\pm$ 0.035 & 3.510 $\pm$ 0.035 & 2.345 $\pm$ 0.036 & 9.9905 $\pm$ 0.0098 & 9.6350 $\pm$ 0.0100 & 9.7998 $\pm$ 0.0101 & 9.8988 $\pm$ 0.0105 & 11.3515 $\pm$ 0.0139 & 10.873 $\pm$ 0.082 & 5.3952 $\pm$ 1.4339 \\
NGC1266 & 28.67 & 0.297 $\pm$ 0.035 & 2.871 $\pm$ 0.038 & 3.169 $\pm$ 0.037 & 3.660 $\pm$ 0.039 & 9.0969 $\pm$ 0.0099 & 8.8174 $\pm$ 0.0102 & 8.7617 $\pm$ 0.0114 & 9.3985 $\pm$ 0.0112  & 10.4591 $\pm$ 0.0140 & 9.899 $\pm$ 0.082 & 1.0374 $\pm$ 0.2615 \\
NGC1316 & 20.88 & -0.026 $\pm$ 0.035 & 0.657 $\pm$ 0.050 & 0.630 $\pm$ 0.049 & 1.259 $\pm$ 0.066 & 10.5457 $\pm$ 0.0099 & 10.1314 $\pm$ 0.0105 & 8.5900 $\pm$ 0.0443 & 8.6443 $\pm$ 0.0218  & 11.9071 $\pm$ 0.0139 & 11.566 $\pm$ 0.082 & 0.3990 $\pm$ 0.0969 \\
NGC1377 & 23.99 & 1.362 $\pm$ 0.035 & 3.434 $\pm$ 0.036 & 4.796 $\pm$ 0.036 & 2.990 $\pm$ 0.036 & 8.9018 $\pm$ 0.0102 & 9.0442 $\pm$ 0.0101 & 9.2442 $\pm$ 0.0105 & 9.5874 $\pm$ 0.0103 & 10.2635 $\pm$ 0.0142 & 9.444 $\pm$ 0.083 & 1.9715 $\pm$ 0.4837 \\
NGC1404 & 20.15 & -0.047 $\pm$ 0.037 & 0.468 $\pm$ 0.082 & 0.420 $\pm$ 0.081 & 0.965 $\pm$ 0.104 & 9.9121 $\pm$ 0.0105 & 9.4988 $\pm$ 0.0109 & 7.6461 $\pm$ 0.0938 & 7.6236 $\pm$ 0.0336 & 11.2734 $\pm$ 0.0144 & 10.929 $\pm$ 0.082 & 0.0708 $\pm$ 0.0178 \\
IC0342 & 3.65 & 0.080 $\pm$ 0.033 & 3.668 $\pm$ 0.045 & 3.748 $\pm$ 0.045 & 2.040 $\pm$ 0.045 & 9.5224 $\pm$ 0.0094 & 9.1360 $\pm$ 0.0097 & 9.4044 $\pm$ 0.0154 & 9.3533 $\pm$ 0.0100 & 10.8821 $\pm$ 0.0137 & 10.394 $\pm$ 0.082 & 2.5307 $\pm$ 0.6219 \\
\enddata
\tablenotetext{a}{W3 band spectral luminosity, Log\,$\nu$L$_\nu$(12$\mu$m), after subtraction of the stellar continuum}
\tablenotetext{b}{W4 band spectral luminosity, Log\,$\nu$L$_\nu$(23$\mu$m), after subtraction of the stellar continuum}
\tablenotetext{c}{W1 in-band luminosity}

\tablecomments{Table 4 is published in its entirety in the machine-readable format. A portion is shown here for guidance regarding its form and content.\\
A null value in the uncertainty of a quantity when the quantity is non-null indicates an upper limit value for the quantity.}


\end{splitdeluxetable}

\end{document}